\documentclass[a4wide,12pt,amstex]{article}
\usepackage{amsmath}
\usepackage{theorem}
\usepackage{makeidx}
\usepackage{fontenc}
\usepackage{amscd}
\usepackage{array}
\usepackage{graphics}

\usepackage{amsfonts}
\usepackage{amssymb}
\usepackage[cp1251]{inputenc}
\usepackage{theorem} 
\usepackage{makeidx}
\usepackage[english]{babel}

\begin{document}
\thispagestyle{empty}
\begin{center}
{\LARGE  Sofia University \\
"St. Kliment Ohridski" \\
\vspace{0.5cm}
\textsl{Faculty of Mathematics and Informatics} \\
\vspace{0.6cm} \normalsize

\verb"master programme:" \\
\vspace{0.2cm}
\textsc{\textbf{\large 'Mathematics and Mathematical Physics'}}\\
\vspace{2.3cm}

\textsf{MASTER THESIS}: \\
\vspace{1.0cm} {\Huge \bf \emph Spectral Geometry and the Semi-classical Approach}}\\
\vspace{1.5cm}
{\Large \textsc{Danail Brezov }  }\\
\vspace{0.3cm}
{
\textsc{M} \verb"21587" }\\
\vspace{2cm}
\verb"scientific advisor:"
\\
\vspace{0.2cm}
\verb"dr." {\Large \textsc{Petko Nikolov} \\
\vspace{2.5cm}
\normalsize
Sofia 2007}
\end{center}

\newpage

\tableofcontents

\newpage

\section{Introduction}
Both Spectral Geometry and Semi-classics seem to be concerned with more or less the same thing - studying the correspondence between classical-dynamical and geometrical quantities, such as the metric, the geodesic flow or the effective potential for example,
on the one hand, and on the other - the spectrum of a Laplace or Dirac-type operator, acting on the space of square-integrable functions on the given manifold. \\
However it seems that each theory has its own independent motivation - the physicists' approach, referred to as 'semi-classical' is, widely speaking, a manifestation of the famous Bohr's correspondence principle\footnote{postulating equivalence between the classical and quantum description in the limit $\hbar\rightarrow 0$ and certain analogy close to that limit - a good example of its validity is a famous theorem stating that the mean values of the quantum probabilities evolve according to classical laws}, while Spectral geometry, representing the mathematician's point of view, is mostly dedicated to resolving inverse problems, or to the contrary, finding counterexamples. The essence of a typical inverse spectral problem is most naturally expressed by the question posed by \emph{Mark Kac} in 1966, 'Can one hear the shape of a drum?', that is usually considered the birth of Spectral geometry. The idea is the following: given a smooth compact Riemannian manifold, one may construct, with the help of its metric, an unique \emph{Laplace-Beltrami} operator, which is elliptic, self-adjoint, with discrete, definite spectrum, that obeys certain growth rate estimates\footnote{in order to have all these properties we assume reasonable boundary conditions in the non-closed case and as for the definiteness, only a finite part of the spectrum can be with inverse sign in the Neumann case}. It is then quite natural that the geometry of the underlying manifold determines the spectrum. The question is to what extend the same is true in the reverse order, or, in other words, can we retrieve from the spectrum a sufficient amount of geometrical data in order to determine completely the metric or the shape of the boundary? It turns out that most typically the answer is negative. Even in dimension two there are examples of iso-spectral domains that are not isometric. Nevertheless, there are few cases of resolved inverse problems, such as low dimensional spheres and tori ($d<7$), hyperbolic surfaces, real analytic surfaces of revolution and the result of \emph{Guillemin} and \emph{Melrose}, that is our main example.
The methods for obtaining such results demand solid background in geometry, mechanics, $PDE$'s and pseudo-differential calculus, as well as several more specific tools. That is why we dedicate the first half of the present work to the introduction of some preliminaries that are crucial for understanding the framework of the problem and apart from that give some physical insight. \\
The second half already goes straight to the topic with the introduction of trace formulae, that appear to be a major tool both in Semi-classics and Spectral Geometry. These equalities relate infinite sums over closed classical paths to traces of functions of the laplacian. Some of these are exact (\emph{Poisson} and \emph{Selberg}), while others (\emph{Gutzwiller}, \emph{Berry-Tabor}) rely on semiclassical asymptotic expansions. This common apparatus is one more reason to regard the two theories as being closely connected - one may treat the on the one hand semiclassical analysis as a tool of Spectral geometry, and on the other - consider a spectral-geometrical problem in the semiclassical context.\\
The heat trace is somewhat exceptional - it expands the heat propagator as a power series in the high-temperature limit. As heat is diffusive, this expansion could not be related to classical hamiltonian dynamics, but is still rather sensitive to geometry - symbolic calculus yields infinitely many terms involving the metric derivatives when constructing a formal inverse and the coefficients appear to be linear combinations of the polynomial curvature invariants of the corresponding degree. This formalism appears to be very fruitful in rather unexpected directions. As we discuss in the following, it gives a regularization scheme and a handy tool for calculating conformal anomalies in quantum gravity, provides a simple procedure for obtaining the \emph{Korteweg-de Vries} hierarchy and the infinite set of first integrals, but finds also an application in Index theory, statistical mechanics and others.\\
Next we introduce quantum billiards as typical model systems in both theories. In spectral geometrical context they are considered as Euclidean domains in the context of the inverse spectral problem, while semi-classical (or geometrical-optical) approach treats them as quantum dynamical systems close to the classical limit. The most familiar interpretation is a particle captured in a box with infinite potential walls (\emph{Dirichlet} boundary conditions). This setting, no matter how conceptually simple, appears to be a powerful tool for quantizing classically chaotic systems. Apart from this, spectral statistics can be used for retrieving qualitative information about the underlying dynamics, as we shall see in the following. \\
In the last section we discuss some strategies and techniques for solving inverse spectral problems and provide concrete examples in which they have found successful application. At the end we review the \emph{Guillemin-Melrose} problem\footnote{The authors show that the function, participating in the homogeneous Robin boundary condition for the region in the inside of an ellipse is completely determined by the laplacian spectrum (more precisely, the resolvent trace built with its help) together with the spectrum of the pure \emph{Neumann} problem} and calculate the spectral variation with respect to varying the function in the homogeneous Robin boundary condition. The corrections, expressed as single-layer potentials, are found by using a slight modification of the standard in quantum mechanics apparatus of perturbation theory.

\newpage

\section{Dynamics and Geometry}
The classical dynamics of a physical system is usually derived from variational principle. According to that formalism, we prescribe for example to a particle, constrained to move on a surface $X$ and sensitive to external potential $V$, a \emph{Lagrangian function}\footnote{in general we define a Lagrangian density, but in the one-dimensional case, both constructions are certainly equivalent}
$$\mathcal{L}(q,\dot{q}) = \frac{m\dot{q}^2}{2} - V(q)$$
where ($q$, $\dot{q}$) are the local coordinates in $TX$.\\
With the help of $\mathcal{L}$ we easily construct the \emph{action functional}
$$S = \int_{t_0}^{t_1}{\mathcal{L}(q,\dot{q})\,d\tau}$$
as s function over the space of paths with fixed endpoints $q_0 =  q(t_0)$ and $q_1 = q(t_1)$ respectively. The classical path is chosen as the critical point of $S$, that is the one for which the action is stationary, that is $\delta S = 0$. Substituting the above definition of $S$ gives, after integration by parts
\begin{equation}
\frac{d}{dt}\left ( \frac{\partial \mathcal{L}}{\partial \dot{q}}\right ) = \frac{\partial \mathcal{L}}{\partial q}
\end{equation}
which is known as the Euler - Lagrange equation of motion.\\
With our particular choice of Lagrangian, the Euler-Lagrange Equations certainly coincide with the Newton second principle
$$\ddot{q} = -\nabla V(q) $$

\subsection{Hamiltonian Dynamics}
Usually, however it is most convenient to describe such systems by means
of Hamilton formalism. In order to switch from Lagrangian to Hamilton formulation of dynamics, we need to consider momenta $p$ instead of velocities $\dot{q}$, that is moving from the tangent to the co-tangent bundle $TX\rightarrow T^*X,\: \dot{q}\rightarrow p$.\\ The Lagrangian $\mathcal{L}(q,\dot{q})$ undergoes a non-trivial change, known as the \emph{Legendre transform} defined as
$$H(q,p) = p\,\dot{q} - \mathcal{L}(q,\dot{q}),\quad \dot{q} \rightarrow p$$
The Lagrangian equations of motion, derived from variational principle are then equivalent to the canonical equations of Hamilton
\begin{eqnarray}\label{ham}
\dot{p}=-\frac{\partial H}{\partial q} \qquad\qquad\qquad
\dot{q}=\frac{\partial H}{\partial p}\label{1}
\end{eqnarray}
where $q$ and $p$ are referred to as the generalized
\emph{coordinate} and \emph{momentum} respectively.\\
One of the reasons to prefer this representation is that it makes
rather explicit some nice properties of the conservative systems
and is very often more convenient for various calculations, including quantization.
This relief is due to the elegant framework of symplectic geometry we are allowed
to use here. The key is that the cotangent bundle $T^*X$, unlike
the tangent, possesses a natural symplectic structure (a non-degenerate
closed two-form), that in the above coordinates is given locally
by $\omega = \rm d p \wedge \rm d q$. Such simple form of $\omega$
is called \emph{canonical} and so is the pair $p, q$. The
canonical pair is however not unique - we can always choose
another pair of canonical variables in which the form $\omega$
looks in the same way. If the system is integrable, we can always
find one pair $J, \phi$, such that the hamiltonian depends only on
the $J$'s, but not on the $\phi$'s and thus the equations (\ref{ham})
look much simpler:
\begin{eqnarray}
\dot{J} = 0 \qquad\qquad\qquad \dot{\phi} = \nu
\end{eqnarray}
The first is obvious and in the second $\nu$ is a constant,
since the $J$ - variables are all constants of motion and $H$
depends only on them.\\
Such variables are called \emph{action-angle variables} and
provide in many cases a powerful tool for integrating the
hamiltonian system - each $J_k$ reduces the phase space to the
hyper-surface $J_k = const$ and the final reduction gives the
unique solution. Of course, one can choose $J_1=H$, since the
'total energy' of such systems is always preserved, and then
follow a standard procedure for construction of canonical basis,
analogous to the method of Gram and Schmidt in the metric case.\\
If, however, one has found a full set of action variables $J_i$,
it is not evidence enough that the system is integrable: for
example, they may appear to be functionally independent
and the reduction in this case is not full. Another problem might
appear if the reduction procedure is not commutative. Therefore we
put the restrictions that the $J_i$'s must be in involution with
respect to the symplectic form, $\omega^*(J_i,J_k)=0$ and
that the differentials $\rm d J_k$ must be all linearly independent.
Under these conditions, the existence of the action-angle basis is
already sufficient for a hamiltonian system to be integrable.\\
Sometimes it is more convenient to write the evolution equations
using the Poisson bracket defined as follows
\begin{equation}
\{F,G\} = \omega^*( d F,  d G) = \sum_k^n{\frac{\partial
F}{\partial p^k}\frac{\partial G}{\partial q^k} - \frac{\partial
G}{\partial p^k}\frac{\partial F}{\partial q^k} }.
\end{equation}
Using this formalism it is easy to express the equations of motion
(\ref{ham}) in terms of this new structure as
\begin{equation}
\dot{p}= \{H,p\}\qquad \dot{q}=\{H,q\}
\end{equation}
and in general, each physical quantity in a system with
hamiltonian $H$ is transformed with respect to the flow defined
infinitesimally by $\dot{x}=\{H,x\}$. Therefore we have that each
quantity that commutes with the hamiltonian the system (with
respect to the Poisson bracket) is preserved by the evolution,
defined by this hamiltonian. Such quantities are called
\emph{conservation laws} and the hamiltonian itself is one of
these due to the fact that our structure is skew-symmetric.
Moreover, if we present another hamiltonian whose action preserves
the initial one, then obviously the new hamiltonian is a
conservation law of the initial system (a version of N\"other's
theorem).\\
Thus the problem of finding a solution of an
integrable hamiltonian system is equivalent to the problem of
finding a complete set of pairwise commuting linearly independent
hamiltonian vector fields. In practice, however, it is often not an
easy task and a lot of techniques have been developed to overcome
technical difficulties. It is out of our aim to go into detail
with this matter, but just as an illustration of a possible hidden
trap for beginners: suppose that we know two of the conservation
laws of an integrable hamiltonian system of dimension $n>2$ and we
wish to generate the others. A straightforward method is to use
the Poisson bracket - since this structure fulfils the Jacobi
identity we have a series of commutators of the two preserved
quantities, that are preserved as well. Nothing, however, can be said
about their functional independence, so there is some extra work to be done.\\
In order to make things clearer, we may think of the dynamical
equations (\ref{1}) as of a vector field $X_H$, associated with the
hamiltonian of the system and therefore, carrying the name
\emph{hamiltonian vector field}. We write instead of (\ref{ham})
\begin{equation}
X_H(\Phi)=\omega^*(dH,d\Phi)=\{H,\Phi\}
\end{equation}
The flow of $X_H$ then defines the evolution of the hamiltonian
system. In particular, this flow preserves the symplectic form as it is easy to be seen
by the expression for its \emph{Lie derivative}\footnote{this
formula, due to Cartan gives the derivative of a quantity in
direction to the flow, defined by $X_H$ (Lie derivative of a
vector field). Here $i_{X_H}$ is an operator, fixing the first
argument equal to $X_H$}
\begin{equation}
L_{X_H}\omega=i_{X_H}\circ d\omega+d \circ i_{X_H}\omega
\end{equation}
The first argument is zero, since $\omega$ is closed. The second
one can be transformed to $d^2 H = 0$ by using that on the flow, defined by $H$ we have $$i_{X_H}\omega = dH$$
that is one way to write the canonical equations.\\
Since $L_{X_H}$ is a linear derivation, the hamiltonian flow
preserves, together with $\omega$, all of its degrees up to $n$,
which means all even dimensional projections of the phase volume,
including the volume itself
\footnote{ note that $\omega^n\sim d{\rm Vol}(X)$ since $\dim {\Omega^n}(X) = 0$ and note also that this imposes one nontrivial restriction on the field $X_H$, namely ${\rm div} X_H=0$}. Therefore, it is not surprising that in hamiltonian
systems asymptotic convergence is not possible (unlike Liapunov),
nor are any types of attractors. All we can have are either
integrable (or quasi-integrable) or
chaotic, usually ergodic types of motion.\\
Another footnote to make here is that the notion of a hamiltonian
vector field makes the analogy between Poisson brackets and
commutators far more explicit. Namely, if a given quantity, say
$F$, commutes with the hamiltonian $H$ of the system (in the sense
that $\{F,H\}=0$), it may be regarded as a hamiltonian
associated to a flow
$$F\rightarrow X_F\rightarrow \phi_s$$
that leaves our initial system undisturbed, which can be written as
$[X_H,X_F]=0$ and vise versa: each time the latter condition
holds, we have a preserved quantity $F$ for $X_H$ (or $H$ for
$X_F$) which can be found by exploiting the differential of the
action, known as the \emph{integral invariant of
Poincar\'{e}-Cartan} \begin{equation} dS=p\,dq-Hdt \label{PC}
\end{equation} Since, on the one hand, $\phi_s$ is a flow that
leaves $H$ and therefore its Legendre transform - the Lagrangian
invariant, and on the other $t-$ and $s-$ derivatives commute (due
to the vanishing of the above commutator), we may write
\begin{equation}
0=\frac{d}{ds}\frac{dS}{dt}=\frac{d}{ds}\left (
p\,\dot{q}-H \right )=\frac{d}{dt}\left(
\dot{q}\frac{dq}{ds}-H\frac{dt}{ds}\right )
\end{equation}
and the expression in brackets on the righthand side is a constant
of motion.\\
Let us go back to the symplectic structure $\omega$. Note that it
splits the $2n$-dimensional phase space into two subspaces on
which it vanishes (in our initial basis it were the coordinate
base and the cotangent momentum subspace). When we change
coordinates the splitting may not be so simple, but each time we
have an $n$-dimensional submanifold, the restriction of $\omega$
on which is identically zero (take for example the submanifold,
defined by the constants of motion $J_k$)
we call it a \emph{Lagrangian submanifold}.\\
Lagrangian submanifods, as we shall see shortly, are crucial both
for dynamics and geometric optics. One way to construct such
manifold is to start with one that we know (say the $q$-space) and
make it flow according to a \emph{canonical change of variables}.
The latter means simply that $\omega$ should be preserved by the
so chosen flow. The group, that is responsible for such
\emph{canonical transformations} is called \emph{real symplectic
group} and denoted usually by $\mathsf{Sp}(2n,\mathbb{R})$.\\

\paragraph{Generating functionals and Hamilton-Jacobi equations:}
In practice, the canonical change of variables, we are interested in, is being made via
generating functional. The main idea roots in the observation that
a coordinate transition $\phi :\, (p,q) \rightarrow (P,Q)$ is
canonical if the co-homological class of the momentum one-form is
preserved under this transformation, that is
$$p\,dq = P\,dQ + d\Phi$$
One can easily assure oneself by differentiating both sides of the
above equality that, since $d^2 \Phi = 0$, $\phi$ is volume -
preserving and hence, canonical. The potential $\Phi$, expressed
in $(q,Q)$-variables\footnote{this can always be done as long as the
Jacobian $\displaystyle{{\rm det} \frac{\partial Q}{\partial p}}$
is non-vanishing, or, equivalently, the Hessian of $\Phi$ is
non-degenerate}, is called a \emph{generating
functional} for the transformation $\phi$ and appears to be a useful tool for the
description of the canonical equations in the new variables.
Moreover, the flow $\phi$ induces its pull-back, acting on the
hamiltonian functions $\phi^*:\, H\rightarrow \Tilde{H}$ which can
be used as an alternative description of the theory. In
particular, we may use this pull-back to find the generating
functional of a transition to action-angle variables and then
express the hamiltonian in these new variables - that would simply
be $\Tilde{H}(Q)$. Since, by the definition of a generating
functional we have
$$p=\frac{\partial \Phi}{\partial
q},\qquad P=-\frac{\partial \Phi}{\partial Q}$$ our aim is to find a
solution of the non-linear $PDE$
\begin{equation}
\Tilde{H}(Q)=H(\frac{\partial \Phi}{\partial q},q,t)
\end{equation}
known as a \emph{Hamilton-Jacobi} type equation.\\
The action variables $Q$ are the constants of motion defined by
$\displaystyle{p=\frac{\partial \Phi}{\partial q}(Q,q)}$ and the
linear flow on the 'angles' is naturally given in terms of the new
hamiltonian:
$$\dot{P}= - \frac{\partial \Tilde{H}}{\partial Q}$$
Thus, resolving the above non-linear $PDE$ is equivalent to
finding the trajectories of the initial Hamiltonian
system. This equivalence can be utilized in both directions, but,
surprisingly enough, it turns out to be the most powerful tool
known so far for integrating the canonical Hamilton equations of
motion.\\
At the end we mention that the classical action also satisfies a
Hamilton-Jacobi type equation. Namely, it is straightforward to
check that
\begin{equation}
\partial_t S + H(\frac{\partial S}{\partial q},q,t)=0
\end{equation}
due to the explicit expression for the Poicar\'{e} - Cartan
integral invariant (\ref{PC}). As we shall see shortly, this leads
to an analogy between the description of the dynamics of a
classical particle and that of phenomena from geometric optics.

\subsection{The Geodesic Flow and Geometric Optics}
A good example of a hamiltonian system can be seen in the problem of a free
particle, constrained to move on a Riemannian surface $M$. It is
well known that all possible trajectories are namely the geodesics
on $M$. Thus the concept of a geodesic flow is somewhat central
for classical mechanics, let alone that even if we have
interaction in the system, its potential might be 'transformed' into
some effective curvature of the underlying manifold, so at the end
we have again the problem of a free particle. This is the case with
the two-bodies problem in physics - the solutions are easy to
obtain as geodesics over the sphere.\\
To put these thoughts in order, let us first consider the metric
Lagrangian $\displaystyle{L=\frac{1}{2}\langle \dot q,\dot q
\rangle }$, which is nothing but the lagrangian of a free particle
of unit mass, constrained to our manifold. The corresponding
hamiltonian is then given by $\displaystyle{H=\frac{1}{2} \langle
p,p \rangle}$. This immediately
yields conservation of momentum and extremal trajectories. \\
Another way to find the equations of geodesics is using the
concept of covariant differentiation on $M$ which is locally given
by $$(\nabla v^i)_j=\partial_j v^i + \Gamma^i_{jk}v^k$$ where the
additional term accounts for the transformation of the local
basis: $\Gamma^i_{jk}=\partial_j {\bf e}^i_k$ is called the
Lie-Christophel Symbol. With a suitable choice of the evolution parameter the equation of the geodesics looks like
\begin{equation}\displaystyle{\ddot v\,^i +
\Gamma^i_{jk}\dot v\,^j\dot v\,^k = 0}
\end{equation}
Both formalisms are certainly equivalent and we have the choice
between the former Lagrangian (or Hamiltonian) representation,
fully in the spirit of extremal principle and the latter, which
accounts for the fact that the tangents to the geodesics are
auto-parallel\footnote{trivially translated along their own integral curves}.\\
Let us consider now a wave function with rapidly
varying\footnote{compared to the alterations of the potential, or
the geometry} phase $\Phi$. When substituted into the
Schr\"{o}dinger equation, semi-classical approximation gives up to
first order in $\hbar$ a Hamilton - Jacobi equation in the form
\begin{equation}
\partial_t \Phi + H(\frac{\partial \Phi}{\partial q},q,t)=0
\end{equation}
This means, on the one hand, that semiclassical 'trajectories' of
photons are given by Hamiltonian canonical equations with
principal function $\Phi$ and, on the other, these trajectories
are all perpendicular to the wavefront $\Phi = const$ (Huigens
principle). There is therefore a kind of dualism between
the description of geometric-optical phenomena by means of their wavefronts, or by
the \emph{light rays}.\\
In the present considerations, we pay much more interested to the
second interpretation, as leading to more obvious optical -
mechanical analogy. For example, by Fermat principle, light
trajectories minimize the time of arrival, which is proportional
to the functional known as \emph{optical distance}
$$ S = c^{-1}\int{n(l)dl}$$
where $n$ is the refraction coefficient of the media and $dl$ -
the length element. In particular, for homogenous media ($n=1$),
those trajectories are nothing but the geodesics. For example, it
is well known from general relativity, that on the large scale,
the light travels through the Universe, following the space-time
geodesics with respect to the metric, 'curved' by the presence of
matter, according to the famous \emph{Einstein - Hilbert
equations}\footnote{stating that the Riemann curvature tensor of
space-time is proportional to the stress-energy tensor of the
classical fields (which is, on the large scale, mostly dominated
by gravitation)}.

\subsection{Morse Theory and Catastrophes}
We take as our main example the geodesic flow or, alternatively,
the geometric optical problem concerning ray families. Both
descriptions happen to give somewhat equivalent picture due to
\emph{Fermat} principle of stationary action.\\
To begin with, let us recall the basics from the classical \emph{Morse
theory}. The first crucial statement of the theory is that on each
compact Riemannian manifold $M$ there exists a smooth function $f$
with only simple, and thus isolated, critical
points\footnote{the point $x_0$ is called \emph{critical} for
$\nabla f=0$ at $x_0$ and \emph{simple}, or non-degenerate if in
addition to that the Hessian $\nabla\nabla f$ is non-degenerate at
this point}, we refer to as \emph{Morse function}. Then at each such critical point the $f$ is locally isomorphic to a quadratic form, so that local extrema are given as
minima, maxima and saddle points only. Define the index of this
quadratic form, called \emph{the Hessian} form, as the
number of its negative eigenvalues. Then the main theorem of the
classical Morse theory states that if to each critical point $x_j$
we assign a cell of dimension equal to the index of ${\rm Hess}(f)$
at $x_j$ and then 'glue' all the cells starting from the lowest
dimension\footnote{the boundary of each cell is glued to the cells
of lower, but maximal possible dimension: for example a $1$-cell
glued to a $0$-cell is isomorphic to a circle, a $2$-to-$1$-gluing
gives a disk and $2$-to-$0$ makes a $2$-sphere etc.} we obtain a smooth
manifold with the same homotopy type as $M$.\\
The simplest examples are, as usual, the two-dimensional sphere
and torus, with the height function. For the sphere we obviously
encounter a $0$-cell (for the minimum) and a $2$-cell (for the
maximum) which is all we may ever expect. On the torus (if we make
it 'stand' properly) we have two saddles plus the above ones,
which gives two circles attached to a common point and then
'dressed' in a $2$-cell skin.\\
\paragraph{Back to the geodesic flow:} Now consider the manifold of all piecewise smooth curves over a
Riemannian manifold $M$ and denote it by $\Omega(M)$. Now
functionals over $M$ are recognized as functions in $\Omega$. In
particular the geodesic action
$\displaystyle{S=\frac{1}{2}\int_{\tau_0}^{\tau_1}{||\dot{\gamma}(\tau)||^2
d\tau}}$ is one such function and we are going to pay special
attention to it in the rest of this section.\\
The variation of the classical functional $S(\gamma)$ is generated
by vectors in $T_\gamma\Omega$. One such vector is the velocity
field $v=\dot{\gamma}$ along the path $\gamma$, or we may choose a
transversal direction, say $u\in T_\gamma\Omega$. The critical
points of the functional with respect to one such variation give
the stationary paths of a free particle, bound to move on $M$. The
Euler-Lagrange equation defines the geodesic connecting
$\gamma(\tau_0)$ with $\gamma(\tau_1)$. This geodesic however does
not have to be unique - what happens in practice is that
uniqueness is fulfilled naturally for close enough initial and
final point, but globally it is no longer so. For instance there
are infinitely many geodesics, connecting the south and the north
pole of a round sphere. In order to understand this phenomenon one
has to fix a geodesic and take into account the second variation
of $S$ around it as well. Uniqueness is required whenever the
Hessian\footnote{for derivation of this formula one may refer to
the famous Milnor's comprehensive book 'Morse Theory'}
\begin{equation}
\delta^2_{uv}S(\gamma)=-\int_{\tau_0}^{\tau_1}{\langle
v,\nabla^2_{\dot{\gamma}}u+R(\dot{\gamma},u)\dot{\gamma}\rangle
d\tau}
\end{equation}
is strictly definite, but as long as one of the eigenvalues
changes sign this uniqueness is broken. To be just a little bit
more precise, at the moment of the sign transition the Hessian is
degenerate, meaning that the field $u$ satisfies the \emph{Jacobi
equation} \begin{equation}
\nabla^2_{\dot{\gamma}}u+R(\dot{\gamma},u)\dot{\gamma}=0
\label{Jac}
\end{equation}
The field $u$ is called a \emph{Jacobi field} and it deforms the
initial geodesic $\gamma$ into a smooth one-parameter family of
geodesics $g(s,\tau)$. We just outline the proof which comes
together with the proof of the above formula (\ref{Jac}). \\
The first variation of the action gives the Euler -
Lagrange equation of the form
$$
\delta_v S(\gamma)=\int_{\tau_0}^{\tau_1}{\langle
v,\nabla_{\dot{\gamma}}\dot{\gamma}(\tau)}\rangle d\tau=0
$$
since the lagrangian is independent of coordinates\footnote{this
is easy to derive by fixing the endpoints and integrating by
parts}. Then, varying second time in the $u$-direction
($u,\,v\in T\Omega(M)$) and assuming that both variations are
independent, we simply arrive at the second order equality
$$
\delta^2_{uv}S(\gamma)=\int_{\tau_o}^{\tau_1}\langle v, \nabla_u
\nabla_{\dot{\gamma}} \dot{\gamma}(\tau) \rangle d\tau
$$
Now, we have to keep in mind that
$$
\nabla_u
\nabla_{\dot{\gamma}}\dot{\gamma}(\tau)=\nabla_{\dot{\gamma}}
\nabla_u+R(u,\dot{\gamma})
$$
provided $[u,v]=0$, which can be easily arranged at least locally.
Secondly, due to equality of the second derivatives, we have
$\nabla_u \dot{\gamma} = \nabla_{\dot{\gamma}} u$ and thus we
easily recover (\ref{Jac}) and besides, if one seeks for a smooth
deformation leaving the geodesic property of $\gamma$ invariant,
that is $\nabla_{\dot{\gamma}} \dot{g}(s,\tau)=0$ preserved in the
$u$-direction, after a while one comes upon the requirement that
$u$ is a Jacobi field and vice versa - to each Jacobi field at
certain geodesic $\gamma$, one may associate a one-parameter
smooth deformations, preserving its geodesic property.\\
Now, given a geodesic $\gamma$, two points on it are called
\emph{conjugate} if there exists a Jacobi field on $\gamma$ which
vanishes at both points. The number of such linearly independent
fields is called \emph{multiplicity} of this conjugacy. This
number is always finite since it must be less than the
dimension of the underlying manifold. Then the \emph{Morse
index} of the geodesic $\gamma$ is defined as the number of
conjugate points along $\gamma$, counted with their multiplicities
and this number is equal to the index of the quadratic form
$\nabla\nabla S$ at the critical point $\gamma$. Furthermore, the
main result of the so extended theory states that chord space of
geodesics with given endpoints is homotopy equivalent to a union
of cells of dimensions, equal to the Morse indices of all possible
geodesics having the same endpoints fixed.\\
First we note that the way we defined the Morse index of the geodesic flow, it happens
to coincide with the \emph{Maslov index} of the Hessian of $S$,
defined as the number of sign changes of the eigenvalues from
negative to positive minus the number of opposite changes.\\
Second, we wish to discuss in brief the Jacobi equation (\ref{Jac}).
It is a second-order evolution equation that describes the
behavior of the geodesic flow
on $M$. Let us take the two-dimensional compact surfaces as
an example. The constant scalar curvature allows for writing
(\ref{Jac}) in the form $\ddot{u}+\kappa u=0$, where $\kappa$ is the
scalar curvature and $u$ is the vector of deformation chosen to
point in a direction normal to the one defined by the evolution
parameter. Then for the sphere the solutions are oscillating,
which is a hint for the presence of conjugate points\footnote{more
generally on $\mathbb{S}^n$ the two poles are conjugated with the
maximal possible multiplicity $n-1$}. On the torus, as it should
be expected, the geodesics are parallel, so the flow is uniform,
while on hyperbolic surfaces it is exponentially divergent. This is a nice
illustration of the fact that whenever the sectional curvature in
all two-directions is strictly negative, all Morse indices are
zero\footnote{this result is far not difficult to prove together
with its almost immediate consequence that closed geodesics on
closed manifolds of negative curvature are isolated, so that there is
only one in each homotopy class - a simple way to do this relies on Bochner technique}.\\
The theory has two quite natural extensions - one of them is to
consider closed orbits, but them one needs to be careful about
periodic conditions imposed on the deformations and possible
complications due to revolutions of simple periodic orbits. The
results, however, are quite analogous to what we have seen so far. \\
Another generalization is to consider manifolds with billiard
types of geodesics, where $\dot{\gamma}(\tau)$ may suffer
noticeable jumps. In this case there are additional terms in the
expressions for the action variations, proportional to these jumps
(and their square). Since the results are quite analogous and nothing conceptually new is obtained, we omit these considerations here, although we shall need the notion of Morse index namely in this case.\\
Although a lot of progress has been done so far in Morse theory - there are generalizations to gauge and super-symmetric theories, we modestly refer to the basics, given in \cite{M1}. For Hamiltonian mechanics and symplectic geometry, a good choice would be \cite{A1} together with the appendix. The Hamilton-Jacobi method can also be found there.
\paragraph{The Poincar\'{e} map} Now is the time to briefly introduce the  \emph{Poincar\'{e} map} for periodic orbits. Given one such orbit $\gamma$ with period $T$, so that $\gamma(T)=G^T\gamma(0)=\gamma(0)$, one takes a point $x\in\gamma$ and considers a symplectic  transversal $S_\gamma$ to $\gamma$ at $x$. Then the Poincar\'{e} map is defined as the first return map to $S_\gamma$ :
$$ \Tilde{P}_\gamma: S_\gamma\rightarrow S_\gamma, \quad \Tilde{P}_\gamma(\zeta) = G^{T(\zeta)}(\zeta)$$
where $T(\zeta)$ is the smallest time it takes $\zeta$ to return to $S_\gamma$\\
The \emph{linear Poincar\'{e} map} is then $P_\gamma = d\Tilde{P}_\gamma$.
Sometimes it is convenient to define ${P}_\gamma$ on the span of complex normal Jacobi fields along $\gamma$. It is a symplectic vector space with respect to the Wronskian $$ w(X,Y)= g(X, D_s Y)-g(D_s X, Y)$$
Then $P_\gamma$ is defined on this space simply by
$$ P_\gamma Y(\tau) = Y(\tau + T)$$
Note that the abstract definition we gave for $P_\gamma$ is often reduced to rather intuitive constructions. For example, in the case of convex billiards, we take the symplectic transversal (or \emph{Poincar\'{e} section}as they usually call it) to be the domain boundary, fibred with the normalized momentum (in more mathematical terms the unit ball bundle $\mathcal{B}\partial \Omega$) and then one defines ${P}_\gamma$ as the \emph{first-return map} of a neighborhood of an initial point $\xi_0$ in $\mathcal{B}\partial \Omega$. It could be interpreted also as an action of monodromy around a given closed geodesic. More precisely $P_\gamma$ maps the variation $\delta\xi_0 = (\delta q_0,\delta p_0)$ into $\delta \xi$ relative to a circuit around the closed path $\gamma$.\\
One major advantage to describe the billiard flow by means of the Poincar\'{e} map is that in this way one reduces the $n$-dimensional dynamical problem to the study of a discrete iterative map on the boundary. As we shall see in the following, this is a great relief.\\
The eigenvalues of the linear symplectic map $P_\gamma$ help us to distinguish between different types of closed geodesics. For example, $\gamma$ is said to be \emph{non-degenerate}, if $\det ({\rm id}-P_\gamma)\neq 0$ or, in other words, unity does not belong to $Spec (P_\gamma)$. Such non-degenerate closed orbits fall in one of the following classes:
\begin{enumerate}
\item \emph{elliptic} if all eigenvalues are of modulus one, so they come in families $e^{\pm i\alpha_j},\: \alpha_j\in \mathbb{R},\: j=1..n $
\item \emph{hyperbolic} if all eigenvalues are of the type $e^{\pm \mu_j},\: \mu_j\in \mathbb{R}$
\item \emph{loxodromic} if $\lambda_j = e^{\pm i\alpha_j\pm \mu_j} $
\end{enumerate}
Since $P_\gamma$ is symplectic, no other possibilities are available.
\begin{figure}\centering\resizebox{9cm}{!}{\rotatebox{0}{\includegraphics{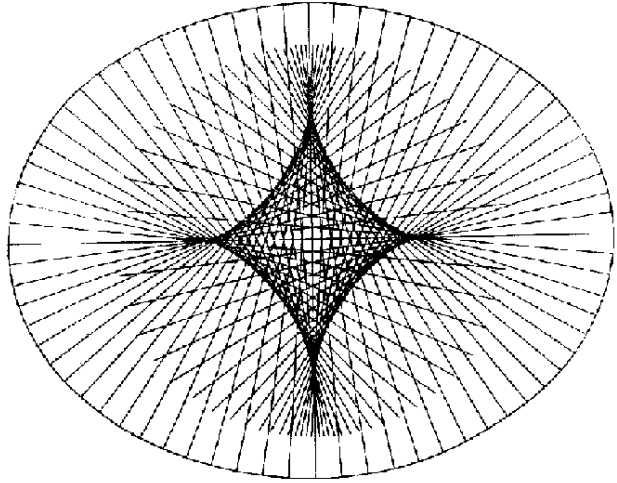}}}\label{evolute}\caption{\small The evolute of the ellipse is built as the envelope of all normals and is thus a caustic for the normal fibration}\end{figure}
\paragraph{Lagrangian singularities} So
far we managed somehow to avoid the question of degenerate
singularities both in our geometrical and dynamical
considerations. This is the point where Morse theory fails and we
need new tools to investigate in depth such structures. These
structures themselves carry the curious name \emph{catastrophes} and
have been somewhat unrevealed mystery of geometry until the second
half of $20^{th}$ century when mathematicians at last came up with
a consistent theory on the matter.\\
To illustrate better the idea of what a catastrophe might be, let
us first give an example. Consider an ellipse in the plane and a
nice Morse function on it, which will be the distance to a fixed point
in the inside region. Now suppose that, by some unlucky
coincidence, one has chosen a point which is a center of curvature.
In this case the distance is no longer Morse function, since the ellipse
looks like a circle up to second order from one such point of view. As it is
well known, all centers of curvature of an ellipse form a
piecewise smooth curve, which is also the envelope of all lines
normal to the boundary, that is called \emph{the evolute} of the ellipse. \\
As we may see from this example, unlike Morse critical points,
degenerate singularities need no longer be isolated, but could
come in various co-dimensions instead.\\
\begin{figure}\centering\resizebox{9cm}{!}{\rotatebox{0}{\includegraphics{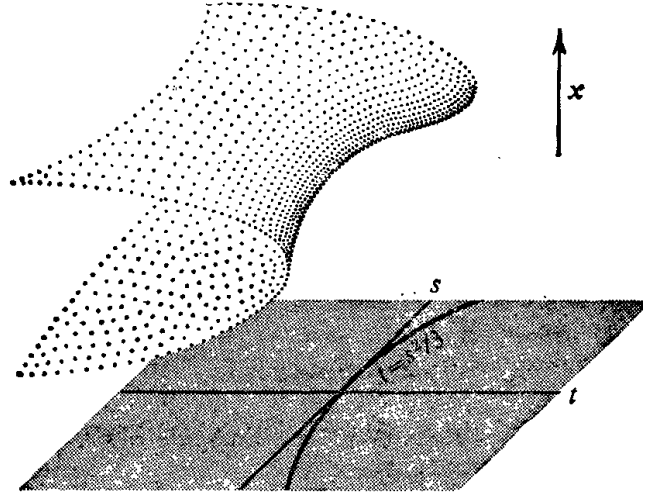}}}\label{projection}\caption{\small The simplest catastrophe - the \emph{fold} in two dimension, obtained by projection}\end{figure}
\noindent Another typical example of a catastrophe are the caustics in
geometric optics. Caustics are, widely speaking, envelopes of ray
families and they are usually visible as a projection of focused
beams. The first example that comes to mind is the focused through the reading-glass
sunbeam that burns the paper. But this type of singularity is
highly unstable - we can easily distort it by a small perturbation. Then it starts to look like a \emph{cusp} - the curve from the previous example.\\
All classes of topologically stable caustics, however, at least in low dimensions\footnote{more precisely, dimensions up to twenty-two and co-dimensions - up to seven}, have been studied and can be locally represented by polynomials. Let us become familiar just with the first few of them - the rest can be easily found in almost every book or article on catastrophe theory.\\
In the maximal codimension we have only two possibilities -
\emph{the fold}, locally given by the potential function
$x^3+\alpha x$. The parameter $\alpha$ is responsible for the
'unfolding' of the singularity.\\
Common examples of a fold catastrophe are the sea's sparkles or the appearance of rainbow via complete reverse refraction of light in small drops of water.\\
A little more complicated is \emph{the cusp}, which looks like
$x^4+\alpha x^2+\beta x$. We already have two good examples of a cusp catastrophe, and
the picture below shows how this singularity is projected in the plain.\\
\begin{figure}\centering\resizebox{13cm}{!}{\rotatebox{0}{\includegraphics{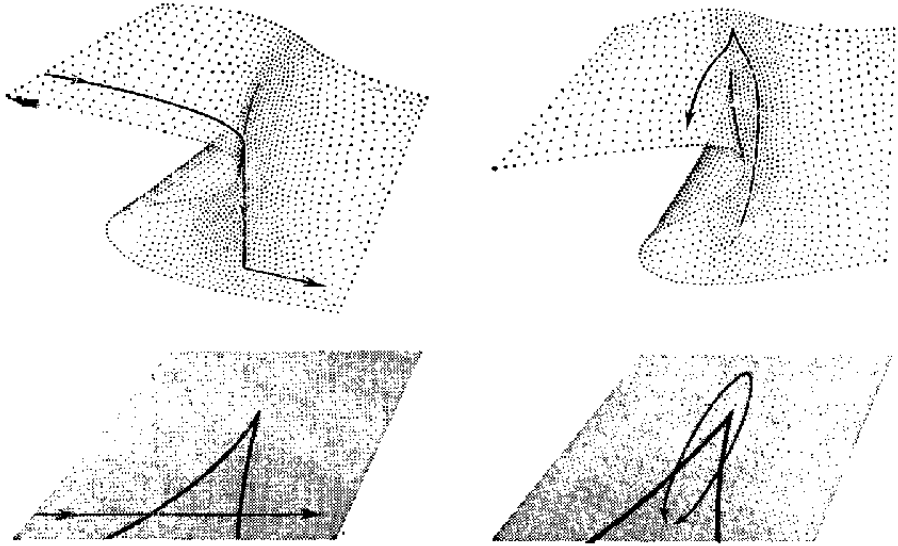}}} \label{cuspa}\caption{\small  The \emph{cusp} catastrophe is the second and last possibility for structurally stable Lagrangian singularity in dimension two (codimension one)}\end{figure}
\noindent \emph{The swallowtail} has co-dimension two and the corresponding polynomial is $x^5 + \alpha x^3 + \beta x^2 + \gamma x$. For examples of the swallowtail and the more complicated catastrophes in practical problems we refer to \cite{SP}.\\
As an easy exercise, one may check how variations of the control
parameters alter the extrema of the potential function. There are
critical values that correspond to bifrucation of different types:
collision and separation of minima and maxima (and in particular
anihilation of a minimum and a maximum), loosing stability
(minimum to saddle point) etc.\\
The whole catastrophe theory classification we talk about is
somewhat special, although it includes almost all possible topics
of interest both in geometry and physics. Its peculiarity
originates in the fact that we are considering Lagrangian
catastrophes only. We remind that a submanifold of a $2n$-dimensional manifold is called
\emph{Lagrangian} if its dimension equals $n$ and the symplectic
form vanishes on it. The most trivial case is the base of a
cotangent bundle, considered as a symplectic manifold (in
particular - the $q$-plane) or, as it was in our geometric optical
settings, the level set of the phase function or the integral
surface of the Hamilton - Jacobi equation. Then we naturally call
Lagrangian the singularities between lagrangian manifolds of equal
dimension. In our case these were the projections of critical
points on the coordinate space. This locus of branch points plays
the role of an invariant curve (or surface, or point) to which
infinitely many rays (or particle trajectories, as far as
classical dynamics is involved) tangate. In the case of the elliptic
billiard table for example, the caustics are namely the confocal ellipses
and hyperbolae with the help of which one integrates the system.\\
An alternative treatment would pay interest in the wave-front singularities, instead of those of the light rays, but we are not going to consider this case here. The reader may look in \cite{A3} and \cite{B1} instead. In the appendices of \cite{A1} one may find a lot of interesting considerations concerning geometric optics, normal forms and catastrophes.

\newpage
\section{Elementary Quantum Mechanics}
In section one we described the dynamics of a particle by a hamiltonian function together with the symplectic structure (or Poisson bracket) in the cotangent bundle. Then integration in this formalism, reduces to finding a complete set of commuting first integrals.\\
In elementary quantum mechanics, as we shall see in an instant,
this picture is somewhat preserved: the reduction of a stationary
quantum mechanical system for example, which is this time
represented by a linear operator, acting in Hilbert space, is
equivalent to finding a generator of symmetry for the system.
Since the corresponding operator commutes with the hamiltonian of
the system, it provides a convenient 'splitting' of eigenstates.
Various problems have been attacked in this way, among which the
hydrogen atom, the oscillator, the Zeeman effect and many
others. A good comprehensive book on the basics is \cite{D}. However, for the path integral formulation we refer to \cite{F} and \cite{F2}.\\
\subsection{Heisenberg and Schr\"{o}dinger Pictures}
We start with a picture in which each physical observable $Z$ is
represented by a linear self-adjoint operator, acting in the
Hilbert space $\mathbb{L}_2$ of square integrable functions. The
evolution of a physical observable given again in terms of a
hamiltonian and a Poisson bracket
\begin{equation}
\dot{Z}=\{H,Z\}
\end{equation}
but this time the hamiltonian function is an operator itself and
the bracket is given by a commutator times a numerical factor
$\displaystyle{\frac{1}{i\hbar}}$, with $\hbar$ being the famous
Plank constant.\\
Straightforward integration of the above formula gives for the
flow of $Z$
\begin{equation}
Z(\tau)=\exp{[\frac{i\tau}{\hbar}H]}\,Z(0)\,\exp{[-\frac{i\tau}{\hbar}H]}
\end{equation}
The above expression is an adjoint action and expands in an
infinite series of commutators according to the famous Campbell -
Housdorff formula.\\
It is sensible to call the operator
$\displaystyle{U(\tau)=\exp{[-\frac{i\tau}{\hbar}H}]}$ \emph{the
propagator} associated to $H$. Since $H$ is self-adjoint by
definition, the propagator is always unitary. Of course, $H$ may
depend on time itself and one should consider $T$-ordered exponents
instead, but for the time being we neglect this case.\\
The reason we have to call an infinite dimensional operator a
\emph{physical observable} is that what we can really observe
and measure in experiments are the eigenvalues of the operator.
Therefore we have to choose a canonical basis in which the
operator is diagonal. One may think of the process of
observation or the experiment itself as being equivalent to one
particular choice of such basis. But we know from simple algebra
that two linear operators share a common canonical basis if and
only if they commute. This is the place we come across the first
crucial difference between classical and quantum mechanics, known
as the Heisenberg uncertainty principle: in the quantum picture we
are 'allowed to observe' only quantities that commute with each
other. In deed, as we shall se in the following we are not even
able to measure simultaneously the initial coordinate $x$ and the momentum
$p$, so what is usually referred to as \emph{quantum
indeterminism} is laid in the very roots of the theory.\\
Nevertheless, the equations of quantum mechanics give rather exact
and in many cases the only known description for various phenomena
from the world of particles, atoms and molecules, to the description of large ensembles such as polymers, crystals, gases and even stars. In order to obtain these equations, however, we need to start with choosing a suitable basis.\\
Let $\{\phi_k\}$ be the basis of eigenfunctions for the observable
$W$. We have $$ W\phi_k = w_k\,\phi_k $$ and since $W$ is
self-adjoint, all eigenvalues are real and the $\phi_i$'s are
orthogonal with respect to the standard scalar product in
$\mathbb{L}_2$:
$$\langle \phi \,|\, \psi \rangle = \int{\phi^*(x)\,\psi(x)dx}
$$
as long as they do not fall in the same degeneracy subspace (have
the same $w$ as an eigenfunction).\\
We use the above to express the matrix elements of another
observable $Z$ as
$$Z_{kl}=\langle \phi_k \,|\, Z\phi_l\rangle$$
Moreover, let $\psi=\sum{c_k\phi_k}$ be an arbitrary\footnote{once
more, for the sake of simplicity we avoid the question of
continual indices, but there might be an additional term of the
form $\displaystyle{\int_k{c(k)\phi(k)dk}}$ in the generic case}
unit vector in $\mathbb{L}_2$. For reasons that were already explained,  we call the vectors in $\mathbb{L}_2$ \emph{states}. Then, in the state $\psi$, the observable $Z$ projects as $$ \langle Z \rangle_{|\psi} = \langle \psi \,|\, Z \psi \rangle $$
In view of the above definition of the scalar product and the
normalization of $\psi$, $\langle Z \rangle_{|\psi}$ can be
regarded as an expression for the expectation value of $Z$ in the
state $\psi$. With this in mind it is easy to interpret
$|\psi|^2=\psi^*\psi$ as a probability density and the Fourier
coefficients in the above expansion give respectively the
\emph{transition amplitudes} $|c_k|^2$ - the probabilities for
transition from a state $\psi$ to $\phi_k$, or the probability to
detect $w_k$ in the experiment.\\
The above description (known as Heisenberg picture or \emph{matrix
mechanics}) gives a good intuitive link between Hamiltonian
dynamics and quantum mechanics. For practical reasons, however, it
is often convenient to consider a dual situation, in which the
observables remain unchanged in time, but states evolve
instead. In explicit terms it follows directly from the above that
if we assume $Z(\tau)=Z(0)=const$ and $\psi(\tau)=U(\tau)\psi(0)$,
we easily obtain the same values for the 'measurable' quantities -
eigenvalues, expectation values, transition amplitudes etc. The
latter leads immediately to
\begin{equation}
i\hbar
\frac{\partial}{\partial\tau}\psi(x,\tau)=H(p,x)\,\psi(x,\tau)
\label{Schr}
\end{equation}
known as the (non-stationary) Schr\"{o}dinger equation. Now it is
time to reveal the explicit form of the operator $H$. In fact,
just as in the classical case, it is a sum of kinetic and
potential term, usually in the form $$H=\frac{p^2}{2m}+V(x)$$
where the classical coordinate and momentum are replaced by the
operators $x\rightarrow$ multiplication by $x$ and
$p\rightarrow i\hbar\nabla_x$ respectively\footnote{in order to be
fair, we add that this choice is only a convention - one may
alternatively choose $p\rightarrow p.$ and $x\rightarrow
-i\hbar\nabla_p$ - the so called \emph{'momentum
representation'}}. Their Poisson bracket agrees with the classical
one, since obviously $$[p,x]=i\hbar $$ Now it becomes clear why
coordinates and momenta are not compatible in measurement.
Moreover, using Fourier analysis\footnote{note that $p$ and $x$
are related via Fourier transform}, it is easy to prove the
estimate for their dispersions $\triangle a = \langle |a-\bar{a}|\rangle$
\begin{equation}
\triangle p \, \triangle x \geq \hbar
\end{equation}
Thus the more precise we measure the coordinate, the less we know
of the momentum and vise versa. The same remark certainly refers
not only to $p$ and $x$, but to any other pair of operators that
satisfy analogous commutation relations, e.g. energy and time,
angular momentum and angle etc.\\
Let us now return to the equation (\ref{Schr}). It is a second-order
differential equation that in most cases is practically
unsolvable. Moreover, regarded as an evolution equation, the
Cauchy problem for which does not exist (due to the uncertainty
principle just discussed), it resembles more the wave equation,
than the equation of conductivity, thanks to the $i$-factor in
front of the time derivative. That is why its solution is
sometimes called \emph{wave function}. In many cases the solutions
really behave like waves.\\
Now consider a special class of solutions, for which the
probability density does not evolve in time. This requirement
allows $\psi$ to depend on $\tau$ only through a factor of unit
modulus, that is
\begin{equation}
\psi(x,\tau)={e}^{-i\omega\tau}\tilde{\psi}(x)
\end{equation}
Then, substituting this into the Schr\"{o}dinger equation, we
reduce the problem to a Helmholtz type equation for the function
$\tilde{\psi}(x)$:
\begin{equation}
H\tilde{\psi}=\hbar \omega \tilde{\psi}
\end{equation}
The eigenvalues of the Hamiltonian operator represent the energy
spectrum $E_k=\hbar\omega_k$ of the system. The spectrum is
discrete whenever there is an appropriate energy bound.
In the generic case we have an discrete and a
continuous part (for example in scattering problems).\\
The representation in which the Hamiltonian is diagonal is called
therefore the \emph{energy representation} and the above equation
that defines it - \emph{the stationary Schr\"{o}dinger equation}.
We are going to use it a lot in the following.\\

\subsection{The Oscillator}
Let us consider one of the simplest classical problems and try to
quantize. The classical hamiltonian of the linear harmonic
oscillator (with unit mass) is given by
$$ H(p,x)=\frac{p^2}{2}+\frac{\omega x^2}{2} $$
where $\omega$ is a constant of elasticity. If we apply directly
the formalism we have built up till now, we the quantum
hamiltonian in the form of a second-order differential operator
with a quadratic potential term. We are interested in the
eigenvalue problem for this operator.\\
Following an idea of Dirac, we express the hamiltonian in terms of
the operators
\begin{equation}
a=\frac{1}{\sqrt{2\omega}}(Q+iP),\qquad
a^\dagger=\frac{1}{\sqrt{2\omega}}(Q-iP)
\end{equation}
where $P$ and $Q$ are the quantum momentum and coordinate
operators, respectively.\\
It is easy to see that $a$ and $a^\dagger$ are not self-adjoint,
but pair-wise adjoint, instead. Moreover, they fulfill the
commutation relation $ \{a, a^\dagger\} = 1 $, that gives, by
induction $ \{a, (a^\dagger)^k\} = k(a^\dagger)^{k-1} $. In this
way $a$ acts on the functions of $a^\dagger$ as formal differentiation.
Now the quantized hamiltonian can be written in the form
$$H = \omega \left ( a a^\dagger - \frac{1}{2} \right ) =
\omega \left ( a a^\dagger + \frac{1}{2} \right )  $$ which allows
for computing easily
$$[H,a]= -\omega a, \qquad [H, a^\dagger] = \omega a^\dagger$$
This construction already resembles very much the standard weight
decomposition of semi-simple Lie algebras. The analogy goes a bit
further with finding the highest (lowest) weight.\\
More precisely, the spectrum of $H$ is bound from below.
$$ E\geq \frac{\omega}{2}$$
To see this, one simply needs to exploit the standard scalar
product in $\mathbb{L}_2$ for an eigen-state, corresponding to the
energy $E$. From the above, we have directly
$$ \langle \psi_E|\, H\psi_E\rangle = E ||\psi_E||^2 = \omega
||\,a\,\psi_E||^2+\frac{\omega}{2}||\psi_E||^2 $$ which proves our
assertion and gives in addition that equality is reached if and
only if $a\psi = 0$. As we shall see shortly, such $\psi_0$ exists
and $\displaystyle{E_0=\frac{\omega}{2}}$ is the lowest state energy,
also referred to as the 'vacuum energy'.\\
In order to show that this lowest energy is reached, we consider
the series of vectors $\psi_{k_E-m}=a^m\,\psi_E$. We claim that if
$\psi_E$ is an eigenvector with eigenvalue $E$ then $\psi_{k_E-m}$
is also such with eigenvalue $E-m\omega$. For $m=1$ we obviously
have $$ H a\psi_E = a H\psi_E - \omega \psi_E = (E-\omega)\psi_E$$
and the rest follows by induction. Since this series is
diminishing and bounded from below, it should be finite, so there
is such a natural number $k_E$ for which $a^{k_E+1}\psi_E = 0$ and
correspondingly such $\psi_0$ for which
$$ a\psi_0 = 0,\qquad \displaystyle{H \psi_0 =
\frac{\omega}{2}\psi_0}$$ On the other hand, the operator
$a^\dagger$ generates infinite, non-vanishing series of solution
$\psi_n = (a^\dagger \psi_0)$ with eigenvalues
$\displaystyle{E_n=\omega \left ( n + \frac{1}{2}\right )}$
respectively.\\
Now it becomes clear why $a^\dagger$ and $a$ are called
\emph{creation} and \emph{annihilation operators} respectively.
With their help we can construct explicitly the basis of
solutions. We start with $\psi_0$ and substitute in the
differential equation $\displaystyle{E_0=\frac{\omega}{2}}$. We
obtain
$$ \psi_0 = Ce^{-\frac {\omega x^2}{2}}$$
where the constant $C$ is defined by the normalization
condition.\\
For the rest of the eigen-states, we obtain, by a complicated and
non-interesting procedure
$$\psi_n\sim H_n(\sqrt{\omega}x) e^{-\frac {\omega x^2}{2}}$$
where $H_n$ are the Hermit polynomials.

\subsection{The Density Matrix} Sometimes, when we are interested in the
practical side of the matter, especially in the study of ensembles
in quantum statistics, it is more sensible to study directly
partition functions (probabilistic distributions) for a certain
problem, rather than the 'non-physical' wave-function $\psi$. As
we already saw, the meaningful quantity that we may get out of
$\psi$ is its squared modulus $\psi^* \psi$. What we do in
practice, however, is to define instead the projector over the
$\psi$ - state $\Pi_\psi$ as acting on a state $\sigma$ by
$$\Pi_\psi\,\sigma = \langle\psi|\sigma\rangle \psi $$
and in Dirac notations we write $\Pi_\psi = |\psi\rangle
\,\langle\psi|$. Note that it is in fact an integral operator with
unit trace since, by normalization, $\langle\psi|\psi\rangle=1$.
This very projector we call a \emph{density matrix} for the state
$\psi$ and denote by $\rho$ or
$\rho_\psi$.\\
The reason for choosing this name is somewhat obvious: once a
complete orthonormal basis of pure states $\phi_i$ is introduced in our Hilbert
space the above integral operator adopts indices, namely
\begin{equation}
\rho_{kl}= c_k^*\, c_l, \qquad c_j = \langle \phi_j|\psi\rangle
\end{equation}
and the diagonal elements $w_i:=\rho_{ii}$ can be regarded as
probabilistic weigths of the 'pure states' $\phi_i$. If the system
is entirely in a pure state, then only one of them equals unity
(all other matrix entries vanish) and $\rho^2=\rho$.\\
With the help of the density matrix it is easier to express
statistical data in covariant terms. For example, the expectation
value of an observable $Z$ is now written as
\begin{equation}
\langle Z\rangle = {\rm tr} ( \rho Z)
\end{equation}
and various thermodynamical functions can also be given in this
way. For example the free energy $F = - k\,T\ln {\rm
tr} \rho$, or the entropy $S=-\sum w_k \ln w_k$.\\
Note that this picture is still Schr\"{o}dinger rather than
Heissenberg and due to equation (\ref{Schr}) we have for the
evolution equation for the density matrix in the form
\begin{equation}
\dot{\rho}=\{\rho, H\}
\end{equation}
which resembles the dynamical equations for observables in
Heissenberg representation, except for the 'minus' sign.\\
In statistical mechanics, we know by general considerations that,
up to a normalization factor, the diagonal elements of $\rho$ are
given in the energy representation by
$$ w_k = e^{-\beta E_k} $$
where $\beta$ is proportional to the inverse
temperature\footnote{equals the inverse absolute temperature times
the inverse Boltzman constant, to be more precise} and $E_k$ is
the energy of the corresponding pure state. In a more covariant
form, we may write
$$ \rho = \exp{[-\beta H]}$$
which leads directly to the famous \emph{Bloch equation} for the
density matrix
\begin{equation}
\frac{\partial}{\partial\beta}\rho + H\rho = 0
\end{equation}
This is a diffusion or heat equation as it is easily seen from the
simplest case - a free particle in Euclidean space. Then the
solution is easily seen to be a Gaussian distribution.
\subsection{Path Integrals}
Now we shall develop an alternative approach to the study of the
density matrix. The so-called \emph{path integrals} involve the
quantum mechanical concept of non-locality, by letting the present
state of the system be influenced not only by a single evolution,
but instead, by all possible paths that link the initial and final points in phase space\footnote{at this stage we are already dealing with a semi-classical method since points in purely quantum phase space are ill-defined due to Heisenberg uncertainty principle}. Following Feynman, we write formally
$$ \rho(x_{in},x_{f},u) = \exp \left (-\frac{\hat{H}u}{\hbar} \right ) $$
which holds for sufficiently close $x_{in}$ and $x_f$. For more general considerations we divide the trajectory into small parts of length $\triangle x$ for which the above expression is still true and then perform a summation (or rather an integration in the limit $\triangle x\rightarrow 0$) over all possible trajectories. \\
In more explicit terms, what we do is to express the propagator kernel (and the Green function) by such path summation. As we know from Huygens principle, the propagator responsible for the evolution of $\psi(q_{in},t_{in})\rightarrow\psi(q_f,t_f)$ can be expressed as a composition of infinitesimal propagators
$$K(q_{in}\rightarrow q_f,\, t_{in}\rightarrow t_f) = \int{dq_1dq_2\ldots dq_{N-1}\prod_{n=0}^{N-1}{K(q_{n}\rightarrow q_{n+1},\, t_{n}\rightarrow t_{n+1})}}$$
with $q_0 = q_{in},\, q_N = q_f$. Since, as we already know, for small intervals
$$ K(q_{n}\rightarrow q_{n+1},\, t_{n}\rightarrow t_{n+1}) = \langle q_{n+1} | {\rm exp}[-\frac{i\hat{H}\triangle t}{\hbar}]|q_n\rangle $$
separating the kinetic and potential term in $\hat{H}$ and neglecting $\mathcal{O}(\triangle t^2)$ terms, we may approximate each of the infinitesimal propagators as $\triangle t \rightarrow 0$ with a factor in the form
$$ \left ( \frac{m}{2\pi i\hbar \triangle t}\right )^{-1/2}{\rm exp} \left [\frac{i\triangle t}{\hbar}\left ( \frac{m}{2}\left (\frac{\triangle q}{\triangle t}\right )^2-V\right )\right ]$$
which gives for the overall kernel
\begin{eqnarray}
K(q_{in}\rightarrow q_f,\, t_{in}\rightarrow t_f) =\lim_{N\rightarrow\infty} \int\ldots\int{\left ( \frac{m}{2\pi i\hbar \triangle t}\right )^{-1/2}}\times \\
\times {{\prod_{j=1}^{N-1}dq_j(2\pi\hbar\triangle t)^{-1/2}}{\rm exp}\left [ \frac{i}{\hbar}\int_{t_{in}}^{t_f}{L(q,\dot{q},t)dt}\right ]} \nonumber
\end{eqnarray}
Note that most 'weighted' paths are namely the classical trajectories, for which the action functional is extremal. On the other hand, a path, participating in the above distribution does not necessarily have to be smooth, as can be seen in the example of Brownian motion.\\
From path integrals one may readily retrieve information about
transition coefficients, that are propagator entries, but also, via
Fourier transform one may recover the Green function. Calculation of path integrals in practice , however, is not that simple. In the trivial case of quadratic Lagrangian, the expression is reduced to the standard Gaussian integral, that is computed exactly. For the more generic case, however, we need to rely on approximation procedures, such as the stationary phase approximation, or the perturbation series expansion, both considered in the following.

\subsection{Perturbation Theory} There are but few known problems in quantum mechanics that allow for exact solutions and for practical matters, we need to be equipped with good approximation methods. Perhaps the most used one (not only in quantum mechanics, but in classical dynamics and QFT as well) is the perturbation method.\\
The simplest illustration of the method\footnote{this is the easiest case with time-independent hamiltonian with discrete spectrum, but also the most relevant to the topics discussed here} starts with an unperturbed Hamiltonian $\Hat{H}_0$, for which the stationary Schr\"{o}dinger equation is solvable by analytical methods, or for some reason we know the spectral decomposition
$$\Hat{H}_0\psi^0_k = E^0_k\Psi^0_k $$
Then imagine that we want to go a bit farther and study the eigenvalue problem of the operator $\Hat{H} = \Hat{H}_0+\Hat{V}$, where the term $\Hat{V}$ is small compared to $\Hat{H}_0$ in the sense that $|V_{kl}|\ll |E_k^0-E_l^0|$ where the matrix coefficients are computed in the initial basis $\{\psi^0_k\}$. This restriction will be justified in the following.\\
Now, as usual, one expands the unknown eigenfunctions $\psi_k$ in the initial basis as
$$ \psi = \sum_j{c_{j}\psi^0_j}$$ and substitutes into the equation for $\Hat{H}$. What one gets is
$$\sum_k{c_j(E^0_j+\Hat{V})\psi_l^0} = \sum_m{E\,c_m\psi^0_m} $$
Multiplying both sides by $\psi^0_i$ and integrating yields
\begin{equation} \label{pert}
(E-E^0_i)c_i = \sum_l{V_{ij}c_j}
\end{equation}
which is a homogenous system for the $c_j$'s which leads to an infinite series of relations for the perturbed energy states $E_j$ in terms of the unperturbed ones $E_k^0$ and the matrix entries of $\Hat{V}$. However, considering the fact that the latter were chosen to be small, we may expand $E_k$ in a series $E_k = E_k^0+E_k'+E_k''+\dots$, where the $j^{th}$ term is a polynomial of order $j$ with respect to the $V_{kl}$'s and cut that series at an appropriate place\footnote{in practice, it usually suffices to take the first two or three terms}. The same procedure we do with $\psi$, or rather, with the $c_k$'s. Then we restrict to the eigen-state $\psi_k$ by taking $c^0_k = 1$ and $c^0_j = 0$ for $j\neq k$ in the above equation. Substituting all this in the equation we may immediately determine the first correction $E_k'$. For $k=i$ it gives
\begin{equation}
E^1_k = V_{kk}
\end{equation}
and for $i\neq k$, obtain the first correction to the eigenfunction
\begin{equation}
c'_i = \frac{V_{ik}}{E^0_k-E^0_i}
\end{equation}
In order to have this correction small, we really need the estimate we imposed in the beginning. Then the perturbed egen-state $\psi$ is given, up to a first-order corrections, by
\begin{equation}
\psi_k \sim \psi^0_k + \sum_{i\neq k}{\frac{V_{ik}}{E^0_k-E^0_i}\psi^0_i}
\end{equation}
Note that the correction term lies in the orthogonal complement of $\psi^0_k$, so the new state $\psi_k$ is normalized up to second order corrections.\\
Following the same procedure, we easily obtain the next few terms
\begin{eqnarray}
E_k'' &  = & \sum_{i\neq k}{\frac{|V_{ik}|^2}{E^0_k-E^0_i}}\nonumber \\
E_k'''&  = & \sum_{j\neq k}\sum_{i\neq k}{\frac{V_{ki}V_{ij}{V_jk}}{(E^0_k-E^0_i)(E^0_k-E^0_j)}}-V_{kk}\sum_{i\neq k}{\frac{|V_{ki}|^2}{(E^0_k-E^0_i)^2}}
\end{eqnarray}
The results obtained here can be generalized to the case when $\Hat{H}_0$ has a continuous part of the spectrum as well. Then summation is simply replaced by integration.\\
In the case when the spectrum of $\Hat{H}_0$ is degenerate we still have to modify our procedure a little, due to the arbitrariness of the states in the degenerate subspace. It happens, that this arbitrariness is eliminated once we impose the condition that states should be altered only by small variation. Such 'correct' states are defined as combinations of the initial ones with coefficients $c^0_{i_1},\ldots\, c^0_{i_s}$ where $s$ denotes the degree of degeneracy of the eigenvalue $E^0_i$. With this substitution, equation (\ref{pert}), restricted on the corresponding invariant subspace gives
\begin{equation}
\sum_k{(V_{ii_k}-E'\delta_{ii_k})c^0_{i_k}} = 0, \quad i,k =  1\ldots s
\end{equation}
which as a linear homogenous system for the $c^0_{i_k}$'s has nontrivial solution if and only if
\begin{equation}
{\rm det}(V_{ii_k}-E'\delta_{ii_k}) = 0
\end{equation}
which on its side is a polynomial of degree $s$ in $E'$ and its $s$ roots give the energy shifts within the degenerate state up to first order. In order to determine the 'correct' eigenfunctions, we substitute these roots back in the matrix equation in order to get all the $c^0_{i_k}$'s. Then, for the perturbation of the eigenfunctions we find
$$ c'_{i_k} = \frac{1}{V_{ii}-V_{i_ki_k}}\sum_l{\frac{V_{i_kl}V_{li}}{E^0_i-E^0_l}},\quad i\neq i_k $$
It may happen that some of the corrections (or even all) are zero. Then one proceeds with higher order perturbations. On the other hand, if the roots are distinct, so are the perturbed levels and this is how perturbation often cancels degeneracy.\\

\noindent Now, for the sake of consistency, we need to consider time-dependent case as well. Let
$$\Hat{H}_0\psi_k^0 = i\hbar\frac{\partial}{\partial t}\psi_k^0$$
in the first place and $\Hat{H} = \Hat{H}_0+\Hat{V}(t)$. Now we expand, just as before,  the corresponding solution in the form $$ \psi_l(t) = \sum_k{a_{kl}(t)\psi^0_k(t)}$$ and substituting into the equation for $\Hat{H}$, we end up with a $ODE$ for the coefficients $a_{kl}(t)$, namely
\begin{equation}
i\hbar\,\dot{a}_{kl} = \sum_m{V_{km}(t)a_{ml}}
\end{equation}
Then, proceeding as before, we get for the first correction to the $l^{th}$ state
$$ i\hbar\,\dot{a}_{kl}' = V_{kl}(t)$$
or
$$ a_{kl}'(t)=\frac{1}{i\hbar}\int{V_{kl}(t)\,dt}$$
There are numerous interesting applications of perturbation theory in diverse branches of classical and quantum Physics. We only point out here one more important aspect that goes far beyond the basics of quantum mechanics, considered here.\\
Let us take the kernel of the propagator for the transition $q_{in}\rightarrow q_f$ of a free particle and denote it by $K_0(q_{in}\rightarrow q_f)$. We already know that this kernel can be represented as path integral in the form
$$ K_0(q_{in}\rightarrow q_f)=\int{\exp{\left (\frac{i}{\hbar}
\int_{t_{in}}^{t_f}{\mathcal{L}_0(q,\dot{q})\,dt}\right )}\mathcal{D}q(t)}$$
where $\mathcal{L}_0(q,\dot{q}) = m\dot{q}/2$ is the classical free-particle Lagrangian and $\mathcal{D}$ denotes the measure of the path integral, as found in the previous section.\\
Now, let us perturb this Lagrangian with a small potential term $ \mathcal{L}(q,\dot{q},t) = \mathcal{L}_0 - V(q,t),\: V\ll \mathcal{L}_0$ and pose the question of finding a reasonable approximation of the perturbed propagator kernel $K_V$.\\
Assuming that $V$ is small compared to $\hbar$ in the whole time interval, we may expand the $V$-dependent exponent factor in the path integral in Taylor series
$$\exp{\left (\frac{i}{\hbar} \int_{t_{in}}^{t_f}{V(q,t)\,dt}\right )} =
1 + \frac{i}{\hbar}\int_{t_{in}}^{t_f}{V(q,t)\,dt} - \frac{1}{2\hbar^2}\left (\int_{t_{in}}^{t_f}{V(q,t)\,dt}\right )^2 + \ldots $$\
which leads to an expansion for the overall kernel in the form
$$K = K^0 + K' + K'' + \ldots $$
For the calculation of the distinct terms it is helpful to note that $K_V$ is a fundamental solution of the Schr\"{o}dinger equation in the sense that
\begin{equation}
\left (i\hbar \frac{\partial}{\partial t_f}+\frac{\hbar^2}{2m}\nabla_{q_f}^2 - V(q,t)\right )K_V = i\hbar\, \delta (t_f - t_{in})\delta(q_f - q_{in})
\end{equation}
and so is $K^0$ with $V=0$. This allows for applying the parametrix method, briefly discussed below in the case of the heat equation, to expand $K_V$ as an infinite convolution product geometric series\footnote{known as \emph{Voltera series}}, which allows for writing the $n^{th}$ correction term in the form
\begin{eqnarray}
K^{(n)}(q_{in}\rightarrow q_f) =
\frac{1}{(i\hbar)^n}\int_{t_{n-1}}^{t_f}\int_{t_{n-2}}^{t_{n-1}}\ldots \int_{q_{in}}^{q_1}{K^0(q_{n-1}\rightarrow q_f)V(q_{n-1},t_{n-1})}\times \nonumber \\
\times {K^0(q_{n-2}\rightarrow q_{n-1})\ldots}\ldots {V(q_1,t_1)K^0(q_{in}\rightarrow q_1)\,dt_n\,dt_{n-1}\ldots dt_1}\nonumber
\end{eqnarray}
This term is interpreted as a propagator of a particle that has been scattered $n$ times by the potential $V$ within the interval $(t_{in}, t_f)$. This interpretation is illustrated by the so called \emph{Feynman graphs}, that are used as a means of calculation as well.\\
The zeroth term in particular describes the probability that the particle remains free without noticing the presence of potential. Then $K_V'$ stands for a one-collision process etc. Resuming, if possible
, all (or at least a sufficient number) of the above terms, we get the demanded propagator $K_V$ (or a good approximation of it). Such summation leads to the integral equation for $K_V$
\begin{equation}
K_V(q_{in}\rightarrow q_f) = K^0(q_{in}\rightarrow q_f)+\frac{1}{i\hbar}
\int_{t_{in}}^{t_f}{K^0(q\rightarrow q_f)V(q,\tau)K_V(q_{in}\rightarrow q)\,d\tau}
\end{equation}
which can be obtained also directly, using the differential equations for the two kernels.

\subsection{Scattering Problems} We end this section with a short discussion
on the stationary Schr\"{o}dinger equation defined on the line
$\mathbb{R}$, which for simplicity we write in the form
\begin{equation}
(H-k^2)\psi=0
\end{equation}
where $\displaystyle{H=\frac{d^2}{dx^2}+V(x)}$ represents the
re-scaled hamiltonian with a finite potential (or 'scattering')
term $V$ and the spectral parameter $k^2$ stands for the energy.\\
The whole space of solution can be expanded over a basis, chosen
in the following way: consider a solution $\psi^r_+$ with that
tends to $e^{ikx}$ as $x\rightarrow\infty$ and another such
$\psi^r_-$ which tends to $e^{-ikx}$. These two independent
solutions, sometimes called \emph{right Jost solutions}, represent, in
physical terms, two plain waves traveling to and from plus infinity,
respectively. There is, however, another option to construct such
a basis by switching from left to right infinity $\psi^l_\pm\sim
e^{\pm ikx}$ as $x\rightarrow \pm\infty$. Since the full
space of solutions of the above second-order ODE is
two-dimensional, these two sets must be linearly dependent. We
shall construct such dependence in an instant. Consider for
example a solution that behaves like an outgoing plain wave as $x$
tends to infinity and in the other limit, it should be given by a
linear combination of the two 'basic' states, or
$$ \psi\sim a(k)e^{ikx}+b(k)e^{-ikx}$$
where $a(k)$ and $b(k)$ are holomorphic in $\mathbb{C}/\{0\}$. The
physical picture represents two asymptotically plane waves - one
initial and one that is 'scattered backward' by $V(x)$.
Alternatively, we may take the initial wave traveling from the
right to the left - a solution $\phi$ that tends to $e^{-ikx}$ as $x\rightarrow -\infty$ and as $x\rightarrow\infty$
fulfills the limit $$ \phi\sim \beta(k)e^{ikx}+\alpha(k)e^{-ikx} $$ The same considerations hold here as well. Moreover,
both $\phi$ and $\psi$ satisfy the property that the Wrontzkian
of the function and its complex conjugate is constant. For
the proof we only use that both functions are solutions of the
stationary Schr\"{o}dinger equation, given above. Take for
instance $\psi$ and substitute then multiply by $\bar{\psi}$. The
same procedure we repeat with inverting $\psi$ and $\bar{\psi}$ -
note that $\bar{\psi}$ is still a solution, since the
Schr\"{o}dinger operator is self-adjoint. Then we subtract the
two terms and finally obtain
$$\frac{d}{dx}\left ( \bar{\psi}\frac{d}{dx}\psi -\psi\frac{d}{dx}\bar{\psi}\right )$$
which means that $$ W(\psi,\bar{\psi}):=\left (
\bar{\psi}\frac{d}{dx}\psi -\psi\frac{d}{dx}\bar{\psi}\right
)=const $$ The same certainly holds for $\phi$ as well.\\
From this single relation we get
$$|a|^2-|b|^2=|\alpha|^2-|\beta|^2=1$$
and because of this neither $a(k)$ nor $\alpha(k)$ is allowed to
vanish for real $k$ \footnote{one may check that their zeroes are
purely imaginary}.\\
On the other hand, in the same way as before, it is obvious that
$W(\phi,\psi)=const$, which immediately leads to $a=\alpha$ and
hence $|b|^2=|\beta|^2=1+|a|^2$. \\
Now let us think in more physical terms and regard the whole problem from the initial wave's point of view - this means, above all, choosing it's amplitude to be unity or, in other words, simply dividing by the non-zero factor
$\alpha(k)$, so that the reflected term adopts a factor
$\displaystyle{r_+=\frac{b}{a}}$ ( respectively
$\displaystyle{r_-=\frac{\beta}{a}}$), and the transited comes
with $\displaystyle{t=\frac{1}{a}}$. The letter are called
\emph{reflection} and \emph{transition coefficients} respectively.
Now we define the \emph{scattering matrix} of the problem as
$$ S(k) = \left(%
\begin{array}{cc}
  t(k) & r_-(k) \\
  r_+(k) & t(k) \\
\end{array}%
\right)
$$
It clearly gives the scattered state (in both directions) expanded
over the basis of Jost solutions. Moreover, the matrix $S(k)$ is
unitary - that is understood most easily if we look at it as
representing an evolution operator, acting on quantum-mechanical states.
It also appears to be \emph{real} in the sense that $S(-k)=\bar{S}(k)$.\\
There is one more significant operator involved in the very root
of our problem. Essentially, this is given by the \emph{ monodromy
matrix} $\mathcal{M}$, defined in the following way. We start with a
solution $\psi$ that has the form of a single plain wave as $x$
tends to minus infinity. Let $\psi^l_0$ be the unique solution of
the free-particle (Helmholtz) equation, that coincides with $\psi$
at this limit. Alternatively, we denote by $\psi^r_0$ the solution
that coincides with $\psi$ in the limit $x\rightarrow\infty$. Then
the matrix $\mathcal{M}$ is defined as the unique linear
isomorphism which satisfies
$$ \psi^r_0=\mathcal{M} \psi^l_0$$
In order to be more explicit, we express a generic monodromy
matrix by means of the transition and reflection coefficients. We
have
$$ \mathcal{M} = \left(%
\begin{array}{cc}
  \bar{t}^{-1} & -\bar{r}_+\bar{t}^{-1} \\
 - r_+ t^{-1} & t^{-1} \\
\end{array}%
\right)
$$
This turns out to be a $\mathsf{SU}(1,1)$ matrix, as can be
derived from a simple consideration: construct a real basis
(involving sine and cosine instead of the complex exponents) in
the space of solutions of the free-particle equation. Then, the
monodromy matrix (in this basis) should be
area-preserving\footnote{that is because it can be thought of as
being adjoint with the phase-area preserving flow, acting in the
space of 1-jets of solutions of the initial equation}, and hence,
$\mathsf{SL}(2,\mathbb{R})$, which in the complex basis, as we
know, takes the form of a $\mathsf{SU}(1,1)$ - matrix. This is
enough for us to prove the above formula. We just note that it
transforms the plain wave $e^{ikx}+r_+ e^{-ikx}$ into
the plain wave $te^{ikx}$ which immediately gives the first
column of the contra-gradient matrix, transforming the basis ${\bf
e}_\pm = e^{\pm ikx}$. For the second one we only have to
take the complex conjugate to obtain finally
$$ ^t \mathcal{M}^{-1} = \left(%
\begin{array}{cc}
  t^{-1} & r_+ t^{-1} \\
 \bar{r}_+\bar{t}^{-1}  & \bar{t}^{-1} \\
\end{array}%
\right)
$$
then we easily obtain the monodromy  matrix, exploiting the fact
that to invert a $\mathsf{SU}(1,1)$ matrix, we only have to
exchange the diagonal elements and take the non-diagonal with
opposite sign.\\
With this formula in mind it is easy to see that the sum of the
transition and the reflection amplitudes is always
unity\footnote{the amplitudes are defined as the squared moduli of
the coefficients and have the physical meaning of probabilistic
densities for transition and reflection, respectively}, which is
the fundamental principle of matter preserving in physics.\\
To sum up, we built in this paragraph a tool through which to each
finite scattering potential one may assign a monodromy matrix in
the group of isometries of the unit disk (or, equivalently, the
upper-half plane, depending on the basis chosen). Alternatively,
one may assign a $\mathsf{SU}(2)$ - scattering matrix, which in
many situations possesses additional symmetries, giving birth
to a means of reducing interaction problems - a common practice in Quantum field theory for example.\\
Sometimes it is sensible to ask weather one can restore the
potential from the scattering data: the $S$-matrix, together with
the finite spectrum (the finite set of negative eigen-energies,
that correspond to bound states) and the divergent points of the
transition coefficient. This is the inverse scattering problem
that turns out to be solvable in some cases and in addition gives
rise to a powerful tool of resolving a special kind of non-linear
differential equations, called for historical reasons \emph{solitons}.
\newpage
\section{Towards the Theory of Integrable Systems} The theory of non-linear integrable
equations (or infinite-dimensional hamiltonian systems) is
relatively new, but has already evolved far and spreads over various
branches of modern mathematics, such as group theory, functional
analysis and others. Our aim here is only to outline the
close relation between this topic and our main concern - spectral
theory.\\
\subsection{Hamiltonian Formulation}
Our base point so far was a quadratic hamiltonian (or
lagrangian) description of a classical dynamical system, naturally
leading to linear ordinary differential equations. The problem of
integrability from standard ODE theory was reduced to finding a
full set of functionally independent and pairwise commuting
integrals of motion. Conceptually the same thing refers to the
non-linear case, but this time the \emph{action-angle} basis is
infinite-dimensional
and the problem of finding one becomes much more complicated.\\
Our only constructive example until the end of this section will
be one of the first well examined non-linear PDE's - the Korteweg
- de Vries (KdV) equation
\begin{equation}
\dot{u}=6uu'-u'''
\end{equation}
where $\dot{}$ stands for time derivative, while $'$ means
differentiation with respect to $x$.\\
First of all, let us introduce our \emph{algebra of observables}
as the space of real-analytic functionals $F[u]$ and then
introduce on it a Poisson structure, defined by the bracket
\begin{equation}
\{F,G\}=\int{\frac{\delta F}{\delta u(x)}\frac{\partial}{\partial
x}\left (\frac {\delta G}{\delta u(x)} \right ) dx}
\end{equation}
where the $\delta$-sign stands for functional derivatives.\\
It is clear that this bracket defines also a symplectic structure
in the usual way and we can easily build a Hamiltonian formalism with its help. In particular, once we have chosen a Hamiltonian functional $F[u]$, we recover the evolution equation of an observable via the Poisson bracket in the standard form
\begin{equation}
\dot{f}=\{f,H\}
\end{equation}
If we choose the hamiltonian in the form \begin{equation}
H[u]=\int_{\mathbb{R}}{\left ( \frac{u'^2}{2}+u^3\right )dx}
\end{equation}
for example, we recover the KdV equation
$$\dot{u}=\frac{\partial}{\partial x}(3u^2-u'')$$
where the last term is obtained via integration by parts.\\
With the help of the Poisson structure, one generalizes the
Liuville-Arnold theorem to the infinite-dimensional case and claims that a
system, defined by the hamiltonian $H[u]$ is integrable if and
only if it allows an infinite set of functionally independent and
pair-wise commuting integrals of motion $I_k[u]$, that are
functionals, commuting with the hamiltonian in the Lie algebra,
defined by the above bracket.\\
In our example we have first $I_0=H$, the second integral
$I_1[u]=\int{u(x)\,dx}$ is concerned with the structure of the
Poisson bracket and more specifically, the presence of a spacial
derivative in it. The third preserved quantity follows from
translation invariance of the hamiltonian functional. One
immediate consequence of this invariance is the vanishing of the
term $\displaystyle{\int_{\mathbb{R}}{\frac{\delta H}{\delta
u}\frac{\partial u}{\partial x} dx}}$ that is the first correction
for the small translation parameter expansion of $H[u]$. Now it is
obvious that if we define $$I_2 = \int_{\mathbb R}{u^2(x)\, dx}$$
then $\displaystyle{u(x)=\frac{1}{2}\frac{\delta I_2}{\delta u}}$ and hence $\dot{I_2}=0$.\\
As we shall see shortly these are really just the first few
representatives of an infinite set of integrals of motion, each of
which can be itself thought of as a hamiltonian of an integrable
system. Therefore these system come in infinite hierarchies
defined by the set of hamiltonians $I_k[u]$. For example, from the
above we easily see that in the hierarchy of the KdV equation we
have also the evolution equations $\dot{u}=u'$,
$\dot{u}=2uu'$ etc.

\subsection{Lax Representation and Iso-spectral Deformations}
Using again the KdV equation, we construct an equivalent operator
representation, known as the \emph{Lax} representation. To begin with, we consider the Sch\"{o}dinger operator for the potential $u$ from the KdV equation
$$ L = -\partial^2_x+u(x,t)$$
that is a special case of \emph{Lax operator} we are going to
investigate here and consider also the equation
\begin{equation} \label{Lax}
\dot{L}=[L,M]
\end{equation}
where the operator ${M}$ is given by
\begin{equation}
M=4\partial^3_x-3(u\partial_x+\partial_x u)
\end{equation}
A simple calculation shows that the left hand side multiplies a
test function by $\dot{u}$, while the remaining factor from the
right-hand side is exactly $6uu'-u'''$ which recovers the initial
partial differential equation for $u$.\\
Now let us see what else we can retrieve from it. First of all,
consider a solution $\phi$ of the scattering equation
$$L\phi=k^2\phi $$
with the asymptotic of a left-going wave at minus infinity. Next
we differentiate the latter with respect to $t$. Substituting into the
Lax equation we obtain\footnote{here we use the fact that the
spectrum of $L$ is preserved with time that is to be explained
below}
$$ (L-k^2)(\dot{\phi}-M\phi)=0$$
so the function $\tilde{\phi}=\dot{\phi}-M\phi$ is also a
solution, corresponding to the same eigenvalue $k^2$. The
asymptotic behavior of $\phi$ completely determines that of
$\tilde{\phi}$. We only substitute to see that
$$ \tilde{\phi}\sim 4ik^3e^{-ikx}\quad {\rm as}\;\; x\rightarrow\infty$$
Since this behavior does not depend on $t$, substituting the
explicit expression for $\tilde{\phi}$ gives the time-dependence
(\ref{pphi}) of $\phi$ exactly as promised in the beginning.\\
Let us now go back to the scattering problem for the
Schr\"{o}dinger equation from the previous section , where the quantum hamiltonian coincides with the Lax operator. Note that we let this time the potential depend on an additional  'time' parameter $t$, so the solutions and all scattering data will
depend on $t$ as well. It turns out that in some cases, a
quite complicated non-linear evolution of $u(x,t)$ may result in a
linear transformation of the $S$-matrix entries. Consider again the KdV equation
\begin{equation}
\dot{u}=6uu'-u'''
\end{equation}
This quite cumbersome evolution equation for the potential naturally induces a quite
simple transformation rule for the scattering coefficients, namely
\begin{equation}
\dot{a}(k,t)=0,\qquad \dot{b}(k,t)=8ik^3b(k,t)
\end{equation}
and the 'scattered wave' evolves like
\begin{equation} \label{pphi}
\dot{\psi}=(4ik^3-{M})\psi
\end{equation}
With this in mind, it is quite natural to use the correspondence
potential-scattering data and transform our non-liner problem
into a linear one. \\
In order to see how the $S$-matrix evolves with time, all that one
needs to do is to substitute in (\ref{pphi}) the right-infinity
limit for $\phi$:
$$\phi(x,k,t) \sim a(k,t)e^{ikx}+b(k,t)e^{-ikx}$$
All these equations are linear and easy to solve. In order to
benefit from this, however, one needs to translate backwards the
information - from evolved scattering data, to potential at time
$t$. This leads to the Gelfand-Levitan-Marchenko integral
equation of the form
\begin{equation}
\displaystyle{K(x,y)+F(x+y)+\int_0^\infty{K(x,z)F(z+y)dz}=0}
\end{equation}
where the function $F$ is defined by the formula
$$\displaystyle{ F(x) = \sum_{n=1}^N{\frac{b_ne^{-\kappa_n x}}{ia'(i\kappa_n)}}+
\frac{1}{2\pi}\int_{-\infty}^\infty{r(k)e^{ikx}dk}}$$
All the functions on the right hand side are completely determined by the scattering data \footnote{including the spectrum, the $S$-matrix and the scattering poles residua} but we are not going into detail with this dependence. Instead we refer to \cite{FT} and especially for the inverse scattering method \cite{Z}.
Then the solution of the above integral equation recovers the
potential function by the simple formula
\begin{equation}
u(x)=-2K'(x,x)
\end{equation}

\noindent One immediate consequence of the the skew-symmetric
nature of $M$ is the unitary evolution for the Lax operator,
which means that at time $t$, we have
$$   L(t) = U(t) L(0) U^\dagger(t) $$
where the unitary operator $U$ has the form $U(t)=\exp(-Mt)$ and
inversely, $\displaystyle{M = U\dot{U}^{-1}}$. This
means that $L(0)$ and $L(t)$ are unitary equivalent and therefore
the spectrum together with all spectral invariants is preserved
with time.\\
This observation has two straightforward consequences. One of
them is the use of the spectral invariants as generators of infinitely
many pair-wise commuting integrals of motion that we need to
resolve the system in Hamiltonian terms. For the special case
of the KdV equation this procedure has a lot to do with our
further considerations on spectral theory. For the time being we
only mention that first few integrals of motion for the KdV equation
are polynomials of $u$ and its spacial derivatives, given by
\begin{eqnarray}
 I_1&=&-\int{u\,dx} \nonumber \\
 I_2&=&\int{u^2\,dx} \nonumber \\
 I_3&=&-\int{(u'^2+2u^3)\,dx} \nonumber \\
 I_4&=&\int{(u''^2-5u^2u''+5u^4)\,dx} \nonumber
\end{eqnarray}
The second consequence we would like to mention here is the following.
Consider the geometry of a smooth manifold with a Laplace operator
$L$ defined on it ($u$ might represent curvature term for example,
and the ordinary derivatives might be replaced with covariant
ones). Then we study all possible smooth deformations on the
manifold and the way they reflect the operator $L$. We call one
such deformation \emph{iso-spectral} if it preserves the spectrum
of $L$. In this context, each such deformation generates an
integrable partial differential equation (infinite-dimensional
hamiltonian system) and vise versa. This dualism between
iso-spectral deformations and integrable system may be used in
several directions but this investigation, as they often say, goes far beyond the scope
of the present work.
\newpage
\section{Calculus}
After providing some physical and geometrical insight, it is time to introduce some of the the basic tools and results used for practical calculation.
\subsection{Laplacian Spectra}
The use of spectral theory in physics is so common that we often put a sign of equality between certain mathematical spectral problem and a whole branch of the physical theory
as it is in the case with the Maxwell, Schr\"odinger,
Klein-Gordon, Dirac, Poisson equations etc.\\
In this section we give the mathematical roots of the beautiful
physical theories derived by the equations, mentioned above. In
order to do this, we introduce the most typical classes of $PDE$'s,
referred to as the \emph{equations of mathematical physics}, and give a first hint how their spectral properties are linked to the geometry of the underlying manifold.\\
This is going to become clear from the examples that follow. We along with them give some facts from the theory of partial differential operators.\\
To begin with, each such operator (of order $k$), acting on a manifold $X$ has the
form\footnote{here and below $\alpha$ is a multi-index, meaning that $ D^\alpha
=\partial_1^{\alpha_1}\partial_2^{\alpha_2}\ldots\partial_n^{\alpha_n}$
($m$ is called 'the length' of the multi-index and
$\displaystyle{|\alpha|=\sum^n{\alpha_i}}$ - its 'volume').}
\begin{equation}
 P(x,D)=\sum_{|\alpha|\leq k}{J_\alpha(x)D^\alpha}
\end{equation}
and can be represented in terms of its symbol
\begin{equation}
\sigma(P)(\xi)_x=\Tilde{P}(x,\xi):=\sum_{|\alpha|\leq
k}{J_\alpha(x)\xi^\alpha}
\end{equation}
This is simply a polynomial in the dual variable $\xi=\xi_i
dx^i\in T^*X$, obtained by a Fourier transform. Its homogenous
part of highest degree $$\displaystyle{\sigma_L(P):=\sum_{|\alpha|=
k}{J_\alpha(x)\xi^\alpha}}$$ is called the \emph{leading symbol} of
$P$ and defines an invariant\footnote{in fact it is the only
covariant part of the symbol} homogeneous map of degree $k$ from
$T^* X$ to the space of coefficients (in the case of matrix-values
coefficients, this is the space of homomorphisms between the
corresponding vector bundles over $X$).\\
Now $P$ is said to be of \emph{Laplace type} if there exists a
metric $g$ on $X$ such, that the leading part of the symbol can be
represented as a scalar square with respect to this metric
$\sigma_L(P)=||\xi||_g^2$.\\
$P$ is said to be of \emph{Dirac type} if $P^2$ is of Laplace type.\\
There are numerous examples of such operators in almost every
branch of modern mathematics. We mention here some of the simplest and most
intuitive ones.
\begin{enumerate}
\item Consider a smooth compact
and orientable manifold $M$ without boundary and regard the spaces
of differential $k$-forms $\Omega^k(M)=\Lambda^k(T^* M)$ and the differential, acting on them \begin{equation}
d:\:\Omega^k(M)\rightarrow \Omega^{k+1}(M) \label{ho}
\end{equation}
The property $d^2=0$ turns (\ref{ho}) into a differential complex with
co-homological groups\footnote{$d_k$ is in this case the
restriction of $d$ to $\Omega^k(M)$}
\begin{equation} H^k(M,\mathbb{R}):={\rm ker}(d_k)/{\rm im}(d_{k-1})\end{equation}
the so-called \emph{de Rham co-homologies}. In the spaces
$\Omega^k$ we have also a well-defined scalar product, thanks to
the Hodge duality
$*:\:\Omega^k\rightarrow\Omega^{m-k},\:m=dim(M)$. Namely for two
$k$-forms $\omega$ and $\phi$ we have the coupling
\begin{equation}
\langle \omega,\phi\rangle :=\int_M{\omega\wedge *\phi}
\end{equation}
which allows for defining the conjugate differential (or
\emph{co-differential}) $\delta:\:\Omega^k\rightarrow\Omega^{k-1}$ by the
equality $\langle \omega,\delta\phi\rangle =\langle
d\omega,\phi\rangle $. Then, since $\delta$ is immediately
nilpotent $\delta^2=0$, the operator
$$\triangle_H:=d\delta+\delta d$$ is by construction self-adjoint, and its leading symbol is given by a scalar square with respect to the
above metric (this is easy to verify using a suitable basis).\\
$\triangle_H:\:\Omega^k\rightarrow\Omega^k$ is called \emph{the
Hodge Laplacian} and the restriction of its kernel to the space of
$k$-forms has the dimension of $H^k(M, \mathbb{R})$ - there is exactly one representative in each cohomological class. This is the statement of the
famous Hodge - de Rham theorem, providing the first example of the
great topological and geometrical significance of the operators we are
concerned with here\footnote{in electrodynamics and fluid
mechanics the operators $d$ and $\delta$ play central role and are
usually denoted by {\bf rot} and {\bf div} respectively - more precisely, they act in this way on one forms, while on $m-1$-forms it is the other way round, since $\delta = \pm *d\, *$ and the Hodge duality on forms holds }.
\item Consider now a (pseudo-)Riemannian manifold with Levi-Civita
connection $\nabla$. It defines the \emph{Laplace-Beltrami} operator, by the formula
\begin{equation}
\triangle_{LB}=\nabla^i\nabla_i=\frac{1}{\sqrt{|g|}}\partial_i\left(
\sqrt{|g|}g^{ij}\partial_j\right )
\end{equation}
where $g^{ij}$ are the entries of the metric tensor and $|g|$ -
the modulus of its determinant. This operator happens to coincide
with the one from the previous example only when acting on scalar
functions, or in the trivial case of flat manifolds.For $1$-forms, they disagree with a term, equal to the scalar curvature of the manifold and for higher rank tensors, this relation becomes more complicated.\\
For physical applications it is often convenient to use the fact that a generic Laplace operator $$P:= -g(\partial,\partial)\circ{\rm id} + a^k\partial_k + b$$ can be represented in the form
\begin{equation}\label{generic}
P:=-g(D,D)-E
\end{equation}
where $D$ and $E$ are a Riemannian connection and an endomorphism, chosen as follows:
\begin{eqnarray}
A_j=\frac{1}{2}g_{ij}\left( a^i + g^{kl}
\Gamma^i_{kl}\right)\nonumber \\
E=b-g^{ij}\left( \partial_i a^j + A_i A_j - A_k \Gamma^i_{jk}
\right ) \label{gen}
\end{eqnarray}
The above note makes it now quite obvious why in classical field
theories we talk about curvatures and and potentials as if they
are equivalent.\\
Exploiting some tricks from differential geometry we obtain  the promised relation between
the two Laplace operators defined so far, at least for $1$-forms and vector fields. Namely, in this case we have
\begin{equation} \triangle^1_H=\left (d\delta+\delta d\right )_{|\Omega^1 } = -\nabla^i\nabla_i +Ric\end{equation}
In fact this is only the first of a series of formulas,
relating the \emph{Hodge laplacian} $\triangle_H$ with the \emph{Laplace-Beltrami} operator $\triangle_{LB}$, known as \emph{Bochner - Weitzenb\"{o}ch} formulas. In a flat space (possibly compactified) certainly both laplacians coincide.\\
One may go a bit farther and prove that
$$ \frac{1}{2}\triangle|\,\xi|^2=|\nabla\xi|^2\pm Ric(\xi,\xi)$$
in which the 'plus' sign refers to harmonic vector fields $\xi$ for which the tensor $\nabla\xi$ is symmetric
(they are locally given by gradients of smooth functions), whereas the 'minus' sign is taken on the space of
\emph{Killing vector fields} - the ones for which $\nabla\xi$ is skew-symmetric\footnote{recall that a Killing field $X$
is by definition a field, generating isometry, and satisfies $\nabla_\nu X^\mu +\nabla_\mu X^\nu=0$}.\\
Integration over $M$ in both cases gives zero and when in the right hand side we have two non-negative summands, they must both
vanish independently. In this way we easily conclude that if $M$ is closed, and the Ricci tensor is non-negative, then all harmonic
one-forms (and respectively vector fields) are parallel\footnote{we call $\sigma\in \Omega^1(M)$ parallel with
respect to the connection $\nabla$ if $\nabla_X(\sigma)=0\:\forall X\in TM$}. In addition to that, if $Ric$ is positive definite at least at one
point (and still non-negative everywhere else), then $\xi$ is bound to vanish there due to non-degeneracy, but since it is parallel, this
means that it vanishes everywhere. In this case there are no non-zero harmonic one-forms or vector fields on $M$ at all. This
result is known as the \emph{Bochner vanishing theorem}.\\
These quite simple considerations provide some useful
information concerning the topology of the underlying manifold.
First of all, if $||Ric||\geq 0$ then the dimension of the maximal
flat subspace imbedded in $TM$ (in the case of symmetric spaces,
this is their rank) equals the first \emph{Betti
number}\footnote{since the above considerations about vector fields naturally refer to one-forms as well and one may easily verify that all parallel sections are harmonic} $\beta_1=\dim{H^1(M,\mathbb{R})}$. Therefore we
have $\beta_1=\beta_{m-1}\leq m=\dim{M}$. Equality holds only for
tori, while on spheres we obviously have no global parallel sections.\\
In the case of strictly negative scalar curvature, the above
equation could be used to verify that there are no Killing vector
fields on $M$. Therefore, the group of isometries
could only discrete and thus, due to compactness, finite. This result may be used directly to prove that all geodesics in this case are isolated. Another possible use wold be in the attempt to obtain hyperbolic manifolds as ball quotients.

\item Now suppose that we have the flat Minkowski space
$\mathbb{R}^{(1,3)}$ and the D'Alambert wave operator
$$\Box=\partial_1-\partial_2-\partial_3-\partial_4$$ acting on it.
This is a well defined Laplace type operator that possesses a
square root. Namely, we may construct a Dirac operator with the
help of the matrices\footnote{usually referred to as \emph{Dirac
matrices}} $\gamma^\alpha$, satisfying the relations
\begin{equation}
\gamma^\mu \gamma^\nu + \gamma^\nu \gamma^\mu = -2g^{\mu\nu}{\rm
id}\label{Cliff}
\end{equation}
Here $g$ is the flat hyperbolic (Lorenzian) metric and our
operator is simply given by
$$ P = \gamma^\mu \partial_\mu+\gamma^0 $$
Note that the existence of Laplace operators is guaranteed by
the metric and the smoothness of the manifold, whereas their square
root might not exist. Speaking in general, the presence of a Dirac operator, acting on $X$ is a matter of whether the vector bundle of interest admits a spin stricture, or a representation of the \emph{Clifford algebra}, which is the universal unital algebra, generated by vectors in $\mathbb{R}^n$ subject to the relations (\ref{Cliff}). If so, we have a double cover of groups:
\begin{equation}
\mathbb{Z}_2\rightarrow \mathsf{Spin}(n)\rightarrow
\mathsf{SO}(n,\mathbb{R})
\end{equation}
The \emph{Spinor} groups in the low dimensions happen to coincide
with the universal covering groups of the special orthogonal
groups. For example $\mathsf{Spin}(3)\cong \mathsf{SU}(2)\cong
\mathbb{S}^3$, or
$\mathsf{Spin}(4)\cong\mathbb{S}^3\times\mathbb{S}^3$ etc. In
order to investigate the problem of existence of spin structures
on manifolds, one needs to study some topological properties of
the corresponding principle bundle with the help of characteristic
classes. Namely, we use the notion of \emph{Stiefel-Whitney}
classes $\omega_i(V)\in H^i(X,\mathbb{Z}_2)$, where $V$ is a real
vector bundle over $X$. For example, if $\omega_1\neq 0$ (and
certainly $H^1(X,\mathbb{Z}_2)\neq 0$), then the bundle $V$ is not
orientable, as we know from elementary algebraic topology. In the
case it is, the second class $\omega_2$ is an obstruction for the
existence of spin structure. Furthermore, the Abelian group
$H^1(X,\mathbb{Z}_2)$ parametrizes the inequivalent spin
structures on $V$. \\
Using this apparatus, we easily obtain that the spheres of all
dimensions are spin as well as odd-dimensional complex projective
spaces. In the real case $\mathbb{RP}^m$ is spin if and only if
$m=4k-1$ and has two inequivalent spin structures as
$H^1(\mathbb{RP}^{4k-1},\mathbb{Z}_2)=\mathbb{Z}_2$. For a spin
manifold we have at least one Dirac operator and a spinor
connection, that agrees with it. In our case it is given by
$D=\gamma^\nu\nabla_\nu$. Then $D^2$ is known as \emph{the spin
Laplacian} and might be shown\footnote{again using Bochner technique - the above formula is due to Lichnerovicz} to be of the form
\begin{equation}
\triangle_{spin}:={\rm Tr}(\nabla^2)+\frac{\kappa}{4}
\end{equation}
with $\kappa(x)$ being the scalar curvature of the underlying
manifold. This formula helps for retrieving results analogous
to these from the previous example. Namely, we have that if the
scalar curvature on $M$ is non-negative, all harmonic spinors are
parallel and if at least at one point it is positive, then there
are no harmonic spinors at all.\\
\end{enumerate}

\noindent The three major classes of second-order linear $PDE$'s, concerned in
mathematical physics in which  Laplace type operators are involved are:\\
\emph{elliptic equations} of the type $\triangle \phi =
\rho(x)$, where the Laplacial is taken with respect to a metric
with positive (elliptic) signature. The function $\rho$ is usually
interpreted as a volume distribution of the charge and $\phi(x)$
is the potential field. Sometimes we regard the homogeneous
equation with $\rho=0$ to investigate stationary processes
(electrostatics, distribution of heat etc.) \\
\emph{parabolic
equations} involve more (for example time) derivatives of lower
order, which are absent in the leading symbol. Therefore, the heat
equation for instance, known also as continuity or diffusion  $\left (\triangle+\partial_t \right ) \phi=0$ is considered to be with degenerate leading symbol. \\
The last class is the class of \emph{hyperbolic equations} of the
above type $\Box \phi = \rho(x)$, but this time with
hyperbolic metric. One of the most popular examples is the
D'Alambert wave equation, considered in the previous section. In
particular problems the save equations is reduced to a spectral
equation of the type $\triangle\phi=k^2\phi$, known as the
Helmholtz equation.

\subsubsection{Boundary Value Problems}
In the case of closed manifolds the Laplacian appeared to be self-adjoint by construction, but this property is no longer guaranteed when introducing a boundary. According to the \emph{Green formula}, one has to encounter boundary terms as well. In order to get rid of them and to keep the symmetry of $\triangle$, one imposes appropriate conditions for the behavior of the solution and/or its normal derivatives at the boundary.\\
As we may see from the following considerations, there are some naturally preferable ways to do this.

\paragraph{The Green Function and Layer Potentials}
For the so called equations of mathematical physics, it is possible to reduce the problem to the boundary, using single and double layer potentials, defined by means of the Green function and thus transform the initial $PDE$ into an integral equation on the boundary.\\
Let $L(x)$ be a general self-adjoint $PDO$ with varying coefficients and $\Lambda(x,\xi)$ represent its integral kernel, given by the Fourier inverse of the characteristic polynomial. Thus we have
$$L(x)\psi(x) = \int{\Lambda(x,\xi)\psi(\xi)d\xi} = F(x)$$
for any solution $\psi$ and we may write formally
$$ \psi = L^{-1}F(x)$$
where the formal inverse is an integral operator. We write
$$ L^{-1}F(x) = \int{G(x,\xi)F(\xi)d\xi} $$
The kernel $G(x,\xi)$, referred to as the \emph{Green function} of $L(x)$ is determined by the sole definition of inverse $L\circ L^{-1} = {\rm id}$, which leads to
$$ L(x)G(x,\xi) = \delta(x-\xi)$$
Let us now take $L$ to be the Laplace-Beltrami operator $\nabla^2$ in a bounded region $\Omega$, and apply the Green theorem for the Green function and an arbitrary solution $\psi$. We have
$$ \int_{\Omega}{\left [G\nabla^2\psi - \psi\nabla^2 G\right ]d\omega} = \oint_{\partial\Omega}{\left [ G\frac{\partial \psi}{\partial\nu} - \psi\frac{\partial G}{\partial\nu}\right ]d\sigma }$$
where $\partial/\partial\nu = {\bf \nu}.\nabla$ is the inward-pointing normal derivative at the boundary.\\
Considering that $\psi$ is a solution and $G$ is the kernel of the formal inverse, we obtain from the above
\begin{equation}
\psi(\xi)=\int_\Omega{G.Fd\omega}+\oint_{\partial\Omega}{\left [ G\frac{\partial \psi}{\partial\nu} - \psi\frac{\partial G}{\partial\nu}\right ]d\sigma }
\end{equation}
In order to solve this integral equation, we need to get rid of some of the terms. This is being done by imposing appropriate boundary conditions.\\
There are three types of boundary conditions used in this situation:
\begin{enumerate}
\item Dirichlet boundary conditions demand the restriction of $\psi$ on $\partial\Omega$ to be equal to some known function $f$. The solution is obtained if we set $G = 0$ on the boundary in order to eliminate the term involving the normal derivative of $\psi$ on $\partial\Omega$. We finally have
    \begin{equation}
    \psi(\xi) = \int_\Omega{G.F\,d\omega}-\oint_{\partial\Omega}{f\frac{\partial G}{\partial \nu}d\sigma}
    \end{equation}

\item Neumann boundary data prescribes the behavior of the normal derivative of the solution at the boundary $\partial_\nu \psi_{|\partial\Omega }= f$. By analogy with the previous case, one may attempt to set $\partial_\nu G = 0$ on $\partial\Omega$, but such solution would be ill-posed, as one may see, substituting into the Green formula. We may easily get rid of this inconvenience, and still have $\partial_\nu G_{|\partial \Omega}= 0$, if we let the Green operator to be inverse just modulo a constant term
    $$ \nabla^2 G = \delta + C$$
    The constant we really need is the inverse volume of $\Omega$ (taken with 'minus' sign). The Neumann BVP then reduces to
    \begin{equation}
    \psi(\xi) = <\psi>+ \int_\Omega{G.F\,d\omega} -\oint_{\partial\Omega}{G. f\,d\sigma}
    \end{equation}
    with $<\psi>={\rm Vol}(\Omega)^{-1}\int_\Omega{\psi\,d\omega}$

\item The Robin case\footnote{called also 'mixed', or 'modified Neumann' } is obtained if we set a combination of the solution and its normal derivative at the boundary to be equal to a known function $\partial_\nu\psi + \mathcal{S}\psi = f$ on $\partial \Omega$. It is reasonable to ask the Green function to satisfy $\partial_\nu G + \mathcal{S}G = 0$ at the boundary. Then, applying directly the Green formula, one gets again
    \begin{equation}
    \psi(\xi) = \int_\Omega{G.F\,d\omega} -\oint_{\partial\Omega}{G. f\,d\sigma}
    \end{equation}

\end{enumerate}
The boundary terms, appearing in the above expressions are usually referred to as \emph{layer potentials}. More precisely, the \emph{single-}, respectively \emph{double-layer} potentials are defined in terms of the integral operators
\begin{eqnarray}\label{layer}
S\ell(k)[f](x) & = &  \oint_{\partial\Omega}{G(k,x,y) f(y)\,dy(\phi)} \nonumber \\
D\ell(k)[f](x) & = &  \oint_{\partial\Omega}{\frac{\partial G}{\partial \nu}(k,x,y) f(y)\,dy(\phi)}
\end{eqnarray}
Now the question of how we build Green functions, satisfying given boundary conditions, naturally comes to mind. To understand this, one first takes a more general fundamental solution, known as the \emph{free Green function} $G_0$ and then adds correcting function in order to fulfill the demanded behavior at the boundary.\\
Note that the non-homogeneous boundary conditions do not constitute a vector space - for example a sum of two such solutions is no longer a solution, so one may think that he has come upon an obstacle. However it is easy to see that all non-homogenous solutions of a given equation of the above type may be obtained form the homogenous ones via modification of Green function, as we already saw, so that it suffices to consider only the homogeneous case\footnote{the same refers to the question of homogeneity of the equation itself - if $\phi$ is a solution of the homogeneous equation, then $\psi = \phi*f$ satisfies non-homogeneous equation with $f$ instead of zero on the right, since the convolution product is compatible with the operators we are concerned with here}.\\
Let us come back to our initial example - the Laplace-Beltrami operator, acting in a bounded domain. In the two-dimensional case, the free Green function is known to be proportional to $\ln r,\: r=\sqrt{x^2+y^2}$, otherwise $G_0 \sim  r^{2-d},\: d\neq 2 $ - for each dimension there is a characteristic divergency, which is rather topological in nature, but easy to find by analytical methods as well. It is obvious that these $G_0$'s are harmonic, except for the singularity at the origin. Adding a regular harmonic function $h$ does not affect that property and that is exactly how boundary conditions are obtained. \\
For the heat equations, it is easy to show via Fourier transform that the fundamental solution has kernel in the form
$$ G_0(x,y,t) = (4\pi t)^{-d/2}e^{-\frac{||x-y||^2}{4t}}$$
A particular interest for us in the following will be the two-dimensional Helmholtz equation $\nabla^2\psi + k^2\psi = 0$. Its free Green function is shown to be given by  a Hankel function of the first kind
$$G_0(\vec{x},\vec{x}') = \frac{i}{4}H_0^{(1)}[k(\vec{x}-\vec{x}')] $$

\subsubsection{Estimates and Properties of the Spectrum}
Here we study some of the properties that a Laplace spectrum on a compact Riemannian manifold should possess. Most of them could be found either in \cite{Gi} or in \cite{Cz}.\\
First of all, let us consider again the eigenvalue problem
\begin{equation}
\triangle \phi = \lambda_k \phi_k
\end{equation}
with the corresponding boundary condition ($D,N$ or $R$). Then we
have for the spectrum growth rate the following estimate:
\begin{enumerate}
\item $\forall\,\varepsilon > 0\:\,\exists \,n(\varepsilon)\,
\in\mathbb{N},\quad n^{\frac{2}{m}-\varepsilon}\leq \lambda_n\leq
n^{\frac{2}{m}+\varepsilon}$, where $m$ is the dimension of the
underlying manifold $M$ \item let $c_k$ denote the Fourier
coefficients in the expansion of any $L_2$ - function $\psi$ over
the basis $\phi_k$ of eigenfunctions of $\triangle$. Then $\psi$
is smooth if and only if
$\displaystyle{\lim_{n\rightarrow\infty}{n^k c_n} = 0,\quad
\forall k\in\mathbb{Z}}$  \item if in $M$ we have a Riemannian
connection and $||\psi||_k$ denotes the sup-norm of the $k^{-th}$
covariant derivative of the $L_2$ - function $\psi$, then for
sufficiently large $n$ there exists $j(k)$ such that
$||\phi_n||_k\leq n^{j(k)}$
\end{enumerate}
Furthermore, one has some estimates for the first eigenvalue for
closed manifolds of dimension $m \geq 2$.  Namely, if $M$ is such
manifold and its Ricci tensor satisfies
\begin{equation}
\rho(X,X)\geq (m-1)\alpha >0
\end{equation}
for all $X\in TM$ and some positive real constant $\alpha$, then
for the first eigenvalue of the Hodge laplacian we have (by
Lichnerovicz formula) $\lambda_1\geq m\alpha$. On the other hand, for
non-negative Ricci curvature, there is a lower bound
$\displaystyle{\lambda_1\geq\frac{\pi^2}{d_M^2}}$, where $d_M$ is
the diameter of $M$. With this newer result, the above estimate
could be improved as follows:\\
Let $M$ be as before and suppose \begin{equation} \rho(X,X)\geq
(m-1)\alpha, \quad \alpha\in \mathbb{R}
\end{equation}
then there are two cases
\begin{eqnarray}
\lambda_1\geq\frac{(m-1)\alpha}{4}+\frac{\pi^2}{d_M^2}\quad
if \:\alpha\geq 0 \nonumber \\
\lambda_1\leq \frac{(m-1)^2\alpha}{4}+\frac{\beta(m)}{d_M^2}\quad
if\: \alpha < 0
\end{eqnarray}
In the latter $\beta(m)$ is a real constant, depending only on the
dimension.\\
In the special case $m=2$ we also have restrictions concerning
multiplicities $\mu_j$ of the non-zero eigenvalues
. For example, we have $\mu_j\leq 2j+1$
for the sphere, $\mu_j\leq 2j+4$ for the torus, for the Klein
bottle and the real projective plane the upper bound is $2j+3$,
and finally $\mu_j\leq 2j-2\chi(M)+3$ for surfaces of negative
curvature\footnote{here and below $\chi(M)$ denotes the
\emph{Euler number} of $M$}.\\
Note that in the case of a Riemannian surface of genus $g$ we also have the classical estimate for eigenvalue multiplicities:
\begin{equation}
\dim{E_{\lambda_k}} =\frac{1}{2}(2g + k +1)(2g + k + 2)
\end{equation}
\newpage
\subsection{The $\Psi$-calculus and Quantization}
\subsubsection{The Fourier Transform, $\Psi D O's$ and Sobolev Spaces}
The Fourier transform $\mathcal{F}$ is a well-defined endomorphism of the space of Schwartz functions\footnote{these are by
definition the smooth functions $f$ for which there exist
constants $C_{kl}$ such that for every choice of the
multi-indices, the estimate $|x^k D^l f|\leq C_{kl}$ hold for all $k$, which means
that the Schwartz functions, together with all their derivatives,
decrease faster at infinity than the inverse of any polynomial
$|D^l f|\leq C_{kl}(1+|x|)^{-k}$} $S(\mathbb{R}^n)$:
$$\displaystyle{\forall f \in S(\mathbb{R}^n) \quad \hat{f}(p)=\mathcal{F}(f)_p=\int_ {\mathbb{R}^n}{f(x){e}^{- i(p,x)}{ d}^n x}} $$ and with its inverse
$$\displaystyle{ \mathcal{F}^{-1}(\phi)_x =(2\pi)^{-n}\int_
{\mathbb{R}^n}{\phi(p) { e}^{ i(p,x)}{ d}^n p}}$$ it is an
involution for even functions due to the property
$\mathcal{F}^2(\phi)_x=\phi(-x)$. The fixed point of this
involution is easily seen to be the Gaussian distribution
$$f(x)\sim(2\pi)^{-n/2}e^{-|x|^2/2}$$ as one may easily
check with a simple calculation.\\
The map $(2\pi)^{-n/2}\mathcal{F}$ is a unitary automorphism of the
Schwartz space of function due to the Perseval-Plancherel theorem
\begin{equation}
||\hat{f}||_{\mathbb{L}_2}=(2\pi)^{n/2}||f||_{\mathbb{L}_2}
\end{equation}
No matter that $S$ is the maximal closed under $\mathcal{F}$
subspace of $\mathbb{L}_2$, it can naturally be prolonged over its
dual space - the distributions in $S'(\mathbb{R}^n)$ by the
formula\footnote{unlike $(\:,\:)$ that we use for the usual dot
product in $\mathbb{R}^n$, here $\langle \:,\:\rangle $ denotes
the scalar product in $\mathbb{L}_2$}
\begin{equation} \langle \mathcal{F}u,v\rangle :=\langle u,\mathcal{F}v\rangle
\end{equation} For example $\mathcal{F}(\delta)=1$ for the
Dirac $\delta$-function etc.\\
One of the main advantages of the Fourier transform for the theory
of PDE is that it turns differentiation into multiplication. More
precisely, for every multi index $\alpha$ we have \begin{eqnarray}
\mathcal{F}(D^\alpha f)=\xi^\alpha f\qquad D= i\frac{\partial}{\partial x}
\end{eqnarray}
which allows for expressing a constant-coefficient differential
operator by means of a polynomial in the $\xi$-variable:
\begin{equation}
P=\sum{a_i
D^i}=\textsc{Op}(\Tilde{P}):=\mathcal{F}^{-1}\Tilde{P}\:\mathcal{F}
\end{equation}
Recall that the polynomial $\sigma(P) = \Tilde{P}=\sum{a_i\, \xi^i}$ is called
\emph{the symbol} of $P$ (often denoted $P\,(x,\xi)$ for simplicity) and we write $\textsc{Op}(\Tilde{P}) = P\,(x,D)$ for an operator with varying coefficients whose symbol is $\Tilde{P}$.\\
The notion of symbol allows for expressing easily the inverse
of various differential operators. For example, the resolvent of
the constant-coefficient operator $1-\triangle$ is simply given by
$$\displaystyle{(1-\triangle)^{-1} = \textsc{Op}\left (\frac{1}{1+|\xi|^2} \right )}$$\\
For the $x$-dependent case there is a natural extension of this
formalism as we shall see later.\\
Another familiar example is the resolvent of the heat equation, which is of particular interest for our considerations. As it not hard to prove by a simple Fourier technique, it has the form $\textsc{Op}\,(\rm {e}^{t|\xi|^2})$, or as we shall formally write, $\rm{e}^{-t\triangle}$.\\
One may alternatively define the action of an operator $P$, acting on $S(\mathbb{R}^n)$ by its integral kernel $K_p\,(x,y)$, given by the symbol via Fourier transform
\begin{equation}
P\,(f)_x=\int_{\mathbb{R}^n}{K_p\,(x,y)f(y){d}y}=(2\pi)^{-n}\int_{\mathbb{R}^n}{
{e}^{i(x-y).\xi}\Tilde{P}(x,\xi)f(y)\, dy\,d\xi}
\end{equation}
We remark that the integral on the right side does not necessarily
converge absolutely, so in the generic case we are not allowed to
interchange the order of integration.\\
Note that for integral kernels the sensible composition law is given by the
\emph{convolution product}\footnote{sometimes called also \emph{star product}}
\begin{equation}
(f*g)(x):=\int{f(x-y)g(y)\rm{d}y}=\int{f(x)g(y-x)dy},\qquad
f,g\in S(\mathbb{R}^n)
\end{equation}
which is a kind of generalization of the usual matrix
multiplication for continuous indices.\\
Although we define the star product for the Schwartz class it
extends to a broader functional space. For example in
$S(\mathbb{R}^n)$ the convolution algebra does not posses unit
element. However, as we know, the $\delta$-distribution is such an
element in the extended case. There are some basic properties of
$*$ we shall take advantage of in the following:
\begin{eqnarray}
\delta* f & = & f \nonumber \\
\hat{f}*\,\hat{g} & = & \widehat{f.\,g},\qquad \hat{f}.\,\hat{g}=\widehat{f*\,g} \nonumber\\
*\,\triangle & = & \triangle* \nonumber \\
f* (g* h) & = & (f*\,g)* h
\end{eqnarray}
From now on we shall use both the symbol and the kernel of the
operators in interest.\\
As we saw in the examples, the inverse of a differential operator
is no more differential, but belongs to a broader class of
operators. For the study of these classes we investigate the
spaces of symbols. Therefore we introduce the notion of the \emph{Sobolev
space} ${\rm H}^s$. For integer values of $s$, this is simply the
space of functions with square integrable derivatives up to order
$s$. The norm in this case is naturally expressed by means of the
$\mathbb{L}_2$-norm with the formula
$$\displaystyle{||f||_s=\sum_{|\alpha|\leq s}{||D^\alpha f||_{\mathbb{L}_2}}}$$
However, the definition is extended to all real values of $s$ as
follows: \begin{eqnarray} f\in{\rm H}^s \Leftrightarrow
f(1+|\xi|^2)^\frac{s}{2}\in{\mathbb L}_2 \nonumber \\
||f||_s:=\int{(1+|\xi|^2)^s|\hat{f}(\xi)|^2\,d\xi}
\end{eqnarray}
Actually, the first follows from the definition of the norm,
thanks to the Perseval-Plancherel theorem.\\
Of course, for integer $s$ both definitions agree and of
course\footnote{therefore from now on we shall write $||\,.\,||_0$
instead of $||\,.\,||_{\mathbb{L}_2}$} ${\rm H}^0\equiv
\mathbb{L}_2$. Moreover, each two positive real numbers $s\leq t$
we have the inclusion
\begin{equation}
S(\mathbb{R}^n)\subset{\rm H}^{-t}\subset{\rm H}^{-s}\subset
{\mathbb L}_2\subset{\rm H}^s\subset{\rm H}^t\subset
S'(\mathbb{R}^n)
\end{equation}
In this hierarchy of spaces we have quite a natural descending map
- derivation. Namely the derivative $D^\alpha$ extends to a
continuous map $$D^\alpha : \: {\rm H}^s\rightarrow {\rm
H}^{s-|\alpha|} $$ Intuitively, by applying this map, we loose the
first $|\alpha|$ square-integrable derivatives of $f$ and fall
down on a lower stair in the
hierarchy.\\
There are several nice properties, concerning the topology of the
Sobolev spaces, which we omit for the sake of simplicity, but
instead we give a good interpretation of ${\rm H}^{-s}$ as the
dual of ${\rm H}^{s}$ with respect to the $\mathbb{L}_2$ scalar
product, which extends to a perfect pairing between ${\rm H}^{s}$
and ${\rm H}^{-s}$.\\
Baring these things in mind, we are ready to give the definition
of a pseudo-differential operator. First let us consider a
function
$p\,(x,\xi):\:\mathbb{R}^n\times\mathbb{R}^n\rightarrow\mathbb{C}$,
such that
\begin{enumerate}
\item $p\,(x,\xi)$ is smooth in both arguments and compactly
supported in the $x$-space \item $\forall \,\alpha,\beta\in
\mathbb{N}^n \:\:\exists \,C_{\alpha,\beta},\quad |D_x^\alpha
D_\xi^\beta p\,(x,\xi)|\leq C_{\alpha,\beta}(1+|\xi|)^{d-|\beta|}$
\end{enumerate}
We call such a function a \emph{symbol of order $d$}  ($p\in
S^d,\:\: d\in \mathbb{R}$) and assign to it an operator in the
familiar way
$$P\,(x,D)f=\textsc{Op}\,(p\,(x,\xi))=(2\pi)^{-n/2}\int_{\mathbb{R}^n}{{\rm
 e}^{i(x-y).\xi}p\,(x,\xi)f(y)\, dy\,d\xi}$$
Such $P$ we call the \emph{pseudo-differential operator}(of order
$d$), associated with (or having as its symbol) $p\,(x,\xi)$.\\
First of all we need a few notes. The well-known differential
operators are a private case of $\Psi D O$'s according to the
above definition - the ones with polynomial symbols. Their order
as differential operators (the degree of the symbol) certainly
coincides with the order given by our new definition. For a wider
class of operators, however, the order is not bound to be an
integer. For example, given a compactly supported smooth function
$f(x)\in C_0^\infty (\mathbb{R}^n)$, for all real $d$ the symbol
$p\,(x,\xi)=f(x)(1+|\xi|^2)^\frac{d}{2}\in S^d$ defines correctly
a $\Psi DO$ of order $d$.\\
Another feature of the pseudo-differential operators is that they
generalize the role of the $PDO$'s in the sense that each $\Psi
DO$ of order $d$ extends to a continuous map $$ P:\: {\rm
H}^s\rightarrow {\rm H}^{s-d}$$ for all real values of $s$.\\
And at the end of this section we give one more very important
definition. We say that $P(x,D)$ is \emph{smoothing} if its symbol
belongs to $S^d$ for all $d$, and that is $C^\infty$, but we write
$p\in S^{-\infty}$. One can show that in this case $P(u)$ is
smooth, regardless of weather $u$ is smooth or not.\\

\subsubsection{Symbolic Calculus, Weyl Quantization and {\emph FIO}'s}
First of all we note that the operators under consideration constitute an
algebra with respect to the operator composition and taking the adjoint.
The infinitely smoothing operators form its ideal and in practice we mod out by it to obtain a nice factor-algebra.\\
We use the notation $P\in \Psi_d$ to show that $p\in S^d$ and
$\displaystyle{\Psi_{-\infty}=\cap_d {\Psi_d}}$ will denote the
space of infinitely smoothing operators. We also use $P\sim Q$ to
show that operators (as well as the associated symbols) fall in
the same equivalence class  - differ only by infinitely smoothing
term. We have the formulae\\
\begin{eqnarray}
\sigma (P^*) & \sim & \sum_\alpha{\frac{1}{\alpha !}d^\alpha_\xi
D^\alpha_x p^*} \nonumber \\
\sigma (PQ) & \sim & \sum_\alpha{\frac{1}{\alpha !}d^\alpha_\xi
p\, D^\alpha_x q}
 \end{eqnarray}
whenever both sides are well-defined. In particular, if we consider only the \emph{leading symbols} (the highest-order homogeneous parts of the symbol functions), we have simply
$$\sigma_L(PQ)=\sigma_L(P)\,\sigma_L(Q)$$
Moreover, the algebra thus defined, is graded, in the sense that
if $P\in \Psi_l$ and $Q\in \Psi_m$, then $PQ\in\Psi_{l+m}$. The
above remains true also in the case of matrix-valued symbols.\\
This construction may be used for example to construct asymptotic
series of operators. More precisely, if $p_j\in S^{d_j},\:
d_j\rightarrow -\infty$ we have $\displaystyle{p\sim \sum_j{p_j}}$
which is unique modulo $S^{-\infty}$. \\
Pseudo-differential operator possess the important property of
\emph{pseudo-locality} which means that if $f\in H^s$ is smooth in
some open set $U$ then $Pf$ is smooth there as well.\\
In practice one has to deal with $\Psi DO$'s defined on manifolds,
rather then on $\mathbb{R}^n$ as we implicitly assumed so far. The
generalization, however, is straightforward using partition of
unity as long as one keeps in mind that the Sobolev spaces the
operators live in are preserved under diffeomorphisms and the
leading symbols transform in a covariant way. This gives a
sufficiently nice transition between coordinate charts, so that
modulo $\Psi_{-\infty}$ everything works fine.

\noindent Now we mention some important classes of $\Psi DO$'s
that appear naturally in calculus and mathematical physics.

\paragraph{Elliptic and Fredholm Operators}
Generally we call $P\in\Psi_d$ \emph{elliptic} if it is invertible
modulo a $\Psi_{-\infty}$ operator. More precisely, this would
mean that there exists $Q\in\Psi_{-d}$ such that $PQ\sim{\rm id}$
and $QP\sim {\rm id}$. Elliptic operators have some very nice
properties compared to the generic case. One of them is the
so-called G\"{a}rding's inequality, which states that if
$P\in\Psi_d$ is elliptic in a region $U$ and $f \in C^\infty_0(U)$, then the estimate
$$ ||f||_d\leq C(||f||_0+||Pf||_0)$$
always holds for some constant $C$.\\
Note that the sum of two elliptic $\Psi DO$'s need not be elliptic. This is
always true, however, if they have both positive symbols. This
allows for defining an elliptic $\Psi DO$ on a manifold again
using a partition of unity. A crucial point in the theory of
elliptic operators is the fact that when defined on closed
manifolds, they are Fredholm. \emph{Fredholm operators} are usually
defined as bounded linear maps between Hilbert spaces ${\rm Fred}(L,M)\in {\rm
Hom}(L,M)$ that are invertible modulo a compact operator, which
means that if $F\in {\rm Fred}(L,M)$ there exist operators $G,H\in {\rm Hom}(M,L)$ such that
$FG-{\rm id}$ is compact in $M$ and $HF-{\rm id}$ - compact in $L$
respectively. Such operators have finite-dimensional kernel and
co-kernel (the null-space of the adjoint operator). Moreover,
their image is closed (and the same refers to the image of the
adjoint). It is not very hard to see that each operator with these
properties is Fredhom and that they survive under composition and
taking the adjoint. The above properties of Fredholm operators are
widely used - for instance, for the derivation of Hodge
decomposition in algebraic topology, or for the definition of
index.

\paragraph{k-$\Psi DO$'s} For a natural generalization of the notion of a pseudo-differential operator, that is often used in physics, we first define the $k$-Fourier transform by the formula
\begin{equation}
\mathcal{F}_k[u](\xi) = \hat{u}_k(\xi)=\int_{\mathbb{R}^n}{e^{-ik\langle
\xi,x\rangle}u(x)\,d^n x}
\end{equation}
and to each symbol function\footnote{in the sense, clarified
above} $p(x,\xi)$, we associate an operator in the familiar way
$$P[u] = \textsc{Op}\,(p)[u] = \mathcal{F}_k^{-1}\,[p\,]\,\mathcal{F}_k\, [u] $$
In more explicit terms what we have is
\begin{equation}
P[u](x)=\left ( \frac{k}{2\pi} \right )^n\int_{\mathbb{R}^n}{e^{ik\langle x,\xi\rangle }p\,(x,\xi)\,\hat{u}_k(\xi)\,d^n\xi}
\end{equation}
The additional $k$-dependence appears in various problems as we
shall be able to make sure very shortly. Apart from the
$k$-dependence hidden in the Fourier transform, one may have to
deal with a more explicit one - in the symbol itself. In this case
it is a standard situation to observe such dependence also in the estimate for $p$ from the definition. Namely, one may be interested in symbol functions for which
$$\forall \,\alpha,\beta\in
\mathbb{N}^n \:\:\exists \,C_{\alpha,\beta},\quad |D_x^\alpha
D_\xi^\beta p_k\,(x,\xi)|\leq C_{\alpha,\beta} k^r
(1+|\xi|)^{d-|\beta|}$$ and then one has an additional
$r$-parameter which appears to be very helpful. For example, if
the above is true for a whole series $p^k_{j}$ with corresponding
$r_j$ than monotonically tend to minus infinity, then this series
is an asymptotic expansion of some symbol $p_k$ in the sense that
$$p^k(x,\xi)\sim\sum_j{p^k_{j}(x,\xi)}$$
The above has the meaning of a power series expansion where the
reminder is bounded and the bound is governed by a term of order $k^{-\infty}$.\\
One significant subclass here is the class of the so-called \emph{classical
symbols} for which the above expansion is in integer powers\footnote{note that the absence of upper index on the right hand side means that the symbols $p_j(x,\xi)$ fulfill only the standard estimate for a generic symbol function, whereas for $p^k(x,\xi)$
it is with an additional factor $k^r$ as previously introduced} of $k$:
$$p^k(x,\xi)\sim\sum_j{k^{r-j}p_j(x,\xi)} $$
In particular, the leading symbol of the above is $\sigma_L = k^r
p_0 (x,\xi)$.\\
The space of operators, associated to such classical symbols is
usually denoted by $\Psi^r_{cl}$. We note that a lot of $\Psi
DO$'s appearing in mathematical physics are of this type. For such
operators we still have a nice-looking composition rule - let
$P_1,P_2\in \Psi^0_{cl}$ and $p_1,p_2$ denote their symbols
respectively. Then if $Q=P_1 P_2$, we have for its symbol
$q=\sigma(Q)$ the power series expansion
\begin{equation}
q(y,\eta)=\exp \left [\frac{i}{k}\langle D_x,D_\xi \rangle \right
] p_1(y,\xi)\,p_2(x,\eta)
\end{equation}
evaluated at $x=y,\,\xi=\eta$.\\

\paragraph{Weyl Quantization}
We already saw that (at least formally) a $\Psi DO$ is defined by
its symbol (and vise versa) via the formula\footnote{we refer to
the more generic case of a $k-\Psi DO$, the usual one is, of
course, the same with $k=1$}
$$ P[u](x)=\left ( \frac{k}{2\pi}\right )^n\int_{\mathbb{R}^n}{p\,(x,\xi)\,u(y)\,e^{ik\langle
(x-y),\,\xi\rangle }d^n\xi\,d^n y} $$ In some cases, however, it
is more convenient to use a slightly different definition of a
symbol, we shall refer to as \emph{Weyl symbol} and denote $p^W$,
in order to have the above in the form
\begin{equation}
P[u](x)=\left ( \frac{k}{2\pi}\right )^n\int_{\mathbb{R}^n}{p^W\left
(\frac{x+y}{2},\xi \right )\,u(y)\,e^{ik\langle
(x-y),\,\xi\rangle }d^n\xi\,d^n y}
\end{equation}
Note that this does not introduce a different operator, but simply
offers another formulation. The relation between the two symbols
is given by the formula\footnote{here and below we use the
notation $D=-i\nabla$}
\begin{equation}
p^W(x,\xi)=\exp \left[ \frac{1}{2} D_x D_\xi \right ] p\,(x,\xi)
\end{equation}
Note that an alternative way to define a $\Psi DO$ is by its
integral kernel. Namely, if $K_p(x,y)$ a matrix-valued function
with compact $x$-support and $f(y)$ is vector-valued, with compact
support, then we define the operator
$$Pf(x)=\int{K_p(x,y)f(y)dy}$$
as a $\Psi DO$ that is $\displaystyle{\Psi_{-\infty}}$ if $K_p$ is
smooth. For the case of Weyl symbols one may use the Fourier
inversion formula to find the one-to-one correspondence between
the two representations.
It is obvious from the above that $$K_p\left
( \frac{x+y}{2},\frac{x-y}{2}\right )=\left (\frac{k}{2\pi} \right
)^n\int{p^W(x,\xi)e^{ik\langle y,\xi\rangle}d^n y}$$ and
then, by Fourier inversion
$$p^W(x,\xi)=\int{K_p\left ( \frac{x+y}{2},\frac{x-y}{2} \right )
e^{-ik\langle x,\xi\rangle}d^n y}$$ One of the main
advantages of the Weyl symbol is that (in terms of operator
multiplication) it is written in explicitly symplectic-invariant form.
In order to see this one needs to regard first $(x,\xi)$ as a point
$z$ in the cotangent bundle over the $x$-space, endowed with a natural canonical symplectic form $\omega$. Then if $PQ=R$, for the corresponding symbols it is true that
$$r(z)=p^W(z-\frac{1}{2}\omega D_z)q(z)=q^W(z+\frac{1}{2}\omega D_z)p(z)$$
which can be interpreted in the following way: let $\gamma\in
\mathsf{Sp}(n,\mathbb{R})$ and $\gamma^*$ denote its natural
pull-back to functions $\gamma^* f (z) = f(\gamma^{-1}z)$. It is
then a matter of computation to show that
$$\gamma^* p\,(x,D)\, \gamma^* q\,(x,D) = \gamma^* r\,(x,D)$$
Using this invariance we can introduce a quantization scheme that
is mostly used for $\Psi DO$'s. Let $\textsc{Op}(p^W_j)\in\Psi^0_{cl},\:
j=1,2$ and let $Q=P_1P_2$ denote their product with Weyl symbol $q^W$. Then
the latter is still a classical symbol and its asymptotic
expansion is given by the formula
\begin{eqnarray}
q(x,\xi)\sim \sum_{k=0}^\infty{ \frac{1}{k!}\left ( \frac{i}{2k}
(D_\eta D_x - D_\xi D_y)\right )^k
p^W_{1,k}(x,\xi)\,p^W_{2,k}(y,\eta)}\mid_{\,\eta=\xi,\:
y=x}\nonumber \\
=\sigma_L(p^W_1)\sigma_L(p^W_2)(x,\xi)-\frac{i}{2k}\{
p^W_{1,1},p^W_{2,1}\}(x,\xi)+\mathcal{O}(k^{-2})
\end{eqnarray}
where the symbols $p^W_{1,k}$ and $p^W_{2,k}$ are those that take
part in the asymptotic power series representing $p^W_1$ and
$p^W_2$ respectively, and $\{ .\, ,.\}$ denotes the Poisson bracket.\\
One may readily see that the above formula leads to a natural
correspondence (modulo terms of order $k^{-2}$) between the classical
quantities (functions on $T^*X$) and their
dynamics, defined by the Poisson bracket on the one hand, and the
quantum observables (operators, acting in Hilbert space) with the
commutator algebra on the other. More precisely, from the above we
easily derive the formula
\begin{equation}
\sigma[P,Q]=\frac{1}{ik}\{\sigma(P),\sigma(Q)\}+\mathcal{O}(k^{-2})
\end{equation}
The Weyl quantization is quite convenient physical considerations as it defines
the operator by its natural action on plane waves. Moreover, it
allows for an easy expression of the expectation value
$$\langle \psi\mid \textsc{Op}(p^W)\mid \psi\rangle =\left ( \frac{k}{2\pi}\right )^n
\int_{\mathbb{R}^n}{p^W(x,\xi)\,W_\psi\, d^n x\,d^n\xi} $$ where the
Wigner function $W_\psi$ is defined as
$$ W_\psi(x,\xi) =  \int_{\mathbb{R}^n}{\bar{\psi}(x-y/2)\psi(x+y/2)e^{-ik\langle y,\xi\rangle} d^n y} $$

\paragraph{A Glimplse On Fourier Integral Operators}
We note that the set of all $k-\Psi DO$'s forms an
infinite-dimensional Lie algebra with the grading
$$[\Psi^{r_1},\Psi^{r_2}]\in \Psi^{r_1+r_2-1}$$
and in particular $\Psi^1$ forms a sub-algebra. It is interesting
to see what are the corresponding Lie groups (if there are any) to
these algebras. It turns out that these groups exist and are
formed by invertible $k$-Fourier integral operators. To introduce
$k-FIO$'s one takes the standard construction of a classical
$k-\Psi DO$ via an oscillatory integral and extends a little.
Namely, the action of a $k-FIO$ is given by an expression of the
form \begin{equation}
\mathcal{F}_k[u](x)=\left ( \frac{k}{2\pi}\right )^d\int_{\mathbb{R}^d\times\mathbb{R}^m}{e^{i k\phi(x,y,\xi)}\,f(x,\xi)\,u(y)\,dy\, d\xi}
\end{equation}
where $f\in S^r_{cl}$ is a classical symbol in the above sense and
the non-degenerate \emph{phase function} $\phi$ is homogeneous map of
degree one in $\xi$ and generates a conic Lagrangian manifold
$\Lambda\in T^*M \times T^*M / \{0\}$. We may also restrict
ourselves to the specific case where $\Lambda$ is a graph of
symplectic diffeomorphism $\eta$ and $\phi$ - its generating
function. With this interpretation in mind it is natural to write
the above oscillatory integral in the form
\begin{equation}
\mathcal{F}_k[u](x) = \left ( \frac{k}{2\pi}\right )^d\int_{\mathbb{R}^d\times\mathbb{R}^m}{e^{i k\left (S(x,\xi)-\langle y,\xi\rangle\right
)}\,f(x,\xi)\,u(y)\,dy\, d\xi}
\end{equation}
With the choice $S(x,\xi)=\langle x,\xi\rangle$ one recovers the
definition of a $k-\Psi D O$. In this case, as it is not hard to
see, $\eta = {\rm id}$ and $\Lambda$ is the diagonal in $T^*M
\times T^*M / \{0\}$. Hence, we have the imbedding $$ DO \subset
\Psi DO \subset FIO $$ Now let us just enumerate some of the nice
properties of the $FIO$'s that make them an often preferred tool for
calculations.\\
First of all, due to their invariance under diffeomorphisms, these
$FIO$'s can be defined on manifolds\footnote{in order to avoid
confusions, let us restrict our attention to compact ones for the
time being} as bounded linear operators. Moreover, they extend to
such maps between Sobolev spaces, just as the $\Psi D O$'s do.
Second, these operators are not pseudo-local, but posses the more
general property to make the wavefront evolve in a fairly 'nice' way, namely
$$ WF(F[u])\subset \Lambda \circ WF(u)$$
It is also a important to see how these operators
behave under composition and conjugation. The answer is quite
satisfactory:
\begin{enumerate}
\item Let $F_i,\: i=1,2$ be two $FIO$'s of order $k_i$ and
$b_i(\xi_i,x,y),\:\phi_i(\xi_i,x,y)$ denote their amplitudes and
phase functions respectively. Then $F_1\circ F_2$ is again a $FIO$
of order $k_1+k_2$ with amplitude $b_1(\xi_2,x,y)b_2(\xi_2,x,y)$
and phase function $\phi_1(\xi_1,x,y)+\phi_2(\xi_2,x,y)$. The
corresponding diffeomorphism is just a composition of the
morphisms, defined by $F_1$ and $F_2$, namely
$\eta_{12}=\eta_1\circ \eta_2$. \item Now let $F^\dagger$ denote
the conjugate of $F$. Then it defines a diffeomorphism $\eta^{-1}$
and is given by an amplitude $b^*(\xi,x,y)$ and phase function
$-\phi(\xi,x,y)$. From here we see that the product of a $FIO$
with its conjugate is actually a $\Psi DO$.
\end{enumerate}

\noindent We give clear account of the fact that the above exposition on pseudo-differential calculus and applications hardly covers the minimum for any decent survey. However the hope is most of all to draw the reader's attention to the topic and to the matter and ensure a sufficient platform for further exploration. Our basic guides in the second half of this section are \cite{Gi2} as well as the appendix of the dissertation \cite{Fur}. Both works have the advantage of clear and plain exposition. A more thorough investigation could be found in H\"{o}rmander's monograph on partial differential equations.

\newpage
\section{Trace Formulae}
Throughout the present work we became familiar with some basic concepts in modern (and not that modern) mathematical physics, starting with integrability, quantization and arriving at the theory of $PDE$'s and $\Psi$-calculus at the end. No matter how significant these things are for the whole theory, we have to admit that nothing yet has been said about spectral geometry itself.\\
The trace formulae introduced below are the basic tools of spectral geometry and reveal the essence of its philosophy. Only in this section we have systematically exposed the theory of the correspondence between spectra, geometry and dynamics.\\
To start with, we recall the classical \emph{Poisson summation} formula:
\begin{equation}\label{Poisson}
\sum_{k\in\mathbb{Z}^n}{f(k)}=\sum_{k\in\mathbb{Z}^n}{\int_{\mathbb{R}^n}{f(\rho)\,e^{2\pi i\langle \rho, k \rangle }d\rho}} = \sum_{k\in\mathbb{Z}^n}{\hat{f}(k)}
\end{equation}
The above formula is crucial for the theory of theta and zeta functions, and it relates the sum over all lengths of closed geodesics (the right hand side) to the spectrum of the self-adjoint operator $\sqrt{\triangle}$ on the torus $\mathbb{T}^n$ (on the left). Thus, it inspires a whole series of formulae, involving operator traces and geometric quantities to be discussed below. \\
Even the simplest case that we consider here can be very helpful in practical computations. It is worth nothing for example to show the extremely valuable identities concerning traces on the unit circle
\begin{eqnarray}
\sum_{k\in\mathbb{Z}}{e^{-k^2\tau}} & = &
\sqrt{\frac{\pi}{\tau}}\sum_{m\in\mathbb{Z}}e^{\frac{m\pi^2}{\tau}}
\nonumber \\ \sum_{k\in\mathbb{Z}}{\frac{1}{x^2-k^2}} & = & \frac{\pi}{x}\cot{\pi x} \nonumber\\
\sum_{k\in\mathbb{Z}}{\delta(x-k)} & = & 1+\sum_{m=1}^\infty{\cos{2\pi m x}}\nonumber
\end{eqnarray}
One of the first generalizations was made by Selberg, who obtained a trace formula for a particle, moving freely on a negatively curved surface $X$. The formula relates the spectrum of the Laplace-Beltrami operator over such hyperbolic surface to the length spectrum of the closed geodesics over it. For a smooth, even test function $h$ on $X$, with compactly supported Fourier transform $\hat{h}$, the formula reads
\begin{equation} \label{Selberg}
\sum_{j=0}^\infty{h(\rho_j)} = \frac{{\rm Vol}(X)}{4\pi}\int_{\mathbb{R}}{ h(\tau)\tanh{(\pi\tau)}\tau\, d\tau}+\sum_{\gamma\in H_*}\sum_{k=1}^\infty{\frac{\ell_\gamma\hat{h}(k\ell_\gamma)}{2\sinh{(k\ell_\gamma/2)}}}
\end{equation}
where $H_*$ is the set of primitive closed geodesics and $\ell_\gamma$ - their lengths.
The quantities $\rho_j$  are related to the spectrum of the Lapacian by
$ \rho_j^2 = \lambda_j + \kappa/4,\: \lambda_j\in Spec(\triangle)$, but we may take the scalar curvature with the normalization $\kappa = -1$. Then the left hand side is, according to Lichnerovicz formula, a symmetric function of the Dirac operator with eigenvalues $\rho_j$.\\
Selberg trace formula appears to be a powerful tool for relating classically chaotic motion to quantum spectra and apart form that has proved useful in scattering theory\footnote{recall that the monodromy matrix constitutes a representation of $\mathsf{SU}(1,1)$, which us a principle bundle over the Poincar\'e disk}.One may attempt for example to restore the whole geometry of a hyperbolic surface $X$ from the spectrum $\{\lambda_j\}$. The idea is to determine one by one the lengths of the primitive closed geodesics on $X$, which are known to be isolated and represent the fundamental cycles generating $\pi_1(X)$. Once $\pi_1$ is determined, $X$ is easily represented as a quotient of the Poincar\'{e} unit disk. \\
There is one rather algebraic formulation of the above trace formula, that relies on the notion of induced representations. Let $G$ be a Lie group, $\Gamma$ - its subgroup and let $L:\Gamma\rightarrow H(L)$ be a representation of $\Gamma$. We use $F$ to denote the space of functions $f:G\rightarrow H(L)$ with the property $\forall \gamma\in \Gamma,\: f(\gamma.g)=L(\gamma)f(g)$. Moreover, the space $F$ is equipped with a scalar product induced by the scalar product in $H(L)$
$$ \langle f_1|f_2 \rangle = \int_{G/\Gamma}{(f_1,f_2)_{H(L)}d\mu_g}$$
where $\mu$ is the Haar measure on $G$.\\
Now the representation we call \emph{induced} by $L$ and denote $U^L$, appears naturally as
$$U^L:G\rightarrow F,\quad U^L(h)\cdot f(g) = f(g\cdot h) $$
A crucial property of induced representations is the following
$$U^{L\oplus M} = U^L\oplus U^M$$
From here it follows immediately that if $L$ is reducible, so is $U^L$ and vice versa.\\
Now let $U^L$ be completely reducible $U^L = \oplus_j n_j M^j $. Then we have for its trace
$${\rm Tr}U^L = \sum_j{n_j {\rm Tr M^j}} = \int_G{M(f)d\mu(M)}$$
The Selberg trace formula then concerns the trace of the induced representation. It states \begin{equation}
 {\rm Tr}U^L = \int_{G/\Gamma}\,\int_\Gamma{f(g.\gamma.g^{-1})\chi_L(\gamma)\,d\gamma\, d\nu(g)}
 \end{equation}
Here $\nu(g)$ is the normalized measure on $G/\Gamma$ induced by $\mu$.\\
In the trivial case when $\Gamma=\{e\}$ and $G$ is abelian, we recover the classical
Poisson summation formula
$$\int{f(\gamma)\,d\gamma} = \sum_{\chi\in\Gamma^\bot}{\hat{f}(\chi)} $$

\subsection{The Heat Kernel and $\zeta$-regularization of the Effective Action}
One of the best known examples of trace formulae in mathematics
and physics is the so called \emph{Weyl expansion}\footnote{sometimes called also
\emph{heat content} or \emph{Seeley-DeWitt expansion}} for the asymptotic fundamental
solution of the heat equation. Here we are going to briefly introduce this technique and give some examples of its applicability from quantum gravity.\\
\subsubsection{High-temperature Expansion}
We consider the heat equation
$$\displaystyle{(\triangle+\frac{\partial}{\partial t})\phi=0}$$
mainly for closed manifolds but also as a boundary-value problem. In both
cases one may construct a \emph{parametrix}, or a fundamental solution modulo
$\Psi_{-\infty}$, given by a power series in $\sqrt{t}$ with locally computable
coefficients $a_k(x)$, depending only on the curvature and its derivatives.\\
To start with, the kernel of the resolvent for $\mathbb{R}^n$ is known to be of the form
$$\displaystyle{K_0(x,y,t) = (4\pi t)^{-\frac{n}{2}}\,e^{-\frac{||x-y||^2}{4t}}}$$
For a curved manifold, however, the Laplacian undergoes certain deformation and
so does its resolvent. Nevertheless, some basic features remain
unchanged: namely, the homogeneity in $|x-y|/\sqrt{t}$, the
translation invariance and the exponential damping for large
'distances'.\\
Now let us consider a perturbative expansion of the resolvent
kernel for a generic closed manifold. It is quite obvious that the
Euclidian laplacian is the zeroth approximation in
$$\triangle=\triangle_0+\epsilon\triangle_1+\epsilon^2\triangle_2+\ldots$$
and respectively $K^0$ is the zeroth order term in the solution. Substitution in
the non-perturbed equation yields
\begin{equation} \label{MP}
\displaystyle{(\triangle+\frac{\partial}{\partial
t})\,K_0=\delta+R} \end{equation}
where $R$ is a lower-order kernel. We claim that the first correction term\footnote{here $*$
stands for the convolution product for the integral kernels defined in the usual way}
is given by $K_0-K_0* R$. This is easy to verify by a
straightforward substitution in the initial equation - here we
take advantage of the fact that $*$ commutes with the heat
operator and $\delta * R = R$ by the very definition of $\delta$. Thus
we have cancelation of the first-order residue $R$, which is  replaced by a lower
order term $-R* R$. To get rid of it, following the same idea, we correct the Green
function by adding the term $K_2=K_0* R* R$, which generates on its side another
residual term of lower order. With this iterative procedure one
obtains a convergent series of the form
\begin{equation}
K\sim \sum_{k=0}^\infty{K_j},\qquad K_j:=(-1)^j K_0\underbrace{* R\ldots *R}_{j} \label{MP}
\end{equation}
and as $j$ tends to infinity, the order of the residual term in
the approximation tends to minus infinity, so the above series
represents by definition the parametrix of the heat equation for a
generic smooth compact manifold without boundary. We note that
this construction is in fact quite natural - this is nothing but a
convergent geometric series in Banach space, where the
usual multiplication is replaced by convolution of Voltera kernels
and the condition $|R|<1$  makes sense once the corresponding
operator norm is introduced.\\
Let us consider now the kernel $K(x,y,t)$. As the general theory
says, it is a well defined smooth function everywhere in its
domain, except on the diagonal $x=y$ where it has a singularity at
$t=0$. The idea of the \emph{Weyl} expansion is to give an asymptotic representation of
$K(x,x,t)$, or more precisely of its parametrix, in the limit $t\rightarrow
0$ as a power series. This is being done with the help of the theorem of \emph{Minakshisundaram}
and \emph{Plejiel}, stating that the parametrix of $K$ is really expandable in the form
\begin{equation}
K(x,x,t)\sim \frac{1}{(4
\pi)^\frac{n}{2}}\sum_{k=0}^\infty{\tilde{a}_{k}(x) t^{k-n/2}}
\end{equation}
modulo infinitely smoothing terms with $d$ being the dimension of the underlying manifold.\\
One immediate consequence of the \emph{MP} theorem is
that the coefficients in this series are spectral invariants, that
locally depend polynomially, as we shall see later, only on the
curvature and its derivatives. The first and simplest ones have very clear
geometrical meaning\footnote{here and below we adopt the notation
$\displaystyle{a_k=\int_X{\tilde{a}_k(x)}}$ for the integrated invariants}. In the case of closed manifolds for example, we have
\begin{eqnarray}
 a_0 & = & {\rm Vol}(X) \nonumber \\
 a_1 & = & \frac{1}{6}\int_X{R\sqrt{g}\,{d}^n x} \nonumber\\
a_2 & = &
\frac{1}{360}\int_X{(2|\hat{R}|^2-2|Ric|^2+5R^2)\sqrt{g}\,{d}^n x}
\end{eqnarray}
In the latter $|\hat{R}|^2$ and $|Ric|^2$ are the square
norms of the curvature and \emph{Ricci} tensors respectively and $R$, as usual -
the scalar curvature of $M$. Now we can easily see that the first
approximation in \emph{high-temperature} (or \emph{ultraviolet}) limit gives us the
opportunity to 'hear' the volume of the under lying manifold, as
\emph{Mark Kac} formulated it. The second is nothing but the
Einstein-Hilbert action. In dimension two it is proportional (via
Gauss-Bonnet formula) to the Euler number of the manifold. The
third accounts for the presence of anomalies in
$4D$ field theory, as we shall explain below etc.\\
Obtaining these quantities in explicit form however is not that
easy. There are several preferable ways to do this - some are based on geometrical invariants and dimensional reduction, others use residual calculus for the resolvent kernel. Here we only give a brief idea of the most popular methods that have found application so far.

\paragraph{Perturbative Ansatz} To begin with, let us consider a
Riemannian manifold with almost flat metric. One may regard the
\emph{MP} theorem as a constructive tool for perturbative calculation of
heat invariants. Supposing that the curvature is relatively small, one may expand the curved metric as a sum of the flat (Euclidean or Lorentzian)
metric $\eta_{\mu\nu}$ and a correcting term $$g_{\mu\nu}=\eta_{\mu\nu}+ h_{\mu\nu}$$
Then one needs to obtain the variation of the Laplacian and by
substituting the Euclidean heat kernel $K_0$ in the formula
(\ref{MP}) for the perturbed Laplacian, obtain (at least the leading
part of) the remainder in (\ref{MP}). Next, using the convolution
procedure described above, one may, in principle, find higher-order
corrections as well. The coefficient in front of the $j^{th}$
power of $t$ in the cut-off Voltera series so obtained is
certainly a polynomial of the entries of $h_{\mu\nu}$ and their
derivatives. These polynomials may look confusingly complex (the
whole complexity coming from the cumbersome-looking variation of
the Laplacian) and yet they are not that difficult to transform in
the form $a_j$ given above by means of Stokes theorem and some
basic identities from vector calculus and differential geometry.
The integral invariants are then obtained by simple integration.
For example the coefficient $a_0$ is trivially given by $$ a_0=(4\pi t)^{\frac{n}{2}}{{\rm
Tr}\,K(x,x,t)}_{|t\rightarrow 0}=
{\rm Vol}(M)$$
However, analogous computation even for $a_2$ becomes complicated. Thus in order to obtain in relatively simple manner higher order coefficients, we need some more efficient method that are described below.
\paragraph{Variational Method}
\noindent In this paragraph we shall be using the so called \emph{smeared} heat kernel
$$K(f,D,t) = {\rm Tr}_{\mathbb{L}_2}(fe^{-tD}) $$
that emphasizes the distributional nature of $K$. Here $f$ is a smooth function and $D$
- an elliptic, self-adjoint operator, acting on the space of square-integrable functions
$L^2$. Accordingly, we refer to $a_k(f,D)$ as to the \emph{'smeared coefficients'}.\\
Then, as this trace is still well defined, one may
write the following variational equations for the corresponding
heat coefficients $a_k(f,D)$:
\begin{eqnarray}\label{var}
\dot{a}_k(1,e^{-2\epsilon f}D)&  =  &(n-2k)a_k(f,D) \nonumber \\
\dot{a}_k(1,D-\epsilon F)&  =  & a_{k-1}(F,D) \nonumber \\
\dot{a}_{n/2-1}(e^{-2\epsilon f}F,e^{-2\epsilon f}D)&  = & 0 \
\end{eqnarray}
where the notation $\dot{x}$ is used for
$\displaystyle{\frac{dx}{d\epsilon}_{|\epsilon=0}}$ and $f,F$ are
two arbitrary smooth functions.\\
To prove the first equation we use that $$\displaystyle{\dot{{\rm
Tr}}[\exp(-e^{-2\epsilon f}tD)]={\rm
Tr}[2ftD\exp(-tD)]=-2t\frac{d}{dt}{\rm Tr}[f\exp(-tD)]}$$ and
expand both sides in power series in $t$. The second one could be
easily checked in a similar way. In order to prove the third
identity, we construct the operator
$D_{\epsilon,\delta}=e^{-2\epsilon f}(D-\delta F)$ for which, by
the first equality, we have
$\dot{a}_{n/2}(1,D_{\epsilon,\delta})=0$. Then we vary the latter
with respect to $\delta$ and since the variations commute, we can
easily apply the second equality to prove what we have in the
third.\\
The strategy we use to construct heat invariants, first used by Gilkey \cite{Gi2}, is the following: first,
according to the \emph{MP} theorem, we construct $a_k$ as a linear
combination of all possible independent invariants of the corresponding degree, one may
build from the metric and its derivatives (and the derivatives of the
potential term, if there is one). Then we use the above
variational equations and some other generic
considerations to eliminate the free parameters. For example, if $n=2$, there are only two invariants - $E$ and the scalar curvature $R$. Note that we can always eliminate the derivatives of $f$ via integration by parts, which is  no longer true in the case of manifolds with boundary. Moreover, all half-integer coefficients equal zero in this case, since one cannot construct an odd dimensional invariant integral
if boundary is absent - integration by parts makes them all vanish.\\
Now, in order to make things a bit more complicated, we let all our considerations
take place in a vector bundle $V$, so that we may have additional indices (gauge, spin etc.)
which we denote by Greek letters, or omit, if possible. Then let us write once more the
Laplacian under consideration, in the generic form (\ref{generic}), which now would be
$$D^{\alpha\beta} = -g^{ik}\nabla_i^\alpha\,\nabla_k^\beta - E^{\alpha\beta}$$
where $\nabla$ and $E$ are respectively a (matrix-valued) covariant derivative and an endomorphism (as chosen in (\ref{gen})), taking values in a vector
bundle $V$ over $X$. We may think of $E$ as of a kind of effective potential, as far as physics is concerned.
In the physical example of a free massless scalar field in curved space-time, it is given approximately by
$$E^{\alpha\beta} \sim -\frac{1}{2}(U''(\bar{\Phi}))^{\alpha\beta} - \frac{n-2}{4(n-1)}R\,\delta^{\alpha\beta}$$
For the construction of the non-vanishing $a_k(f,D)$ we consider several quite special cases. Assume, first of all, that the underlying manifold has a direct product structure
${M} = {M}_1\otimes{M}_2$ and the operator $D$ is given respectively by $D = D_1\otimes 1 + 1\otimes D_2$. Since the bundle indices are also independent, one may write symbolically  $\exp(-tD) = \exp(-tD_1)\otimes \exp(-tD_2)$ and use that the smearing function factorizes as well $f(x_1,x_2) = f_1(x_1)f_2(x_2)$ to obtain finally
\begin{equation}
a_k(x,D) = \sum_{p+q=k}{a_p(x_1,D_1)\,a_q(x_2,D_2)}
\end{equation}
The above equation may be used to show that the dependence of the heat coefficients on the dimensionality $d$ is fairly trivial - they are all simply multiplied by an overall factor $(4\pi)^{-n/2}$. In order to prove this, one takes ${M}_1\cong S^1$, for which we know, from Poisson summation formula, that the high-temperature expansion has only one non-zero term. More precisely
$$ K_{S^1}(1,-\partial_x^2) \sim \sqrt{\frac{\pi}{t}}$$
and applying the product formula for $a_k$ one obtains the recursion relation needed.\\
Therefore, it is sensible to write
\begin{eqnarray}
a_0(f,D) & = & (4\pi)^{-n/2}\int{d^n x\sqrt{g}{\rm tr}_V\{f\alpha_0\}} \\
a_1(f,D) & = & (4\pi)^{-n/2}\int{d^n x\sqrt{g}{\rm tr}_V\{f(\alpha_1\,R+\alpha_2 E)\}}
\end{eqnarray}
and so on. We already know that $\alpha_0 = 1$, and can prove it once more by using our new approach. The coefficient $\alpha_2 = 1$ is easily determined by the second variational equation. For $\alpha_1$, however, we need to include higher-order coefficients in the system of equation which would give in the end $\alpha_1 = 1/6$.\\
When boundary is introduced, we have in general
$$a_k(f,D,\mathcal{B}) =  \int{d^n x \sqrt{g} f(x) \tilde{a}_k(x,D)}+\sum_{j=0}^{k-1}{\int_{\partial M}{d^{n-1}\xi\sqrt{h}f^{(j)}\tilde{a}_{k,j}(\xi,D,\mathcal{B})}}$$
where $h$ is obviously the induced metric and $f^{(j)}$ denotes the $j^{th}$ normal derivative at the boundary. The dimensional dependence of the constant coefficients is shown in complete analogy to the previous case.\\
In the following we need asymptotic expansion on the interval $x\in [0,\pi]$, namely
\begin{equation}
K(1,-\partial_x^2,\mathcal{B}^{\pm}) \sim \sqrt{\frac{\pi}{4t}}\pm\frac{1}{2}
\end{equation}
where $\mathcal{B}^{\pm}$ denotes the Neumann (resp. Dirichlet) boundary conditions. For the modified Neumann (Robin) case we have
\begin{equation}
K(1,-\partial_x^2,\mathcal{B_S}^+) \sim \sqrt{\frac{\pi}{4t}}+\frac{1}{2}+\mathcal{S}\sqrt{\frac{t}{\pi}}+\mathcal{O}(\mathcal{S}^2)
\end{equation}
Since $a_0$ is absolutely the same as before we omit it and compute the next two in the expansion. We have
\begin{eqnarray}\label{bound}
a_{1/2}(f,D,\mathcal{B}^\pm) &  = &  (4\pi)^{-\frac{n-1}{2}}\int_{\partial M}{d^{n-1}x\sqrt{h}\, \beta^{\pm}_1{\rm tr}_V f} \nonumber \\
a_1(f,D,\mathcal{B}^{\pm}) & = & (4\pi)^{-n/2}\frac{1}{6}\left \{ \int_M{d^n \sqrt{g}\,{\rm tr}_V(fR+6fE)} \right. \nonumber \\
& + & \left.\int_{\partial M}{d^{n-1}x\sqrt{h}\,{\rm tr}_V(\beta^\pm_2 f L_{kk}+\beta^\pm_3 \partial_\nu f+\beta^\pm_4 f \mathcal{S})} \right \} \nonumber
\end{eqnarray}
where $L_{ik}$ denote the components of the extrinsic curvature of the boundary, or the second
fundamental form, defined as $L(Y,Z)=\langle \nabla_Y Z,\partial_n\rangle $ ($\partial_n$
denotes, as usual, the unit inward normal to the boundary).
The unknown coefficients $\beta^{\pm}$ can be determined with the help of the above result for the interval. For example, it immediately gives\footnote{when $M$ is an interval, the boundary integral reduces to a sum of contributions form its endpoints} $ \beta^+_1 = -\beta^-_1 = 1/4$ and $\beta^+_4 = 12$ ($\beta^-_4$ is certainly meaningless).\\
In order to determine $\beta_2^\pm$ and $\beta_3^\pm$, we use the conformal variational equation (the first formula) in (\ref{var}).\\
The variation of the second fundamental form $L$ is given by the variation of the metric and has the form
$$\frac{d}{d\epsilon}_{|\epsilon = 0}L_{kk} = -fL_{kk} - (n-1)\partial_\nu f $$
Now, in order to keep the Robin boundary conditions conformally invariant (in the Dirichlet case this is always so), we demand that the boundary function $\mathcal{S}$ fulfills
$$\frac{d}{d\epsilon}_{|\epsilon = 0}\mathcal{S} = -f\mathcal{S} + \frac{1}{2}(n-2)\partial_\nu f$$
and after finding a bunch of explicit expressions for the other terms in (\ref{bound}), we arrive at a linear system for the $\beta$'s, satisfied by
$$\beta^+_2 = \beta^-_2 = 2,\quad \beta^+_3 = -\beta^-_3  = 3 $$

\noindent
For the sake of generality we suggest below an explicit expression
for the first few heat kernel coefficients of $K_D(1,D,t)$
\begin{eqnarray}
a_0 & = & \int_X{\dim{V}} \nonumber \\
a_{1/2} & = & - \frac{\sqrt{\pi}}{2}\int_{\partial X}{\dim
V}\nonumber \\
a_1 & = & \frac{1}{6}\int_X{{\rm Tr}(6E+R\,\mathbb{I}_V)} + \frac{1}{3}\int_{\partial X}{{\rm Tr}(L_{ii})}\nonumber \\
a_{3/2} & = & -\frac{\sqrt{\pi}}{190}\int_{\partial X}{{\rm
Tr}(96E+16R-8Ric^a_{a}+7L^i_{i}L^j_{j}-10L^i_{j}L_{i}^j)}\nonumber \\
\end{eqnarray}
For the computation of more complicated terms we refer to \cite{GB} and \cite{Gi2}.
In \cite{Av} and \cite{P2} one may find alternative methods that allow in principle  for
computing infinitely\footnote{this should not be taken literally, as it would mean
'with infinite complexity', since both procedures are iterative} many $a_k$'s. We also refer
to \cite{Va} for a survey on the famous DeWitt procedure, that is not mentioned here.\\

\noindent One more use of the heat kernel expansion can be found in
obtaining semi-classical distribution functions of various systems from
quantum statistical mechanics. For examples
\begin{eqnarray}
Z_{\mathbb{S}^1} & \sim & \frac{1}{\sqrt{4\pi
\beta}}\nonumber\\
Z_{\mathbb{S}^2} & \sim & \frac{1}{4\pi \beta}\left(
1+\frac{\beta}{3}+\frac{\beta^2}{15}+\ldots \right)\\
Z_{\mathbb{S}^3} & \sim & \frac{1}{(4\pi \beta)^{3/2}}\left(
1+\beta+\frac{\beta^2}{6}+\ldots \right)\nonumber
\end{eqnarray}
are the high-temperature asymptotics\footnote{recall that
$\displaystyle{\beta\sim\frac{1}{kT}}$, where $T$ is the absolute
temperature and $k$ - the Boltzman constant} for the partition
functions of the rotators with two, one and zero symmetric axes respectively.
Using this expansion we are able to compute the thermodynamical
functions in Quantum Statistics and QFT in the high-temperature
(respectively ultraviolet) limit. For example the inner energy is
given by $\displaystyle{U=-\frac{\partial \ln{Z(\beta)}}{\partial
\beta}}$, the Helmholtz free energy - simply by
$\displaystyle{F=-\frac{\ln{Z(\beta)}}{\beta}}$, the specific
heat - by \\
$\displaystyle{C_v=\frac{\partial}{\partial
T}\left(kT^2 \frac{\partial \ln{Z(\beta)}}{\partial
\beta}\right)}$  and so on.\\

\noindent The quantities $a_k$ find also one very curious application - in the
theory of integrable systems. As we already saw, each such system
is equivalent to an iso-spectral deformation of its \emph{Lax} operator.
In some cases we can regard the \emph{Lax} operator as a Laplacian acting
on a curved manifold. Then the spectral invariants in the
heat-kernel expansion for this operator are first
integrals for the corresponding hierarchy of dynamical systems.
In the case for the KdV equation for example, the Lax operator is known
to be of the form $\displaystyle{L=-\frac{\partial^2}{\partial
x^2}+u(x)}$, which is the Hamiltonian
of a one-dimensional scatterer with potential $u(x)$. The spectral invariants
in this case generate the \emph{Korteweg - de Vries} hierarchy via
the first-order $PDE$ system
\begin{equation}
\frac{\partial u}{\partial
t}=\frac{(2n)!}{2n!}\frac{\partial}{\partial x} a_n([u]_n)
\end{equation}
where $[u]_n$ denotes the $n$-jet of $u$ over the point $x$. For
$n=0$ we obtain the trivial $u=const$, for $n=1$ we have $u_t=u_x$, and
$n=2$ gives the well-known KdV equation $u_t=u_{xxx}+6uu_x$.\\
Using the above formula, one may in principle generate
non-linear $PDE$'s from the same hierarchy - as many as heat kernel coefficients $a_k$
are available and on the other hand, reconstruct these coefficients from the explicit
form of the integrable equations in given KdV hierarchy. This interesting observation is left
to the reader for further development.

\subsubsection{The $\zeta$-function Approach} In the following we suggest an interesting
illustration of the way the \emph{MP} theorem may lead to rather unexpected benefits for modern physical applications. The particular case, considered here is still rather simplified, but the technique used, is very powerful and somewhat conceptual, with far reaching consequences, some of which we discuss here as well.\\
In order to get started, let us first introduce the notion of spectral
$\zeta$-function of an elliptic self-adjoint operator.\\
We remind that the definition of the classical \emph{Riemann} $\zeta$-function involves a Dirichlet
L-series of the form
\begin{equation}
\zeta(s)=\sum_{n=1}^\infty{n^{-s}}
\end{equation}
which is convergent for $\Re{(s)}>1$. However, \emph{Euler} managed to
introduce an analytic continuation in the whole complex plane
(excluding $s=1$) with the following functional equality
\begin{equation}\label{feq}
\displaystyle{\zeta(s)=\frac{\pi^{s-\frac{1}{2}}
\Gamma(\frac{1-s}{2})}{\Gamma(\frac{s}{2})}\zeta(1-s)}
\end{equation}
For the proof we use the famous \emph{Poisson summation formula} (\ref{Poisson})
and then the
\emph{Mellin transform} for $\zeta$:
\begin{equation}\label{Mellin} \displaystyle{
\zeta(s)=\frac{1}{\Gamma(s)}\int_0^\infty{ \,t^{s-1}\sum_{k\in
\mathbb{N}}{e^{-tk}}dt}}
\end{equation}
To obtain this formula one only needs to expand $\zeta(s)\Gamma(s)$, using
their definitions\footnote{recall that $\Gamma(s) = \displaystyle{\int_0^\infty{t^{s-1}e^{-t}dt}}$} and then interchange the summation and integration.\\
In this way (by means of formula (\ref{feq})) $\zeta$ is defined as
a meromorphic function with simple poles at ${s=1, 2\ldots \frac{n}{2}}$ for even $n$, and ${s=-\frac{1}{2}, \frac{1}{2}\ldots \frac{n}{2} }$ for odd $n$,
as well as simple zeroes for $s\in -\mathbb{N}$ - the poles of the Euler $\Gamma$ - function. There
are also non-trivial zeroes, that are proved to be bounded in the
strip $0\leq\Re{(s)}\leq 1$ (none of them lies on the real axis) and
spread symmetrically with respect to the line $\Re{(s)}=1/2$. The
famous and still unproved Riemann hypothesis consists of the
assumption that all of them lie on this line.\\

\paragraph{One-loop Regularization} Let us now take advantage of the nice $\zeta$-formalism developed above. Namely, in
physics we use $\zeta$-functions for regularization of infinities.
To give an example what this is, we consider once more the
functional equality (\ref{feq}). In the case $s=2$ we have the
result\footnote{again a result due to Euler - in fact one may relate $\zeta{2k}$ to the non-zero \emph{Bernoulli} numbers, using a consideration about the harmonic oscillator given later in this section}
$$\displaystyle{\sum{\frac{1}{k^2}}=\frac{\pi^2}{6}}$$ and via (\ref{feq}) this defines correctly the value
$$\displaystyle{\zeta(-1)=-\frac{1}{12}}$$ which is quite an amazing
result: we sum all natural numbers and obtain a negative
fractional number. The key to understanding this is to accept the value $-1/12$ as the regularized value of the above sum, the value we obtain when we get rid of divergencies via analytic continuation.\\
In Physics we resort to the trick of regularization quite often in order to
compute measurable quantities - vacuum energies, functional determinants and self-interacting potentials
are some of the things that often need to be regularized.  In the above example, we obtained
a regularized expression for the resultant force, acting on a
particle in an infinite chain with potential of interaction $U\sim
-x^2$ and lattice constant equal to one.\\
In order to solve more serious problems, we need first to build a more powerful formalism. In complete analogy with the above
case, we define the \emph{Spectral $\zeta$ - function} of an
elliptic self-adjoint operator $D$ with the formal
series\footnote{later we shall skip the ${\rm dim}E_\lambda$ -
term, assuming that in the degenerate case multiplicities are
always encountered}
\begin{equation}
\zeta_D(s)=\sum_{\lambda\in Spec(D)}{\lambda^{-s}{\rm
dim}E_\lambda}
\end{equation}
which is correctly defined at least for $\Re{s}>n/2$.
The question of convergence in the elliptic case is ruled by considered
above estimates for the spectrum of $D$.  Note also that the Mellin transform (\ref{Mellin}) applies here as well with the only difference that the sum over natural numbers is replaced by a sum over the spectrum\footnote{in other words, $\zeta(s)$ is the Mellin transform of the heat kernel} and $\zeta(s)$ is so defined as a meromorphic function with zeroes, determined by the poles of $\Gamma(s)$.\\
For a more detailed study of the spectral $\zeta$-function and its analytical properties we refer to \cite{Hu}.

\noindent Now we shall focus on a more 'physical' application of $\zeta$-regularization. Namely, in the following example we regularize the determinant of an elliptic, self-adjoint operator,
acting on functions in $\mathbb{L}_2$. The determinant in question appears naturally in
the path integral formalism of - when we calculate Gaussian
integrals for the partition functions in QFT.
In our example we regard a free massless scalar field $\phi(x)$ in
curved space-time $(X,g)$.  The classical action is well known to be given in this case by the expression
\begin{equation}
 {S}=\frac{1}{2}\int_X{\mathcal{L}}=\frac{1}{2}\int_X{\nabla^\mu{\phi}\nabla_\mu{\phi}\sqrt{|g|}\,d^n x}
\end{equation}
Variational principle yields $\nabla^2\phi=0$
inside the domain of interest\footnote{and with the corresponding boundary
conditions as long as boundary is concerned}. The kernel of the propagator is given by the path integral
\begin{equation}
K(q_{in},q_{f};t_{in},t_{f})=\int_{q_{in}}^{q_f}{D_p\,D_q
\exp{\left( \frac{{ i}{ S}}{\hbar}\right )}}\end{equation}
where $q_{in}=q(t_{in})$ and $q_f=q(t_f)$ are the initial and
final positions respectively,
$$\displaystyle{D_q\,D_p=\lim_{n\rightarrow\infty}\left [
\prod_{k=1}^n {\frac{dp_k \,dq_k}{2\pi\hbar}}\right ]
\frac{dp_{\,0}}{2\pi\hbar}}$$ - the integral measure,
defined as usual, and $\displaystyle{{
S}=\int_{t_{in}}^{t_f}{\mathcal{L}(q,\dot{q})dt}}$ is the classical
action along the trajectory starting from $q_{in}$ at time
$t_{in}$ and arriving at $q_f$ at time $t_f$. It is clear,
however, that apart from the classical one, there are infinitely
many possible trajectories and $K(q_{in},q_{f};t_{in},t_{f})$ is
interpreted as a partition function in the space of such paths.\\
In QFT due to the hyperbolic signature of the metric, the integral
measure above is somewhat ill-defined. Therefore the simple trick
of analytic continuation in the complex plane is performed -
the so called \emph{Wick rotation}. This is nothing but rotation
of the time coordinate from the real to the imaginary axis
($t\rightarrow i\tau$), so that we end up with an Euclidian or Riemannian,
rather than pseudo-Riemannian metric and imaginary time. Due to
the covariant transformation law we obtain the transformed action
(we refer to as \emph{Euclidean action} from now on) as $S_E=iS$
and the path integral from above in the form\footnote{the
measure $D_\phi$ is chosen in such a way that the final result is
free of multiplicative constants}
\begin{equation}
Z[\phi]=\int{D_\phi\exp{(-S_E/\hbar)}}=\det{D}^{-1/2}
\end{equation}
Let us write down the above partition function in the form $e^{-W}$. The functional
$$\displaystyle{W=\frac{1}{2}\ln{\det D}=-\frac{1}{2}\zeta'(0)}$$
in this expression we call from now on \emph{the one loop
effective action} for the corresponding theory. The effective
action gives the one-loop quantum correction to the classical one
and is a powerful tool for various calculations. In our somewhat
simplified considerations the path integral representing $K$ is
Gaussian (the classical action contains only a quadratic term) and
we obtain immediately $K\sim {\rm det}(D)^{-1/2}=\exp(\frac{1}{2}\zeta_D'(0))$, so
at least here $W$ recovers one-loop contribution to the path integral exactly. In more
complicated settings, where there are several quantum corrections
to the effective action. Namely, we expand the Lagrangian up to
the quadratic order in quantum fluctuations $\phi$
\begin{equation}
L\simeq L_{cl}+\langle \phi,J\rangle +\langle \phi,D\phi\rangle
\end{equation}
where $J$ is an external current and $L_{cl}$ - the classical
Lagrangian, to obtain the more generic form of the path integral
\begin{equation}
Z[J]=e^{-L_{cl}}\det{(D)}^{-\frac{1}{2}}\exp{[\frac{1}{4}JD^{-1}J]}
\end{equation}
and in this case the effective action gives only the determinant
contribution to the overall propagator.\\
We are now close to obtaining a regularized expression for the
effective action in terms of the spectral $\zeta$-function. What
we have to do is simply express $\zeta_D(s)$ as the Mellin
transform of the heat kernel and then differentiate at $s=0$. When
we deal with ultraviolet divergences it is useful and legal to
expand the heat kernel as a power series in $t$ and neglect the
higher order terms, as the \emph{MP} theorem teaches us. Then, after
introducing a cutoff $\tau\rightarrow 0_+$ and exploiting the formulae for the first
spectral invariants, differentiation term by term gives:
\begin{equation}
W[g_{\mu\nu}]=-\int{\frac{d^4 x\sqrt{|g|}}{32\pi^2}\left [
\frac{1}{2\tau^2}+\frac{R}{6\tau}+\left(
\frac{1}{120}R^2+\frac{1}{60}R_{\mu\nu}R^{\mu\nu}\right
)\ln{|\tau|}+\ldots\right ]}\nonumber
\end{equation}
where the remainder consists of finite terms only. Next we introduce a
modified action for the background field (gravitation) so that its
singularities cancel the singularities of $W$:
\begin{equation}
S^b_{grav}[g_{\mu\nu}]=\int{d^4 x\sqrt{|g|}\left [ -
\frac{R+2\Lambda_b}{16\pi G_b}+\alpha_b\left (
\frac{R^2}{120}+\frac{R_{\mu\nu}R^{\mu\nu}}{60}\right )\right ]}
\end{equation}
The quantities with subscript $b$ are called \emph{bare coupling
constants} of the theory. They are functions of $\tau$ chosen to
cancel the divergent terms in $W$. Their presence can never be
observed since the back-reaction can never be switched off.
However, with their help, one obtains the expression of the
regularized effective action as $$W_{reg}=W+S^b_{grav}$$
For the sake of honesty we note that there are several different ways
to get rid of divergencies in physical theories, but we describe only the one that corresponds most directly to the topic of this work and at the same time is usually referred to as the most elegant regularization scheme known so far.\\
We refer to \cite{MW} for a systematic exposition of the $\zeta$-regularization procedure from a physicist's point of view. Another good source of information for the application of this technique in QFT is \cite{Va}. The latter is in fact suitable reference for the most part of this section.

\paragraph{The Oscillator} There is a nice example that allows for exact expressions for the heat kernel and the operator $\zeta$-function. This example is one of the very few exactly solvable problems in quantum mechanics, but also one of utmost significance for the whole theory.\\
The partition function of the harmonic oscillator can be written in the form
$$ Z(\beta) = \frac{1}{2\sinh{(\hbar\omega\beta/2)}}=\frac{1}{\hbar\omega\beta} +
\sum_{j=1}^\infty{\frac{B_{2j}}{(2j)!}(1-2^{1-2j})(\hbar\omega\beta)^{2j-1}}$$
with $B_{2j}$ being the non-zero Bernoulli numbers.\\
For the relevant $\zeta$-function, we have already a number of ways to retrieve the result
$$ \zeta_{HO}(s) = \frac{2^s-1}{(\hbar\omega)^s}\zeta_R(s)$$
with $\zeta_R$ standing for the classical Riemann $\zeta$-function.\\
One may find exact values of thermodynamical quantities in this and just a few other cases. However, in quantum statistics, the oscillatory degrees of freedom play central role, so the result, no matter how simple, is not to underestimate.

\paragraph{Residual Calculus} Here we give one more idea how to
calculate heat invariants. The trick is to use the integral
representation (\ref{Mellin}) of $\zeta_D(s)$ and its inverse
\begin{equation}
{\rm Tr}K(x,x,t)=\frac{1}{2\pi
i}\oint{t^{-s}\Gamma(s)\zeta_D(s)ds}
\end{equation}
where the contour of integration encircles all poles of the
integrand. Then, baring in mind that
$\displaystyle{a_k=\frac{\partial}{{k!}\,\partial x^k}{\rm
Tr}K(x,x,t)}$, we easily obtain the expression
\begin{equation}
a_k={\rm Res}_{s=n/2-k}\Gamma(s)\zeta_D(s)
\end{equation}
and in particular $a_{n/2}=\zeta(0)$. Curiously enough, this gives $\zeta(0)=0$
for odd-dimensional closed manifolds.\\
The power-series form of the heat kernel, obtained above, allows
for expressing the high-temperature asymptotic and all thermal
functions once we know explicitly the $\zeta$-function in certain
quantum-mechanical problem. The latter is, unfortunately, very
rarely so. At least, the above expression gives one more curious
connection between the analytic properties of a Laplace-type
operator, and geometric quantities of the underlying manifold. For
example it is far non-trivial that the regularized range of the
Laplace-Beltrami operator on two-dimensional manifolds is
proportional to their Euler number.\\

\subsubsection{Conformal Anomaly and the Polyakov Action }
Now we briefly discuss the case of conformal anomaly.
In the classical conformal field theory we have action, which
is invariant under conformal transformations of
space-time, that are, roughly speaking, all isometries plus
dilatations\footnote{and more precisely speaking, a pseudo-Riemannan space with metric of signature $(p,q)$ has $\mathsf{SO}(p+1,q+1)$ as a conformal group - for example the conformal group of the Gaussian sphere $\mathbb{CP}^1$ is easily shown to be $PSL(2,\mathbb{C})\sim \mathsf{SO}(3,1)$} of the type $$ g_{\mu\nu}\rightarrow e^{2\rho(x)}g_{\mu\nu}$$ Such classical theories are
characterized by traceless energy-momentum tensor
\begin{equation} T_{\mu\nu}=\frac{2}{\sqrt{g}}\frac{\delta W}{\delta
g_{\mu\nu}}
\end{equation}
since the variation of the action with respect to dilatations is given by
\begin{equation}
\delta W=-\int_M{\delta \rho T^\mu_\mu d\omega_g}
\end{equation}
However the conformal symmetry is broken by quantization even in
the one-loop approximation\footnote{recall that in this case the laplacian appears as a fluctuation operator of the theory}, as one may easily verify, substituting
the effective action, constructed above and obtaining finally
\begin{equation}
T^\mu_\mu=a_{n/2}(x)
\end{equation}
So, the one-loop quantum theory possesses an energy-momentum
tensor, that is not traceless anymore. \\
There is a very famous physical example in which the explicit form of the effective action may be obtained via integration of the conformal anomaly. The model considers a free scalar field on compact Riemannian surface, described by the classical action
\begin{equation}
{S} = \int_M{d^2 x\sqrt{g}\,\partial^{\,\nu}\phi\,\partial_\nu\phi}
\end{equation}
It is well known on the other hand that all two-dimensional metrics are conformally flat, so we may choose a gauge in which
$$ g_{\mu\nu} = e^{2\rho(x)}\eta_{\mu\nu}$$
and calculate in this gauge the trace of the stress-energy tensor and thus the conformal anomaly. Varying $S$ we obtain
\begin{equation}
\delta W = \frac{1}{12\pi}\int_M{d^2x\,\delta\rho\,\eta^{\mu\nu}\partial_\mu\partial_\nu \rho}
\end{equation}
Baring in mind that in the so chosen gauge
$$\sqrt{g}\, R = -2\eta_{\mu\nu}\partial_\mu\partial_\nu \rho $$
we obtain after integration the conformal part of the effective action in the form known as the \emph{Polyakov action} in string theory
\begin{equation}
W = \frac{1}{96\pi}\int_M{d^2 x\sqrt{g}\, R \triangle^{-1}R}
\end{equation}
The inverse of the laplacian is naturally obtained via integration over the Euclidean Green kernel.\\
We note that this results may in principle be generalized to higher dimensions but, unless $M$ is conformally flat, the integrated anomaly represents only a part of the effective action. For more thorough investigation on this matter and further applications of spectral methods in quantum field theory we refer to \cite{BM}, \cite{NO} and \cite{Va2}, together with the references therein.

\subsubsection{A Note On Index Theory}
Now we mention a rather peculiar appearance of the spectral invariants from the heat kernel expansion in index theory.\\
To begin with, define the index of a Fredholm operator as
$${\rm ind}\,F=\dim \ker (F) - \dim {\rm coker} (F)$$
From this definition it is easy to see that the index is additive
(for a composition of operators) and ${\rm ind}(F^*)=-{\rm
ind}(F)$, therefore the index of a self-adjoint operator is
always zero.\\
The index has a very strong topological meaning in the elliptic
case - it depends only on the homotopy type of the leading symbol
of the operator within the class of elliptic $\Psi DO$'s of the
given order. This observation has some deep consequences
connecting analytic theory of elliptic operators on manifolds to
algebraic geometry.\\
A good basic example is obtained if we \emph{roll up} the
whole de Rham complex, distinguishing only
between even and odd dimensional forms. Then the index of the
Dirac type operator $P:=d+\delta:
\Omega^{[0]}\rightarrow\Omega^{[1]}$ defined as ${\rm ind}_P:=dim
\ker(P)-dim \:{\rm coker}(P)=dim \ker(P)-dim \:{\rm im}(P)^\bot$
gives exactly the Euler number of $M$.\\
There are several similar observations that inspired the famous index theorem of
Atiyah and Singer, relating the analytic index with the topological index, given by the
$A$-genus\footnote{the $A$-genus is defined as
$\hat{A}=\prod_i\frac{c_i/2}{sinh(c_i/2)}$, where the $c_i$'s are
the \emph{Chern classes} of the bundle} of the corresponding
bundle. Namely, for an elliptic operator, acting on section of this
bundle, we have
\begin{equation}\label{AS}
{\rm ind}\, P=\int_M(\hat{A})
\end{equation}
Suggesting that the integral on the right is integer-valued is already
quite non-trivial. The approach of Gilkey to this matter
uses the construction of the two operators $P^+=PP^*$ and
$P^-=P^*P$. It is easy to see that their non-zero eigenvalues
coincide, so when we consider the difference of the their
heat-kernels, only the terms, corresponding to the zero modes
survive, and these are the terms, independent of $t$ in the series
expansion. Thus, what we have in the end is
\begin{equation}
{\rm ind}\,P=a_{n/2}(P^+)-a_{n/2}(P^-)
\end{equation}
One may consider once more the previous example - the \emph{rolled de Rham  complex}
 examine the heat kernels of $d\delta$ and $\delta d$, or more precisely, the coefficients $a_1$ in both cases, to obtain the same result. Moreover, by the above construction and \emph{ Gauss-Bonnet }theorem, we have
$${\rm ind}(d) = \chi(M)  = \frac{1}{2\pi}\oint_M{R}\, d\sigma $$
which is a special case of (\ref{AS}).

\subsubsection{The $\eta$-function and the Berry Phase} We conclude with a short discussion on $\eta$-functions. Consider first an operator $P$ that is not positive
definite. Then the $\zeta$-function is in general
ill-defined. We define a new spectral
function instead by the series
\begin{equation}
\eta_P(s)=\sum_{\mu\in Spec(P)/\{0\}}{{\rm sign}(\mu)|\mu|^{-s}}
\end{equation}
For this new quantity we have an integral representation again
thanks to the Mellin transform
\begin{equation}
\eta_P(s)\Gamma(\frac{s+1}{2})=\int_0^\infty{t^{\frac{s-1}{2}}{\rm
Tr}[P\exp{(-tP^2)}]dt}
\end{equation}
In particular, for the $\eta$-invariant we have
\begin{equation}
\eta_P(0)=\int_0^\infty{\frac{1}{\sqrt{\pi t}}{\rm
Tr}[P\exp(-tP^2)]dt}
\end{equation}
that is a regularized expression for the spectral asymmetry of $P$
(recall that the $\zeta$-invariant $\zeta_D(0)$ gives in a similar
way a regularized expression for the range of $D$).\\
Another useful applications of $\eta$-functions is the
computations of functional determinants for operators of the type
under consideration. First of all we use that $P^2$ possesses a
well-defined $\zeta$-determinant and thus
\begin{equation}
\det{|P|}=\exp{(-\frac{1}{2}\zeta_{P^2}'(0))}
\end{equation}
Next we easily decompose $Spec(P)/\{0\}$ to purely positive and
purely negative eigenvalues in order to obtain a formal expression
for $\zeta_P'(0)$. The final result is
\begin{equation}
\det(P)=\exp[-\frac{i\pi}{2}(\zeta_{P^2}(0)-\eta_P(0))]\det|P|
\end{equation}
The argument of the oscillating exponent to the right is sometimes
called \emph{Berry phase} of the determinant and plays an
important role in path integral formalism.\\
The spectral invariants considered in this section possess also a
topological meaning due to certain index theorems. They are of
different sensibility and some of them are rather subtle for
investigation. A good survey on this matter is \cite{Ro}
\subsection{Path Integrals and the Gutzwiller Trace Formula}
In this section we consider the non-stationary
Schr\"{o}dinger equation
\begin{equation}
\left( \frac{\hat{p}^2}{2m}+V(q)+{ i}\,\hbar\, \partial_t
\right ) \psi(q,t)=0
\end{equation}
where the first two terms represent the quantized hamiltonian
$\hat{H}(p,q)$, consisting of potential $V(q)$ and kinetic energy
$\frac{\hat{p}^2}{2m}=\frac{-\hbar^2}{2m}\nabla^2$. It is easier
to expand the solutions over the basis of solutions of the
stationary equation - the eigenfunctions of the Hamiltonian
$\phi_k:\: \hat{H}\phi_k=E_k\phi_k$. For bounded systems this
basis is known to be orthonormal and complete. The solution
$\psi(q,t)$ we are interested in is then expanded as
\begin{equation} \psi(q,t)=\sum_n^{\infty}{c_n{ e}^{-\frac{{
i}E_n t}{\hbar}}\phi_n(q)}
\end{equation}
where the Fourier coefficients are given by
$\displaystyle{c_n=\int_{\mathbb{R}^m}{ \bar{\phi}_n(q)\psi(q,0)\,d^m
q}}$. We can write it also as
\begin{equation}
\psi(q,t)=\int{K(q,q',t)\psi(q',0)d^m q}
\end{equation}
where the integral kernel of the propagator
$\displaystyle{K(q,q',t)=\sum{\bar{\phi}_n(q')\phi_n(q)e^{-\frac{iE_n t}{\hbar}}}}$ carries the whole dependence
on $q$ and $t$ so it is itself a solution of the initial
Schr\"{o}dinger equation. Moreover, it is a fundamental
solution\footnote{that is $K(q,q',0)=\delta(q-q')$} due to the
completeness of the
basis.\\
The Fourier transform of $K$ in the time variable is a
well-defined distribution depending on the dual variable - the energy.
This is nothing but the resolvent of $\hat{H}$ and is usually
introduced in a regularized way as follows:
\begin{eqnarray}
G(q,q',E+i\varepsilon) & = & \frac{1}{i\hbar}\int_0^\infty{dt\,e^{-\frac{it}{\hbar}(E+i\varepsilon)}K(q,q',t)} \nonumber \\
& = & \frac{1}{i\hbar}\sum_n
{\bar{\phi}_n(q')\phi_n(q)\int_0^\infty{dt\,e^{-\frac{it}{\hbar}(E-E_n+i\varepsilon)}}}  =
\sum_n{\frac{\bar{\phi}_n(q')\phi_n(q)}{E-E_n+i\varepsilon}}\nonumber
\end{eqnarray}
Once we have guaranteed the existence of the above integral, we may get rid of the imaginary shift by taking the
limit $\varepsilon\rightarrow 0$. In the following we shall be
interested in the trace of the green function
\begin{equation}
{\rm Tr}G(E)=\int_{\mathbb{R}^m}{G(q,q,E)\,d^m
q}=\sum_n{\frac{1}{E-E_n}}
\end{equation}
which is connected to the spectral density $\displaystyle{d(E)=\sum_n{\delta(E-E_n)}}$ with the famous \emph{Sohotski
formula}
\begin{equation}
\pi d(E)=-\Im\,{({\rm Tr}G(E))}
\end{equation}
\subsubsection{Semi-classics for the Sch\"{o}dinger Equation}
For a relatively slowly varying potential\footnote{the
semi-classical approximation is valid only in the case when the
potential does not change considerably within the de Broiglie
wavelength} we are allowed to apply the Wentzel-Kramers-Brillouin
(WKB) ansatz for the wave function:
\begin{equation}
\psi(q,t)=A(q,t)e^{\frac{i}{\hbar}R(q,t)}
\end{equation}
We note that this $\hbar\rightarrow 0$ asymptotic solution cannot be
prolonged trivially for large times due to the formation of caustics at time $t\sim\hbar^0$.
Therefore, we need to let it evolve with clear account of the singularities it passes through.
Substitution into the initial equation yields (after separating the real
and imaginary part and taking the limit $\hbar\rightarrow 0$)
\begin{eqnarray}
\partial_t\rho+\nabla (\rho{\bf v}) & = & 0 \nonumber \\
\partial_t R+{H}(\nabla R,q) & = & 0
\end{eqnarray}
The former is interpreted as a continuity equation for the flow
$\rho=A^2(q,t),\quad {\bf v}=\frac{1}{m}\nabla R(q,t)$, referred
to as \emph{the Madelung flow}, while the latter is simply the
classical Hamilton-Jacobi equation for $R(q,t)$.\\
Now it is easy to obtain the evolution of $R$ with respect to this
flow, replacing $\nabla{R}\rightarrow p$
$$\dot{R}(q,t)=p\dot{q}-H(p,q)=\mathcal{L}(q,\dot{q},t)$$
where $H$ and $\mathcal{L}$ are respectively the classical Hamiltonian and
Lagrangian of the system. Thus for a classical trajectory, starting
at the point $(q',p'),\:R$ evolves with time as
$$R(q(t),t)=R(q',0)+\int_0^t{\mathcal{L}(q,\dot{q},t')dt'}$$
the integral to the right, being the Hamiltonian principal
function. Denoting it by $R(q,q',t)$ we obtain the classical
initial and final momenta
\begin{eqnarray}
p' & = & -\nabla_{q'}R(q,q',t)\nonumber \\
p & = & \nabla R(q,q',t)
\end{eqnarray}
The conservation of matter demands covariant transformation of the
volume form so we have
\begin{equation}
A(q,t)=\sqrt{{\rm det}\left ( \frac{\partial q'}{\partial q}\right
)}A(q,0)=e^{-\frac{i\pi\nu}{2}}\sqrt{\left |{\rm
det}\left ( \frac{\partial q'}{\partial q}\right )\right
|}\,A(q,0)
\end{equation}
where the quantity $\nu(q,q',t)$, called \emph{the Maslov index},
counts the sign changes of the Jacobi determinant along the path,
connecting $q'$ with $q$. Strictly speaking, there are usually
many classically allowed paths $\gamma_k$, ending up in $q$ at
time $t$, and each such path has its own starting point, principal
function and Maslov index, and gives its own contribution to the
evolution of states. Thus we need to sum over all such trajectories:
\begin{equation}
\psi(q,t)=\sum_k{e^{\frac{iR_k(q,q_k',t)}
{\hbar}-\frac{i\pi\nu_k}{2}}\sqrt{\left |{\rm det}\left (
\frac{\partial q_k'}{\partial q}\right )\right |}\,\psi(q_k',0)}
\end{equation}
Now we are ready to express the propagator, but first, to get rid
of the singularities due to the point-like contribution of initial
states in configuration space, we switch to momentum space, where
the Dirac delta-distribution is transformed to a constant factor
by simple Fourier transform. We have
$$\rho(q,t)d^m q = |c|^2 d^m p'$$
and changing from $p'$ to $q$ in the righthand side, we easily obtain
$$\rho(q,t) = |c|^2{\rm det} \left ( - \partial_{q'}\partial_{q} R(q,q',t)\right ) $$
For normalization reasons, the constant is easily shown to be
$c=(2\pi i\hbar)^{-m/2}$ so, expressing the propagator as
before, in terms of the evolved states, one obtains the famous
expression for the semi-classical \emph{Van-Vleck propagator}
\begin{equation}
K(q,q',t)=(2\pi i\hbar)^{-m/2}\sum_k{e^{\frac{iR_k(q,q_k',t)} {\hbar}-\frac{i\pi\nu_k}{2}}\sqrt{\left
|{\rm det}\left ( - \partial_{q'}\partial_{q} R(q,q',t) \right
)\right |}}
\end{equation}
\subsubsection{The Method of Stationary Phases and the Green
Function} From the above formula, it is not difficult to obtain
the Green function as $\displaystyle{G(q,q',E)=\sum_k G_k}$ with
each $G_k$ being the Fourier image of the $k^{-th}$ term in the
above sum, representing $K(q,q',t)$. First of all, we need to know
how to compute integrals of the type
\begin{equation}
I=\int_{\mathbb{R}^m}{u(x)e^{ik\,f(x)}d^m x}
\end{equation}
in the limit $k\rightarrow\infty$, where $u(x)$ and $f(x)$ are
sufficiently smooth and $\Im{(f)}\geq 0$. The idea for this method
is to take advantage of the fact that the main contribution to
the integral comes from the set of points where the phase function is
stationary - that is $f'=0$, and $\Im{(f)}=0$. \\
Around a stationary point $x_0$ we have, up to the leading term
\begin{equation}
\int_{\mathbb{R}^m}{u(x)e^{ik\,f(x)}d^m x}=e^{ik\,f(x_0)}\left ( \frac{2\pi}{k} \right
)^{m/2}\!\frac{u(x_0)}{\sqrt{{\rm det}\left ( -  if''(x_0)\right)}}+\mathcal{O}(k^{-m/2-1})
\end{equation}
With this formula in mind it is not hard to obtain a good
approximation for some of the integrals, representing
$G_k(q,q',E)$. The stationary phase condition\footnote{the
condition, under which the approximation is legal} is given in our
case by $\partial_t R_k + E = 0$ since $\nu_k(q,q',t)$ is constant
on the trajectory. This means that the time of the saddle
point $t_c$ is the time it takes to a particle with energy $E$ to
move from $q'$ to $q$ following $\gamma_k$. Thus the stationary
phase method is good only for relatively long trajectories. A
direct application of the above formula (or rather to its direct multidimensional generalization) yields
\begin{eqnarray}
G^{long}_k(q,q',E)\thickapprox \frac{(2\pi i\hbar)^{\frac{1-m}{2}}}{i\hbar}\,e^{\frac{iS_k(q,q_k',E)} {\hbar}-\frac{i\pi\nu_k}{2}}\times \\
\times\left [{\rm det}\left(
\partial^2_t R(q,q',t_c)\right )\right]^{-\frac{1}{2}}\sqrt{\left
|{\rm det}\left ( - \partial_{q'}\partial{q} R(q,q',t_c) \right
)\right |}\nonumber
\end{eqnarray}
where
$\displaystyle{S(q,q',t)=R(q,q',t_c)+Et_c=\int_{q'}^q{p\,dq}}$ is
the classical action functional for the trajectory $q'\rightarrow
q$.\\
We may, on the other hand, separate the spatial derivatives of $R$
in normal and tangential parts with respect to the flow and thus
represent the latter determinant as
\begin{equation}
{\rm det}\left ( - \partial_{q_i}\partial_{q'_j} R \right
)=\frac{1}{|\dot{q}|\,|\dot{q'}|}\, \partial^2_t R(q,q',t)\,{\rm
det}\left ( - \partial_{q_i}\partial_{q'_j} S\right)
\end{equation}
which leads an integral expression for the long-time contribution
to the trace in a form, suggesting once more a stationary-phase
approximation. Here the stationary condition is
\begin{equation}
\nabla R(q,q',E)=\nabla_q S|_{q=q'}+\nabla_{q'} S|_{q=q'}=p-p'=0
\end{equation}
which means, together with the energy-preserving condition, that
the orbits, giving the main contribution to the trace are not only
periodic in the spatial variable, but also in momentum, so they
end up in the same point in phase space.\\
Introducing the stability (or \emph{monodromy}) matrix ${P_\gamma}$ of an orbit, as the
matrix relation between the initial and final variations of the
phase space variables in direction, perpendicular to the
flow\footnote{note that this is by definition the matrix of the
linear \emph{Poincare\'{e}} map for this orbit}
\begin{equation}
\delta (q,p)_\perp={P_\gamma}\,\delta(q',p')_\perp
\end{equation}
and using the definitions of the initial and final momenta and
their variations in terms of the action derivatives, one may prove
the matrix relation
\begin{equation}
{\rm det}({\rm Id}-{P_\gamma})={\rm det}\left (
\partial_{q}\partial_{q} S (q,q,E)\right){\rm det}\left (
\partial^2_{q'q} S \right)^{-1}
\end{equation}
which allows for the expression of the long-time contribution in
somewhat completed form. Namely, after all substitutions and
integrating over the diagonal $q=q'$ we obtain
\begin{equation}
{\rm Tr}G^{long}_k(q,q',E)\thickapprox\frac{1}{{ i}\hbar}\oint
{\frac{d\,q_\parallel}{\dot{q}_\parallel}\frac{{ e}^{\frac{{
i}S_k(E)}{\hbar}-\frac{{ i}\pi\nu_k(E)}{2}}}{\sqrt{ |{\rm
det}\left ( \mathbb{I} - {P}_{\gamma^k} \right ) | }}}
\end{equation}
where summation is taken over all closed classically allowed paths through $(p,q)$.\\
Now realizing that the integral to the right is simply the period
$\displaystyle{\oint
{\frac{d\,q_\parallel}{\dot{q}_\parallel}}}=\oint{dt}=T_k$ and
taking into account the revolutions of simple orbits, that also
give contribution to the trace\footnote{if $l$ is the winding
number of the orbit, then we simply have $T\rightarrow
lT,\:S\rightarrow lS,\:\nu\rightarrow l\nu$ and ${P_\gamma}\rightarrow P_\gamma^l$} we finally obtain the famous Gutzwiller
trace formula
\begin{equation}{\rm Tr}(G)={\rm Tr}(G_0)+\frac{1}{i\hbar}\sum_{k,\,l}{\frac{T_l}
{|{\rm det}(\mathbb{I}-J_k^l)|^\frac{1}{2}}e^{\frac{il}
{\hbar}S_k - \frac{il\pi}{2}\nu_k}}
\end{equation}
The contribution from short-time trajectories on the other hand, for which the
stationary phase method is not applicable, is approximately given
by
\begin{eqnarray}
G^{short}_k(q,q',E) & \thickapprox & \frac{1}{i\hbar}\int_0^\infty{\left ( \frac{m}{2\pi i t\hbar}\right
)^{\frac{m}{2}}e^{\frac{i}{\hbar}\left (
\frac{m(q-q')^2}{2t}-V(q)\,t+E\,t\right )}dt} \nonumber \\
& \thickapprox & -\frac{im}{2\hbar}\left (
\frac{\sqrt{2m(E-V)}}{2\pi\hbar|\,q-q'|}\right
)^{\frac{m-1}{2}}H^{(1)}_{\frac{m-1}{2}}(S_0/\hbar)
\end{eqnarray}
where $S_0(q,q',E)=\sqrt{2m(E-V)}|q-q'|$ is the short-distance
approximation of the classical action. Moreover the Hankel function of the first
kind is defined as $$H^{(1)}_\nu(x)=\frac{1}{\pi i}\int_0^\infty{\frac{e^{\frac{x}{2}
(t-\frac{1}{t})}}{t^{\nu-1}}\,dt}$$ Now one simply takes the imaginary part
of the trace and substitutes in the integral the asymptotic
expansion for the Bessel function $$
J_\nu(x)\thickapprox \frac{1}{\Gamma(\nu+1)}\left (
\frac{x}{2}\right )^\nu ,\quad |x|<<1$$ which is the real part of
part of $H^{(1)}_\nu (x)$.\\
Then, after integration over a ball of radius $|p|=\sqrt{2m(E-V)}$ in momentum space, one  obtains the famous \emph{ Weyl law } for the asymptotic distribution of energy levels
 $$\displaystyle{\lim_{\lambda\rightarrow\infty}{N(\lambda)}=
 \frac{Vol(M)\lambda^{\frac{d}{2}}}{2^d
 \pi^{\frac{d}{2}}\Gamma(\frac{d}{2}+1)}}$$

\subsection{The Berry-Tabor Trace for Integrable Systems}
In the derivation of the previous result, we implicitly assumed that the closed orbits are isolated so that the stationary phase method for the Green function could be applied to each orbit separately. In practice, however, this is not always a reliable assumption. There are cases when the closed orbits come in continuous families filling whole areas of the phase space. One such case is a bound integrable system, with multiply periodic classical motion. In this case, as we know, phase-space trajectories fill invariant tori, given by action-angle variables. The idea for the derivation of the following trace formula is to use these variables in order to simplify the trace expansion in the integrable case.\\
We start with the familiar expression for the semiclassical propagator
\begin{equation}
K(q,q',t)=(2\pi i\hbar)^{-m/2}\sum_k{e^{\frac{iR_k(q,q_k',t)} {\hbar}-\frac{i\pi\nu_k}{2}}\sqrt{\left
|{\rm det}\left ( - \partial_{q'}\partial{q} R(q,q',t) \right
)\right |}}
\end{equation}
where the summation in our case is taken over all closed paths, and without writing additional subscripts, it is understood that the principal function $R(q,q',t)$ is evaluated separately at each such path.\\
Next, we use the fact that the action variables ${ I}_k$ are constants of motions and the angles ${ \theta}_k $ evolve linearly with time according to $\dot{{ \theta}} = \nabla_{ I} H({{I}})={ \omega(I)}$, so that we have the classical action in the form
$$S = { I .(\theta - \theta')}-H({ I})t$$
and thus for the determinant in the propagator we can write
$${\rm det}\left ( - \partial_{q'}\partial_{q} R(q,q',t) \right)=\left ( \frac{d{ \theta }}{d{ I}}\right )^{-1} = \frac{1}{t^m \det (\partial { \omega }/\partial { I})}$$
When we take the propagator trace, we use the multiple periodicity with respect to the angle variables and expand the integral as a sum of integrals over the lattice, defining the $m$-torus with generators ${ I}$ and coordinates ${ \theta}$:
$$ \int_{\mathbb{R}^m}{d^m q}K(q,q,t) = \sum_{{\bf L}\in\mathbb{Z}^m}{\int_0^{2\pi}{K({ \theta} +2\pi{\bf L, \theta},t)}\,d\theta}$$
In fact, the integer vector ${\bf L}$ accounts for the different topologies of closed paths \footnote{in the two-dimensional case for example, we sum over toric knots}. Under the assumption that no vector ${\bf L}$ can give paths in more than one $m$-torus simultaneously, all summations considered so far can be labeled by the corresponding $\mathbb{Z}^m$- coordinate ${\bf L}$. For instance, the classical action over a closed  ${\bf L}$-path is given by
$$ S_{\bf L} =  2\pi { I_{\bf L}}(t).{\bf L} - H({ I_L}(t)).t$$
The above expression is independent of ${ \theta}$ and so should be the semiclassical propagator in these coordinates. Hence, integration over the angle variables is simply reduced to a factor $(2\pi)^m$.\\
After substituting it in the initial expression for $K$, we need to extract the zero-length orbits with ${\bf L = 0}$ - they give the smooth term in the spectral density, just as in the Gutzwiller case. To the rest we apply stationary-phase method and after that  evaluate the amplitudes by introducing curvilinear coordinates $\xi_j$ in the ${ I}$-space. Let $\xi_0$ point in the $\nabla H$ direction, and the rest $m-1$ $\xi_i$'s be local parameters on the energy contour $H( I_L)= E$. Then we finally obtain
\begin{equation}
n(E)=\bar{n}(E)+\frac{2}{\hbar^{(m+1)/2}}\sum_{\bf L\neq 0}{\frac{\cos[2\pi{\bf L}({ I_L}/\hbar - {\bf \nu_L}/4)+\pi{\bf \beta_L}/4]}{|{\bf L}|^{(m-1)/2}|{ \omega(I_L)}|\sqrt{|\kappa( {I_L})|}}}
\end{equation}
where $\kappa$ is the scalar curvature of the energy contour in the $\xi$-coordinates, $\beta_{\bf L}$ is the excess of positive over negative eigenvalues in the matrix
$$ \omega .\frac{\partial^2{ I}}{\partial\xi_j\partial\xi_k}$$
and the vector-valued Maslov factor $\bf{\nu_L}$ encounters the number of caustics met by the orbit at each cycle of the corresponding torus.\\
The above proof of the Gutzwiller formula follows \cite{Lu} and for the Berry-Tabor trace we refer to the original paper \cite{BT}.

\subsection{The Wave Trace}
 The wave trace is defined as the distributional trace or as
asymptotic series for the kernel of the wave resolvent
\begin{equation}
U(\tau)={\rm Tr}\,\exp{(i\tau
\sqrt{\triangle})}=\sum_{\lambda\in Spec{\triangle}}{e^{i\tau\sqrt{\lambda}}}
\end{equation}
The very crucial fact that makes the above a powerful tool in
inverse problems, known as Poisson relation, is  that the singular support (wave front) of $U(\tau)$ is contained in the length spectrum (the set of lengths of closed geodesics)
both for closed manifolds and manifolds with boundary\footnote{an interesting observation in the case of real-analytic manifolds shows that we may have an equality rather than inclusion if we consider the analytic wave front (defined as the complement of the
set where $U(\tau)$ is real analytic) instead of the ordinary
one}. Moreover, as to each length in this set always correspond at
least two geodesics (say $\gamma$ and $\gamma^{-1}$, where the
latter is understood as 'time reverse'), we may, with no loss of generality, restrict ourselves to the even part of the trace
\begin{equation}
U_+(\tau)={\rm Tr}\,\cos{\tau\sqrt{\triangle}}
\end{equation}
What we have to deal with in practice is the singularity expansion
of the above trace around closed geodesics. More precisely, as
long as all closed geodesics are non-degenerate, one has (modulo
$\Psi_{-\infty}$)
\begin{equation}
U(\tau) = \alpha_0 + \sum_{l\in
Lsp(M,g)}{\alpha_l(\tau)}
\end{equation}
where the distribution $\alpha_0$ is given by a power series
expansion in $\tau$ with coefficients that actually coincide at
$\tau = 0$ with the heat invariants. The more complicated terms
$\alpha_l$ involve infinite series with logarithmic terms, but the
principal part that measures the most severe singularity - that is
the coefficient in front of $(t-l+i\epsilon)^{-1}$, is given by
the familiar formula
\begin{equation}
\alpha_{l,-1}=\sum_\gamma \frac{e^{\frac{i\pi}{4}\mu_\gamma}l^\#_\gamma}{|\det(\mathbb{I}-P_\gamma)|^{\frac{1}{2}}}
\end{equation}
where the sum is taken over all closed geodesics with length
$l_\gamma = l$ and $l^\#_\gamma$ denotes the primitive length in
each case. This formula works both for closed manifolds and
domains with Neumann boundary conditions. In the Dirichlet case
one has to modify it with a factor $(-1)^r$, where $r$ is the
number of reflections of the given orbit (note that this is the
well known from geometric optics phase inversion from a mirror).
In the Robin case this factor is more complicated and will be
discussed below.\\
From the above it is already quite visible that the wave trace is
a more general case of the Gutzwiller trace formula that we
derived in connection to the Schr\"{o}dinger equation. The
information contained in the latter will often be sufficient for
our considerations, but knowledge about the full wave trace (if
possible at all) is much more profitable in inverse spectral
problems.\\
There are several distinct approaches aiming the computation of
the wave trace coefficients. The first one is to construct a
micro-local parametrix just as in the case of the heat equation,
and then to use stationary phase method to evaluate the singular
terms in the expansion. This appears, however, quite a cumbersome
procedure even in the simple case of closed manifolds. Instead of going into detail, we give in the last section just an idea of several alternative methods. For more consistent exposition we refer to \cite{Ze}, \cite{Ze2} and \cite{Ze3}.

\newpage
\section{Billiards}
Billiards are model systems that for many reasons have earned great popularity in the investigation of various problems in mathematical physics. Among their major advantages are their conceptual simplicity and the good behavior in the semiclassical limit, that makes them easy to quantize. Moreover, they are so far the best candidate in the attempt to  quantizing chaos.\\
There are many books and articles that for some reason incorporate this item - starting with \cite{B}, \cite{Gal} and the appendix of \cite{A1} for the classical dynamics of a billiard, through the Berry articles \cite{Ber} and \cite{BH} (and many others) and arriving at some modern aspects, found in \cite{Fur}, \cite{Zee}, \cite{An} and \cite{OK}.
We specially refer to \cite{Waa} for the elliptic case and to \cite{Par}, \cite{Sie} - for a thorough review of various mathematical aspects, concerning ergodicity as well.\\

\noindent The notion of a (classical) billiard system is quite an intuitive one: given a connected domain $M$ in $\mathbb{R}^n$ with piecewise smooth
boundary $\displaystyle{\partial M = \cup_i N_i}$
($N_i$ being smooth boundary arcs), the billiard dynamics is defined
as the dynamics of a free particle, constrained in $M$ and
reflected according to the familiar from geometric optics Snell law of
reflection\footnote{according to which the angle of reflection is equal to the angle of incidence}. That is how the \emph{billiard ball map}
is constructed: given an initial point $x\in M$ together with the
initial direction (or equivalently a point $\xi$ in the unit ball
bundle over $M$), we parameterize the unit geodesic $\gamma_\xi$
passing from $\xi$ with some parameter $\tau$ (usually chosen to
be the distance, measured along the geodesic arc - we shall call
it 'time' here), and then the 'transformed point' $G_\tau (\xi)$
we define as the position and normalized velocity of the same free
particle after time $\tau$.\\
We note that billiard trajectories can also be defined as the the
trajectories  along which the singularities of the wave equation
are propagated. Most of them are simple $n$-bounce orbits ($n\geq
2$), governed by the Snell law. However if the trajectory intersects the boundary
tangentially rather then transversally (forming a so-called 'gliding ray'), one has to deal with computational complications. Therefore, it is common practice to
consider only convex domains for which all closed trajectories are $n$-bounce orbits plus  closed geodesics on the boundary (it two dimensions there is only
one such 'glancing' trajectory that is the boundary itself). \\
Another important class are the two-fold trajectories called also
\emph{bouncing ball} orbits. Each convex billiard has at least two
of them, one being the diameter. For this special type of orbits, the notion of \emph{ellipticity} is reduced to the constraint that the length $l_\gamma$ should satisfy either $l_\gamma < \min (\rho_0,\rho_1)$ or $\max (\rho_0,\rho_1) < l_\gamma < \rho_0 + \rho_1$, where $\rho_{0,1}$ denote the radii
of curvature at each of the two endpoints respectively.
Alternatively, $\gamma$ is said to be \emph{hyperbolic} if
$l_\gamma > \rho_0 + \rho_1$ or $\min (\rho_0,\rho_1) < l_\gamma
<\max (\rho_0,\rho_1)$. In particular, if $\gamma$ is a local
minimum diameter, then $l_\gamma < \rho_0 + \rho_1$ and for local
maxima we always have $l_\gamma > \rho_0 + \rho_1$.\\
In the elliptic case the eigenvalues of $P_\gamma$ are of the form
$e^{\pm i\alpha}$, where the real parameter $\alpha\neq
k\pi$ is given by $$ \cos{\frac{\alpha}{2}}=\sqrt{\left (
1-\frac{l_\gamma}{\rho_0}\right )\left (
1-\frac{l_\gamma}{\rho_1}\right )}$$ At the same time, the
eigenvalues in the hyperbolic case, which are of the form $e^{\pm \lambda},\:\lambda\in \mathbb{R}$ are calculated with the
help of the formula $$ \cosh{\frac{\lambda}{2}}=\sqrt{\left (
1-\frac{l_\gamma}{\rho_0}\right )\left (
1-\frac{l_\gamma}{\rho_1}\right )} $$
No matter the simple definition, billiards might become very
complicated in terms of dynamics. Fortunately, there are special cases when they appear much easier to study, thanks to some exceptional features, such as
integrability, ergodicity, or discrete symmetries. In this paper
our main concern is  the elliptic billiard table, which is a typical example of an integrable system (its integrability follows from the integrability of the geodesic flow
of the ellipsoid, by a limiting procedure), but we also pay
attention to the chaotic case, as it is rather conceptual.\\
Polygonal billiards on the other hand are interesting as they illustrate the role of
corners in the boundary. They can exhibit both integrable (triangle, rectangle) and chaotic motion. Nevertheless we omit this topic in order to be as concise as possible.\\

\subsection{Integrable and Chaotic Billiard Dynamics}
Now we investigate a little bit more thoroughly both types of billiards we are interested in.\\
In the integrable case trajectories typically come in families, parameterized by \emph{KAM tori}\footnote{the tori that foliate phase space, given by action-angle variables} while in the latter they fill regions of the phase space everywhere dense - the dynamics is said to be \emph{ergodic}.\\
We consider below the two most typical integrable billiard tables, shown in the picture - the rectangle and the ellipse (and its limiting case - the circle, usually studied separately).\\

\begin{figure}\centering\resizebox{13cm}{!}{\rotatebox{0}{\includegraphics{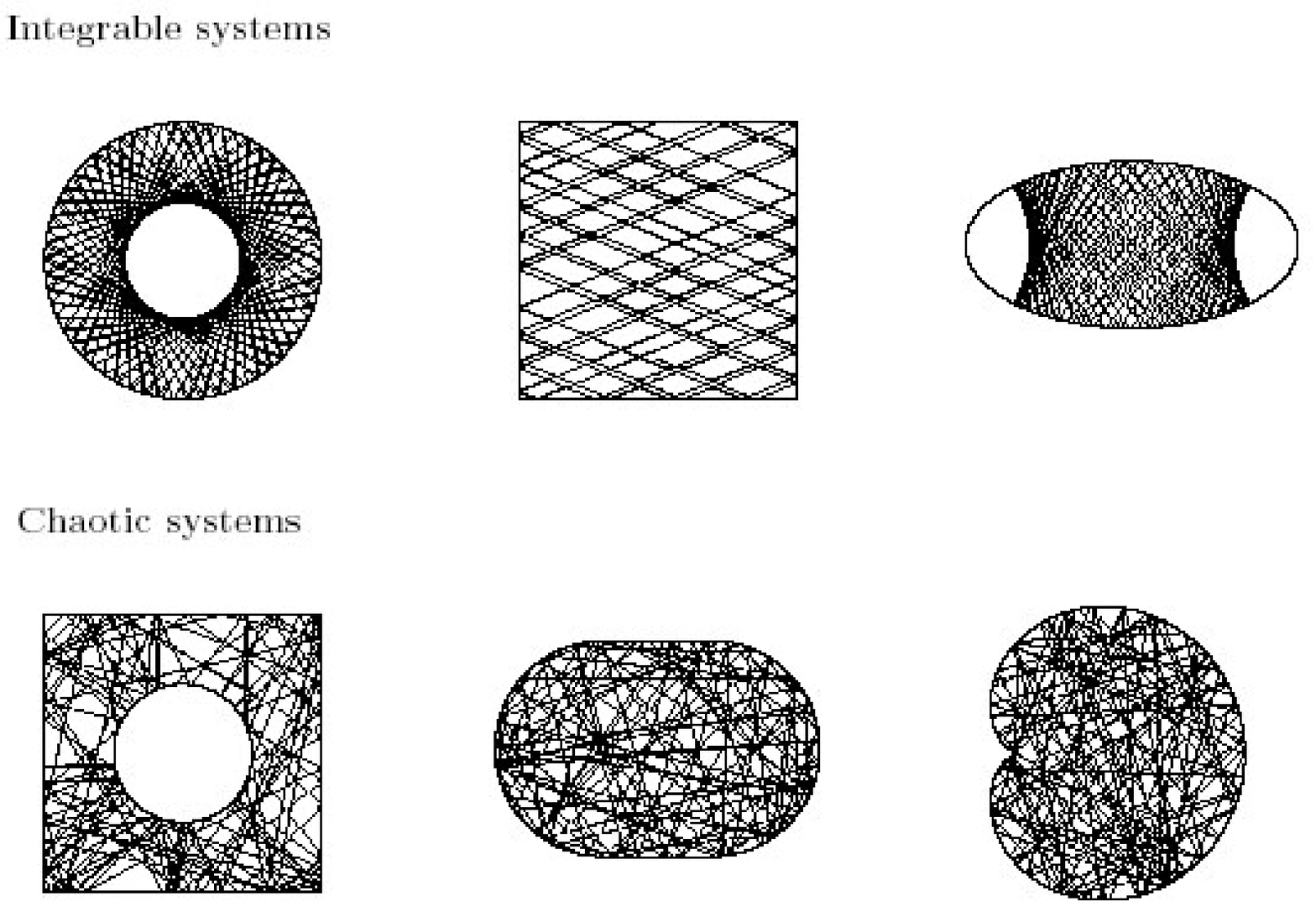}}}\label{billiards}\caption{\small Here we see plots of several typical billiard tables - the integrable \emph{circle, rectangle} and \emph{ellipse}, and the chaotic ones - \emph{Sinai, stadium} and \emph{cardioid}}\end{figure}
\noindent For the rectangular table integrability is obvious - after a suitable folding, the table is transformed to a torus and the billiard flow, up to a parametrization, coincides with the geodesic flow on the torus, which is integrable by definition.\\
For the ellipse we may deduce the integrability from the integrability of the geodesic flow over the three-dimensional ellipsoid, which projects exactly on the billiard flow after shrinking one of the axes to zero size. This procedure can be shown to preserve the integrability.\\
In order to show that the $n-$dimensional ellipsoid has integrable dynamics, one may introduce Jacobi elliptic coordinates $\lambda_i$, defined as the roots of the algebraic equation
$$\sum_{k=1}^{n}{\frac{x_k^2}{a_k-\lambda}} = 1 $$
where $a_k$ are the semi-axes and $x_k$ - the cartesian coordinates of $x$.
However, by doing so one may attempt to show directly that the elliptic table is integrable. In the two-dimensional case these coordinates reduce to the sum and difference  of the distances from a given point to the foci $f_1$ and $f_2$. So, let for example $r(f_1,f_2)=2c$, and  $r_i(x) = r(x,f_i)$ for each point $x$ on the table. Then the elliptic coordinates are defined, according to the above definition by
$$\xi = r_1+r_2, \quad \eta = r_1 - r_2$$
In this way the whole table is knitted by an orthogonal net of confocal ellipses $\xi=const$ and hyperbolae $\eta = const$, as shown on the picture. Now, using the fact that these are actually caustics for the closed billiard orbits and each orbits remains tangent to the same ellipse or hyperbola of the family, one may treat them as first integrals and thus show that the billiard is integrable.\\

\begin{figure}\centering\resizebox{9cm}{!}{\rotatebox{0}{\includegraphics{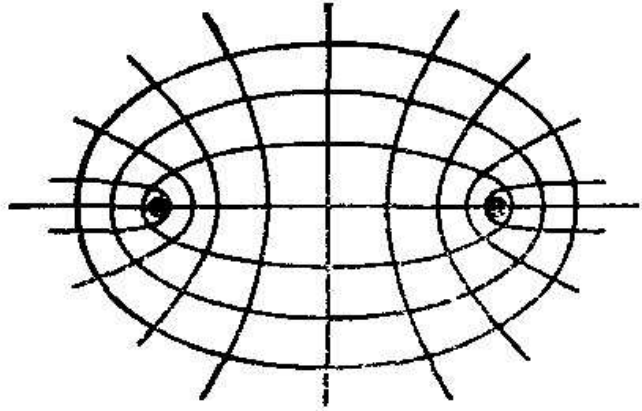}}}\label{ellipse}\caption{\small The \emph{elliptic coordinates} allow for the \emph{integration} of the billiard table and the \emph{geodesic flow} of the ellipsoid in the higher dimensional case. At the same time they describe most naturally the \emph{caustics} - confocal ellipses and hyperbolae.} \end{figure}

\noindent For chaotic billiards, in many cases the same limiting procedure can still be used. Planar convex tables with additional scatterers in the inside for example \footnote{ that are somewhat natural generalization to the Sinai billiard} may be regarded as projected hyperbolic surfaces. The latter are known to posses chaotic geodesic flow as seen from the Jacobi equation. Note also that the scatterer's boundary is concave from the inside of the table, which is reason enough for a chaotic motion. Other typical cases of chaotic tables are represented by the \emph{stadium billiard} and the \emph{cardioid}, shown above, as well as various \emph{Sinai billiards}, that are usually constructed by removing a disc from the skin of the torus. In fact, a slightly modified Sinai billiard - a circular scatterer in a rectangular box, instead of a torus, would be the simplest known so far proof of the ergodicity of hard-sphere gasses - even for two particles, the system is chaotic, as the additional scatterer (representing the second particle) looks concave from the inside.\\
In the case of chaotic billiards what one usually does in attempt to study the spectral statistics is to retrieve information about the mean level spacing, clustering etc., using ergodic theory. These estimates are often the only relevant data to predict and test by experiment. However in many cases they reach to amazingly rich results.\\
It would be unfair not to mention the large class of billiards, known as \emph{pseudo-integrable}. The pseudo-integrable systems in general are given by a hamiltonian
$$H = H_0(p,q) + \epsilon H'(p,q) $$
where $H_0(p,q)$ is a completely integrable hamiltonian and $\epsilon$ - a small parameter. \\
It is known from KAM theory, that for $\epsilon$ close enough to zero, most tori survive and the set measure of totally destroyed tori grows with $\epsilon$, so the system is by no means ergodic, nor integrable for a generic $H'(p,q)$.
\newpage
\subsection{Billiard Quantization}
Roughly speaking, to quantize a billiard, means to study the wave (or Schr\"{o}dinger) dynamics instead of the geodesic (or billiard) flow. Due to the compactness of the domain in interest (the billiard table $\Omega$), the wave dynamics is reduced to the stationary equation
\begin{equation}
\left ( \triangle + \lambda \right)\psi(x) = 0 \qquad x\in \Omega
\end{equation}
However, since $\Omega$ is not anymore a closed manifold, one needs to impose suitable boundary conditions on $\partial\Omega$. The most naturally appearing in quantum mechanical setting boundary conditions are homogenous Dirichlet. They are interpreted as infinite potential walls, surrounding the domain. Neumann boundary data is mostly applicable to problems concerning electrodynamics, acoustics and fluid dynamics.\\
The Robin case is also relevant, as observed by Ballain and Bloch, in the investigation of a particular problem of nuclear physics.

\subsubsection{A Toy Model for the Toy Model}
The first explicit example to consider here will be the circular billiard table - one of the few that exhibit exact solutions. The dynamics is rather simple - it is governed by the conservation law $p=const$ (all reflection angles are equal). Nevertheless, there is a conceptual difference between orbits with angles that are rational, respectively irrational sub-multiples of $\pi$. The former close after $n$ bounces while the latter fill a whole annulus\footnote{these are certainly projections of rational and irrational windings of the torus, defined by the integrability of the system - the inner radius of the torus projects onto the caustic for the corresponding family of orbits}. Note that even rational orbits are non-isolated, but come in families instead - we may rotate each $n$-bounce orbit to an arbitrary angle $\phi\in (0,2\pi/n)$ that actually labels the family. In addition, all closed orbits are neutrally stable.\\
The reason to consider this case in the first place is that the Helmholtz equation for the circle has an exact solution given by
$$ \psi_{lm} = c_{lm}J_l(k_{lm}r)\exp {il\theta},\quad l\in\mathbb{Z},\:m\in\mathbb{Z}/\{0\}$$
and the free Green function can be expressed in terms of modified Bessel functions as
$$ G_0(x,x',s^2)=-\frac{1}{2\pi}K^0(s|x-x'|)$$
What one needs to do next is to adapt this kernel to the boundary conditions and then, using the classical Sohotski formula, express the spectral density in the large $k$ limit.\\

\subsubsection {The Beauty of Ellipses}
Elliptic billiard tables posses some noticeable properties, in addition to integrability, that make them an interesting item in themselves. For example, it is a familiar result that a trajectory, passing trough one of the foci, passes trough the other after the first reflection. Moreover such trajectory asymptotically tends to the major axis - the line, determined by the two foci. This is the stable bouncing-ball orbit. There is only one more two-fold trajectory - the minor axis, which is unstable. Note that these two lines are also the symmetry axes of the ellipse - it is preserved under reflection with respect to them. The fact that the table is parameterized by a caustic net is rather geometrical, even if we leave aside its crucial role for the integrability of the system. The two types of caustics determine two types of trajectories - the first, sometimes referred to as \emph{rotation}-like trajectories are those, that never cross the segment between the foci - the ones that are tangent to confocal ellipses. Orbits in the second class are said to be of \emph{oscillatory} type, as they bounce near the minor axis, each time crossing the segment between the foci and always touch the same confocal hyperbola. Note that according to our previous definitions these are exactly the elliptic and hyperbolic orbits in the billiard. We also make a point here that the fixed set of this billiard is \emph{clean}, meaning that all fixed orbits are sub-manifolds, with their tangent spaces being the fixed set of the differential map, or, equivalently, the length functional is \emph{Bott-Morse} on the free loop space.
\begin{figure} \centering \resizebox{8cm}{!} {\rotatebox{0} {\includegraphics{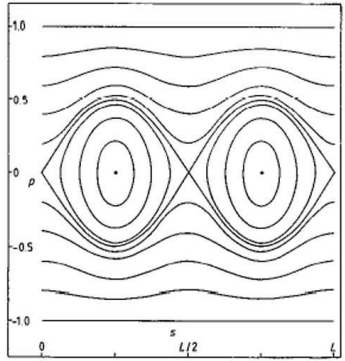}}}\label{berry_deform}\caption{\small The \emph{phase portrait} of the elliptic billiard in $s,p$-coordinates: the two-bounce orbits may be regarded as degenerate cases of oscillatory and rotation-type motion, respectively. The stable one begins at $p=0,\: s = L/2$, and the unstable (the minor axis) - at $p=0,\: s=0$} \end{figure}
The whole chord space has the topology of \emph{M\"{o}bius} strip, since closed orbits are periodic and invariant under time-reversal. When unfolded, this strip is seen to be divided into regions of elliptic and hyperbolic type of motion - note that due to the additional symmetries it looks rather specific. The separatrices are given by the two axes of the ellipse.\\
As far as classical dynamics is concerned, we remind that the billiard map is integrable with the help of the elliptic coordinates, we previously introduced. If the boundary ellipse with foci, positioned at $f_{1,2}=(\pm\epsilon,0) $, is defined by the equation
$$x^2+\frac{y^2}{1-\epsilon^2}=1, \quad 0 \leq \epsilon < 1 $$
then the elliptic coordinates $\eta = \cosh{\rho}$ and $\xi = \cos{\phi}$ are introduced via the transformation $$ (x,y) = (\epsilon\cos{\phi}\cosh{\rho},\epsilon\sin{\phi}\sinh{\rho})$$
The confocal ellipses are the curves $\rho = const,\:\: \phi \in [0,2\pi)$ and the hyperbolae - $\phi = const,\: \rho\in [0, {\rm arcosh}(\frac{1}{\epsilon})]$.
In these coordinates the classical hamiltonian of a free particle of unit mass is given by $$\displaystyle{H=\frac{p^2_\rho+p^2_\phi}{2\epsilon^2(\cosh^2(\rho)-cos^2(\phi))}}$$
This hamiltonian is now most easily seen to be integrable - one just needs to multiply by the denominator and introduce the separation constant $K$ to reduce the problem to a system of one-dimensional equations
\begin{eqnarray}
p_\rho^2 & = & 2E\epsilon^2\cosh^2{\rho}-K  \nonumber \\
p_\phi^2 & = & K - 2E\epsilon^2\cos^2{\phi}
\end{eqnarray}
Then the two integrals of motion are the energy $E$ and $\kappa^2 = K/2E$. The above equations can be regarded as describing hamiltonian dynamical systems with effective energy $E_\rho = -E\kappa^2,\: E_\phi = E\kappa^2$ and effective potentials $V_\rho = -E\epsilon^2\cosh^2{\rho}$ and $V_\phi = E\epsilon^2\cos^2{\phi}$, respectively.\\
We note that the above representation allows also for exact quantization - the system is transformed to two identical Mathieu equations of the form
$$ (1-x^2)f'' - x f'+(\lambda -c^2x^2)f  =  0 $$
with $x=\xi$ and $x=\eta$ respectively.\\
The eccentricity $e$ of the ellipse is given in these coordinates by $e =
(\cosh^2 {\rho})^{-1}$. If we consider the slope angle of the
ellipse $\displaystyle{\psi}$ (the angle, between the counterclockwise oriented
tangent and the $x$-axis at each point), it is obviously connected
with the elliptic coordinates by the formula
$$\tan{\psi} = \frac{dy}{dx} = -\tanh{\rho}\cos{\phi} $$
and the radius of curvature is given correspondingly by
$$R(\psi)=\frac{ds}{d\psi}= \frac{\epsilon \cosh{\rho}\sinh{\rho}}{(\cosh^2{\rho}\sin^2{\psi}+\sinh^2{\rho}\cos^2{\psi})^{3/2}}$$
In these coordinates one may find explicit expression for the
caustics in the billiard as
$$F(s,p)=\frac{p^2-e^2\cos^2{\psi(s)}}{1-e^2\cos^2{\psi(s)}} $$
In certain considerations, however, it is more convenient to work in different coordinates. Namely, let $s$ denote the arc-length coordinate of the initial bounce of the orbit and $p=\cos{\theta}$ - the normalized momentum\footnote{since the hamiltonian is homogenous function of $p$, it can always be re-scaled, restricting the momentum to a unit ball sub-bundle of $T^*(\partial\Omega)$}. Then the canonical symplectic form is given by the area element $\omega = \sin {\theta}\, d s\wedge d\theta $ and it is easy to see for example that the billiard map preserves it.\\

\begin{figure} \centering \resizebox{13cm}{!} {\rotatebox{0} {\includegraphics{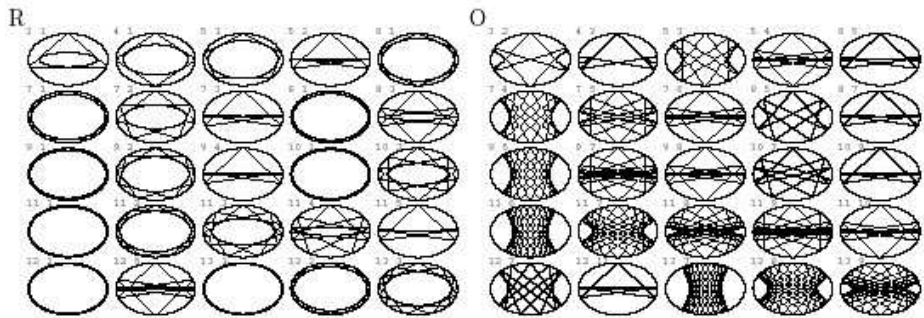}}}\label{billiards_classic}\caption{\small Standard plots of \emph{closed geodesics} of different type for the elliptic billiard. For higher number of \emph{reflections} caustics become quite visible.}\end{figure}
\subsection{Back to Semi-classics}
Various techniques have been involved so far in the semi-classical study of billiards. One can attempt to apply any of the methods discussed above as well as many others that have been invented for specific cases. \\
For instance, one may go for the heat kernel expansion for the billiard table, considered as a manifold with boundary. Here are the first few heat kernel coefficients for a
Dirichlet quantum billiard in terms of the boundary curvature
$\kappa$
\begin{eqnarray}
a_0  =  Area(\Omega), \qquad a_{1/2}  & = &  -\sqrt{\frac{\pi}{2}} L(\Sigma)\nonumber \\
a_1  =  \frac{1}{3}\int{\kappa\,d\sigma}, \qquad a_{3/2} & =  & \frac{\sqrt{\pi}}{64}\int_\Sigma{\kappa^2\,d\sigma} \nonumber \\
a_2  = \frac{4}{315}\int_\Sigma{\kappa^3\,d \sigma}, \qquad
a_{5/2} & = & \frac{37\sqrt{\pi}}{2^{13}}\int_\Sigma{\kappa^4\,d\sigma}-
\frac{\sqrt{\pi}}{2^{10}}\int_\Sigma{(\kappa')^2\,d\sigma}\nonumber
\end{eqnarray}
where $L(\Sigma)=\int_\Sigma{d\sigma}$ denotes the length of the
boundary $\Sigma =
\partial \Omega$.\\
We note that for many purposes, as long as billiards are concerned, the heat trace approach is insufficient as it does not account for the dynamics in any explicit way - unlike the wave group, the heat semi-group is dissipative and could not be related directly to a conservative dynamical system.\\
A more adequate approach in this direction would involve the wave trace, or semiclassical propagator traces. In the case of two-dimensional convex billiards for example, we have the two parts of the spectral density $d(E)= \bar{d}(E)+ d_{osc}(E)$ in the form
\begin{eqnarray}
\bar{d}(E) & \sim & \frac{2m\Omega}{2\pi\hbar^2}J_0(0) \nonumber \\
d_{osc}(E) & \sim & \left (\frac{m}{2\pi^2\hbar^2}\right )^{3/4}E^{-1/4}
\sum{\frac{\Omega_i}{\sqrt{l_i}}\cos{\left ( \frac{l_i}{\hbar}\sqrt{2mE}+2\nu_i - 1/4\right )} }\nonumber
\end{eqnarray}
where $l_i$ denotes the length of the $i^{th}$ orbit and $\Omega_i$ - the area of its band. The summation is performed over all primitive periodic orbits and their repetitions.\\

\begin{figure} \centering \resizebox{12cm}{!} {\rotatebox{0} {\includegraphics{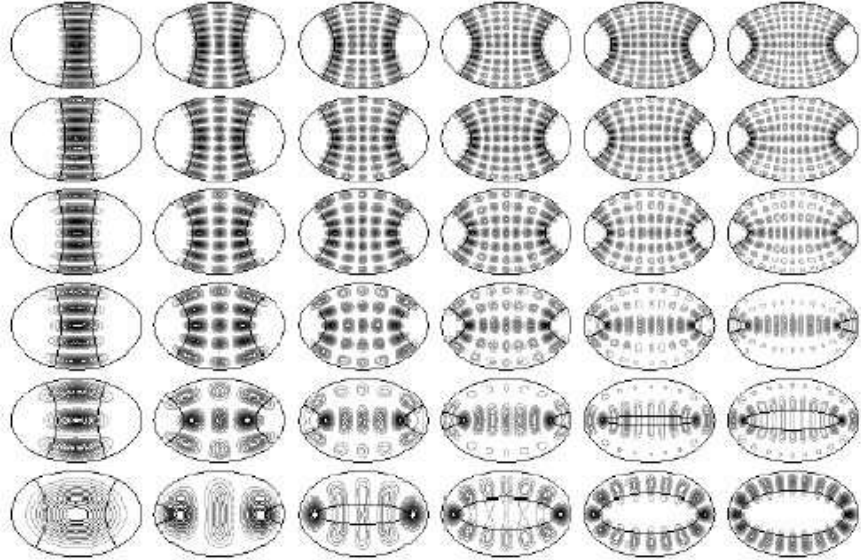}}}\label{billiard_quant}\caption{\small Plots of the \emph{probability density} of different states for the elliptic table. The accumulation of intensity close to \emph{classical caustics} is obvious.}\end{figure}

\subsubsection{Integrability vs. Ergodicity}
The problem of quantizing classically chaotic systems has drawn a lot of attention of both physicists and mathematicians. Unlike the integrable case, exact quantization here is inapplicable since uncertainty principle and discrete spectrum lead to somewhat regular behavior. The only way known so far for quantizing chaos is to consider instead the semi-classical limit $\hbar\rightarrow 0$ of a given chaotic system. Only then the typically chaotic classical properties such as ergodicity or mixing are preserved\footnote{which means that the small $\hbar$ and large $t$ limits do not commute, or the semi-classical estimate is not uniform in time}.\\
One of the fundamental systems in quantum mechanics - the helium atom (that is actually the three-body \emph{Coulomb} problem) has been successfully quantized only with the help of the Gutzwiller trace formula, that is already a serious achievement. For detailed study on this problem we refer to \cite{CW}.\\
Now let us turn our attention to the chaotic billiard tables, which are the main tool to model quantum chaos on the semiclassical level. Such systems differ from the integrable ones in several aspects. Firstly, their closed orbits are exponentially unstable and this usually leads to ergodicity in the phase portrait. Secondly, most orbits are isolated and pass trough almost every point of the $2N-1$ energy surface in phase space, due to the lack of first integrals. Nevertheless there are exceptional families of orbits, but these certainly do not fill resonant $N$-tori as in the integrable case.\\
There is significant difference between isolated and non-isolated orbits in terms of eigenvalue clustering as we shall see further. An adequate way to describe the first case is the Gutzwiller trace that has proved a fairly powerful tool. \\
Nevertheless we follow here a slightly different approach. First we suggest an alternative description of quantum mechanics that is very popular, although not taught in the under-graduate course, and appears to be quite convenient for semi-classical considerations. Then we briefly compare the spectral statistics in the integrable and the chaotic case.\\
\paragraph{The Wigner Function Approach}
The idea of the Wigner function for a quantum state $\psi$ is expanding in a basis of plain waves. More precisely, it is defined as the distribution
$$ W_\psi(x, p) =  (2\pi\hbar)^{-m}\int_{\mathbb{R}^n}{\bar{\psi}(x-y/2)\psi(x+y/2)e^{-ik\langle y, p\rangle} d^n y} $$
The quantum-mechanical description of a system in terms of $W_\psi$ allows for representing in a convenient form basic physical quantities, such as the density
$$|\psi(q)|^2  = \int{ d^m p\, W_\psi(q,p)}$$
or in momentum representation
$$|\psi(p)|^2  = \int{ d^m q\, W_\psi(q,p)}$$
but it also allows for a very 'nice-looking' formulation of the semiclassical description. Namely, we may insert the \emph{de Broigle} wave approximation of $\psi$ in the definition of $W_\psi$ and compute the integral for small $y$. What we get is the form
$$ W_\psi \sim \left | \det \left ( \frac{\partial^2 S(q,P)}{\partial q_i \partial q_j}\right ) \right |\delta (p-p(q,P))$$
which, apart from other things, restricts the dynamics (in the roughest approximation) to classical Lagrangian manifold, given by the Hamilton-Jacobi equation.\\
In the integrable case we have a quantized torus, labeling functions by quantum numbers ${ k} = (k_1\ldots k_m)$ and
$$W^{k}_\psi (q,p) = \frac{1}{(2\pi)^m}\delta(I(p,q) - I_{ k}) $$
where the action variables are given by the WKB quantization condition
$$I_{ k} = \hbar\left ( { k}+ \frac{ \nu_k}{4}\right ) $$
In the analogous situation concerning ergodic systems, we have similarly
$$ W_\psi(q,p) \sim \frac{\delta(E-H(q,P))}{\int\int{d^m p\, d^m q\, \delta(E-H(q,p))}}$$
Using this approach one may find some qualitative differences between the two cases in terms of spectral statistics. We shall only enumerate some basic results, concerning spectral clustering and degeneracies. A more detailed study on that topic may be found for example in \cite{Be}.\\
To begin with, we make a brief discussion on degeneracies. We shall be interested not in degeneracies, coming from symmetries for the Hamiltonian, but only those, obtained by smooth varying of parameters in a family of 'similar' Hamiltonians. Thorough investigation shows that for a 'generic' system, crossing of level curves demands varying at least two parameters (see \cite{Be} for details), so it is of codimension two\footnote{in two dimensions these singularities are projections of the diabolic points at which cones in phase space meet}. This makes it highly unusual to have such situation in ergodic or pseudo-integrable systems, although it is quite common for ergodic systems to have levels that become very close to each other and then repulse again (these may be viewed as hyperbolae on the touching cones, passing close to the singularity).\\
However, in the case of torus-quantized systems, codimension one usually suffices (lattice planes always cross each other when perturbed), so degeneracy of such kind may be observed and a lot of examples are available.\\
The \emph{level spacing} $S$ is another quantity that differs a lot between the ergodic and the integrable case. Let us consider the distribution $P(S)$ defined as the differential probability for the spacing
$$S(\lambda) = \frac{\triangle\lambda}{<\triangle\lambda>}$$
to lie in the interval $(S, S+dS)$ and its asymptotic behavior as $S\rightarrow 0$. If it tends to zero, this can be thought of as effective 'repulsion' of levels, whereas if the limit of $P$ is a non zero constant, then spectral levels become thicker and form a cluster at $\lambda$.\\
    It has been proven that, broadly speaking, $P\sim S^{k-1}$ where $k$ is the minimal codimension of the singular set of spectral degeneracies in the hamiltonian family, considered above. Thus, we have level repulsion for ergodic systems and clustering for those that exhibit torus quantization (integrable or pseudo-integrable). For integrable systems with more than one degree of freedom one has an even better estimate of the form $P\sim e^{-S}$.\\
One more point to make here is that the mean level density in the Gutzwiller trace formula is of order $\hbar^{-m}$ so one might quite naturally expect to obtain it by smoothing $d(E)$ over this range. However, this is not a good idea, since the oscillatory contribution has wavelength of order $\hbar^{-1}$ which is much larger scale in dimensions higher than one. Therefore $\bar{d}$ and $d_{osc}$ should be treated separately. Let us focus for the time being on the latter.\\
Note that for isolated orbits the amplitudes in $d_{osc}(E)$ oscillate with each repetition $l$ for stable orbits and decay exponentially with $l$ for unstable ones. In the integrable case, where orbits come in $m-1$ parameter families, these amplitudes are shown to behave like $l^{\frac{1-m}{2}}$\\
On the other hand each closed orbit contributes to the level density an oscillation of wavelength given by
    $$\tilde{\lambda}_j = T_j(\lambda)h/p $$
    where $T_j$ is the period and $p$ - the momentum. According  to this formula, closed orbits describe level clustering of order $\hbar$, whereas the mean level spacing, as we already know, forms on the much finer scale $\hbar^{-m}$.\\
    In the case of billiards, where the classical action along the $j^{th}$ orbit is given by $$ S_j = \hbar k\ell_j$$
    we have an analogous rule for the wave-number range
    $$\tilde{k} = 2\pi/\ell_j$$
    Note also that in both cases each closed orbit does not give individual levels, but collective properties of the spectrum\footnote{this non-locality can be partly explained by the Fourier transform used in the derivation of the Poisson type summation formulae and seems to be a major property of classical-to-quantum phenomena}, namely, clustering of range given by the above equations. Thus the determination of an individual level in the spectral density involves summation over infinitely many (or at least sufficiently many) closed orbits in order to generate a single delta function via interference of oscillatory terms.\\
    The sense of 'sufficiently many' varies from integrable to ergodic systems and for the former this number behaves like $\hbar^{-m(m-1)}$ as $\hbar\rightarrow 0$. It is even bigger in the ergodic case. For the Sinai billiard for example, as a typical representative, it has been evaluated to $n \sim \hbar^{-3}e^{\alpha/\hbar}$.\\
    In both cases there are much more practical methods for determining individual levels, but nevertheless the path summation is appreciable in certain cases - for instance, if we only want to know the 'smoothed' over a certain range $\triangle\lambda$ level density, it suffices to include much less paths as the range increases. This can be of practical interest for acoustics, where certain resonances (in a concert hall for example) dominate the fine structure of spectra and thus introduce a natural 'smoothing' criterion.\\
Now let us give some explicit examples of particular results obtained in the ergodic case.\\
The first one is the Sinai billiard, that is a typical classically chaotic system. The smooth part of the spectral staircase function is given by
    $$ \bar{\mathcal{N}}(E) = \frac{\lambda}{32\pi}(1-\pi r^2)-\frac{\sqrt{\lambda}}{4\pi}[1+\sqrt{2}/2-r(2-\pi/4)]+\frac{31}{96}$$
where $r$ is the radius of the scattering disk and the torus is taken to be double-folded unit square. The first term above stands for the area contribution, the second - the one of the perimeter, and the last represents the contribution from curvature and corners.\\
A more thorough investigation would reveal the presence 'ghost' orbits - these are non-physical trajectories that pass through the obstacle (or, as in the case of non-convex tables, leave the interior of the billiard domain). One of the advantages of path summation in this case is that such orbits 'cancel' each other, after adding sufficiently many terms in the trace. For details we refer to \cite{Be}.\\
For the case of Robin boundary conditions we refer to \cite{Sie} where the boundary function $\mathcal{S}$ is taken into account to obtain for the smooth part of the spectral density\footnote{see the article for much more precise results}
$$ \bar{d}(\lambda) = \frac{Area(\Omega)\sqrt{\lambda}}{2\pi} - \frac{L(\Sigma)}{4\pi}\left [ 1 - \frac{2}{\sqrt{1+\frac{\mathcal{S}^2}{\lambda}}}\right ]+\ldots$$
whereas the oscillatory contribution comes with an additional phase (compared to the Dirichlet case)
$$\Phi_\gamma \sim 2\sum{\arctan(\frac{p_i\sqrt{\lambda}}{\mathcal{S}(\sigma_i)})}$$
where $p_i=\cos{\theta_i}$  are the tangential projections of the momenta in the points of reflection of the corresponding closed orbit $\gamma$.\\

\noindent We leave this quite interesting topic of quantum chaology for the moment, realizing that there are many aspects and that we consider here only a few examples. For the most part we follow \cite{Be}. One may, however, decide to use more statistical methods, as it is done in \cite{Bo}, or study more thoroughly the quantization of discrete maps, as shown in \cite{CW}. We refer to the latter for a thorough investigation on the matter.\\
There is one more case we only mentioned in the previous section - the geodesic flow on a hyperbolic surface that also exhibits chaotic motion, but this time our research is armed with the more exact Selberg trace formula and the whole machinery of harmonic analysis on homogeneous spaces. For a review on this special case we refer to \cite{Mr} and other articles by the same author.\\

\newpage
\section{Inverse Problems}
Solving an inverse spectral problem meas to determine geometrical data (metrics, boundary conditions etc.) from spectral data. If isospectral deformation is impossible, then the metric or the boundary data is said to be \emph{spectrally rigid}. If it is the unique geometry with this spectrum\footnote{usually it is understood 'within the same class', meaning that each time we try to determine a metric or domain, we do it relative to other representatives of some restricted class of metrics and domains}, then we call it \emph{spectrally determined}.\\
There are only few positive inverse spectral results known
so far and a vast number of counterexamples. For instance, it has
been proved that the standard metric on the sphere is spectrally
determined only in dimensions up to six. The ellipses in the plane
are known to be spectrally determined for the Robin boundary
conditions, as we shall see in the following, but it is not known
whether they are in the Dirichlet case. Flat tori are spectrally
determined only in the low dimensions and in the higher-dimensional case this is so only locally - there are isospectral pairs that are not isometric, but no isospectral
deformations are available, since each metric is determined in a
sufficiently small neighborhood of the superspace. And in the end
we know almost nothing about the hyperbolic manifolds. In
dimension two we know that isospectral pairs contain only
hyperbolic representatives, which cannot be said for sure in the
generic case.\\
However we have spectral determination within the corresponding class of simple real analytic surfaces of revolution with one critical distance from the axis and convex
analytic plane domains with the symmetries of the ellipse and a bouncing ball orbit of fixed length.\\ 
Spectral problems may be understood as revealing the relation between dynamics, geometry and spectrum, hidden in the trace formulae. Note that in these formulae we cannot in general distinguish between different geodesics, but encounter just what is usually referred to as \emph{length spectrum}.\\
The length spectrum is defined as the set of all lengths of
simple closed geodesics. Moreover, when multiplicities of geodesic
lengths are also taken into account, we refer to this as the
\emph{extended length spectrum}. Since the length spectrum gives
no information about which length belongs to which closed
geodesic, in practice it is often preferable to deal with the so
called \emph{marked length spectrum} which classifies the closed
geodesics in homotopy types (in the case of closed manifolds) or
in terms of \emph{rotational numbers}\footnote{this quantity is
defined as the ratio of the winding number and the number of
reflections} (as long as convex domains are involved).\\
It is curious to note that equality of the
marked length spectra implies isometry at least for locally
symmetric spaces and so does symplectic conjugacy between the
geodesic flows. \\

\subsection{Basic Strategies}
How do we find a positive or a negative inverse spectral
result?\\
The negative one is certainly easier to prove, since one may just find a counterexample. For a positive result, however, a lot of work needs to be done. Perhaps the simplest inverse result is the spectral determination of Dirichlet discs. It uses only the two leading terms in the heat asymptotic expansion and the property that for any planar domain the inequality $$ Area(\Omega)\leq \frac{L(\partial\Omega)}{4\pi}$$ holds and turns to equality if and only if $\Omega$ is a disc.\\
This case is, however, rather exceptional - there are many others in which the heat invariants remain constant under deformation which does not in general preserve the spectrum (for example the 'bumpy metrics' considered in \cite{Me}) so one needs to resort to more subtle spectral data, such as the wave trace.\\
To sum up, solving a generic inverse spectral problem is far from being trivial, but if there is any chance for it to be done, it would be best to have certain procedure to follow. The first step is usually defining as much spectral invariants as possible (from heat or wave trace, zeta function etc). Next one needs to express these invariants in geometrical or dynamical terms, as in the case of the heat kernel coefficients. In the wave case we have the coefficients in the asymptotic expansion arising in an analogous way from singularities (or non-commutative residua). Under certain simplifying assumptions (for instance a simplicity of the length spectrum or analyticity of the domain) one may find a very 'nice' set of geometrical and dynamical spectral invariants, such as the length spectrum itself, or the Birkhoff normal form. But then comes the third crucial step when we need to completely determine
the geometry (or dynamics) out of these invariants, and this is, first of all, rarely possible, and second - there is plenty of hard work to be done.\\

\paragraph{Livsic Cohomology}
Let us now go a little bit more into detail. Up to first order
in perturbation theory, we have for the variation of eigenvalues
$$ \dot{\lambda}_k= \langle \dot{\triangle}\phi_k,\phi_k \rangle $$
where $\phi_k$ are the unperturbed eigenvalues, and
$\dot{\triangle}$ - the variation of the laplacian, which is,
generally speaking, a $\Psi DO$ with a leading symbol $\dot{g}$. The
linearized problem suggests defining the space of variation of the
metric (the tangent to the supers-pace of $M$ at $g$) for which $$
\int_\gamma{\dot{g}}=0 \Longleftrightarrow \dot{\lambda}_k = 0$$
for all indices $k$ and closed geodesics $\gamma$. At this stage
a special class of co-homologies, the so-called \emph{Livsic cohomologies}, are already involved. The Livsic cohomological problem studies whether each cocycle $ \int_\gamma{f}= 0\: \forall \gamma$ necessarily has a potential in the sense $f=X_h(F)$ where $X_h$ is
the generator of the geodesic hamiltonian. This is connected to inverse theory in the following way: let the variation of the metric tensor satisfy $$ \int_\gamma{\dot{g}}= 0 \quad\: \forall \gamma$$ so that it also preserves the
extended length spectrum. When the cohomology is trivial one has
by all means the above representation of $\dot{g}$ and then one
has to study the harmonic analysis of the underlying manifold in
order to determine whether such representation is possible at all.
For negatively curved surfaces the answer is 'no' and thus no
iso-spectral deformations are available in this case. Therefore
such manifolds are spectrally rigid. This technique works fairly well in
the case of closed manifolds - especially when all geodesics are
closed or in the case of integrable systems, foliated by flat tori
- the Livsic equation then involves integration over these level
sets and Fourier analysis suffices. However this approach is
generally ill-defined or too complicated for billiards.\\

\paragraph{The Parametrix}On the other hand one may choose to construct a parametrix of the wave equation in analogy with the heat kernel. However such construction becomes in this case rather  complicated\footnote{nevertheless it has been successfully used for obtaining certain inverse results before new schemes were developed} even for simple geometries. Therefore we prefer to  briefly review two other more up-to-date methods in the context of concrete problems they can be successfully applied to.\\
For a more consistent exposition we refer to \cite{Ze2} and \cite{Ze3}. Among the few surveys on inverse problems, we adhere mostly to \cite{Ze} and \cite{Me}.

\subsection{Quantum Birkhoff Normal Forms and Surfaces of Revolution}

Let us first note that the wave group is a quantization of the
geodesic flow and certain correspondence-principle relations hold.
For instance, if two laplacians are conjugated with respect to a
unitary $FIO$, the geodesic flows are conjugated with respect to a
homogeneous symplectic transformation. At least locally,
isospectrality implies (for simple length spectrum) symplectic
equivalence around closed orbits (equal Birkhoff normal forms). In the next paragraph we remind the notion of a classical normal forms, that is somewhat central for the theory of dynamical systems, and then obtain formal quantization.\\

\paragraph{Classical Birkhoff Normal Forms}
The so-called \emph{Birkhoff Normal Forms} treat hamiltonians close to a stationary point (such as periodic orbits, that will be our main concern). Roughly speaking, putting an equation or a hamiltonian function into normal form means finding simple approximate expression (linear, polynomial etc.)\\
Compared to the generic case, however, here there are certain principle difficulties due to the symplectic nature of the problem. First of all, the transformation that brings $H$ into normal form should be canonical, so usual cutting the Taylor series up to the quadratic term, as we are used to in the context of $ODE$ close to equilibrium, is not sufficient. By doing so, in this case one may lose qualitative information - for instance the linearized equation may predict Lyapunov stability, that is observed in practice and can be easily destroyed if one takes a  higher-order correction. If the system is close to integrability, it could be possible to find a set of action-angle variables for which its normal form is integrable up to the higher-order remainder. However its influence may accumulate over a sufficiently large time scale and may end up in a chaotic motion. \\
There is one more major obstacle - the appearance of resonances. Suppose that one has managed to put $H$ into normal form close to stable equilibrium. Then its quadratic term (for $n$ degrees of freedom) will give an expression of the type
\begin{equation} H\sim \frac{1}{2}\sum_{k=1}^n{\omega_k\tau_k}\end{equation}
where the canonical actions are expressed in terms of the new symplectic coordinates $P$ and $Q$ as $\displaystyle{\tau_j  = \frac{1}{2}(P_j^2+Q_j^2)}$ and $\omega_j$ are the corresponding frequencies. The system is said to be resonant of order $s$ if there exists a linear combination of the frequencies with integer coefficients $k_j$ such that $\sum{|k_j|} = s$. The main theorem, due to Birkhoff, states then, that if there are no resonances of order $s$ and smaller, then $H$ can be put in normal form which is polynomial of order $s$ in the $(P,Q)$-variables (that is $[s/2]$ in the $\tau$'s) and the reminder is of order $\mathcal{O}(\tau^{\left [\frac{s+1}{2}\right ]})$.\\

\paragraph{Quantization and relation to wave invariants} The strategy of normal forms
works for integrable dynamics and we illustrate it in the case of simple real analytic surfaces of
revolution. The first step is to use the Laplace spectrum to
construct the quantum Birkhoff normal form of the wave group, that
is essentially, expressing the operator $\sqrt{\triangle}$ as a
polynomial of the action operators $I_k$ and the tangential
derivative along $\gamma$. Then, one hopes to reconstructs the
metric from its normal form. After imposing some additional
assumptions on the geodesic flow, in many cases this turns out to be possible, although the procedure is far non-trivial. Here the aim is only to give an idea of it, avoiding technicalities as much as possible. \\
First of all one conjugates $\sqrt{\triangle}$ into the maximal
abelian algebra of $\Psi DO$'s, generated by the action
variables around a closed non-degenerate geodesic and the
tangential derivative $\displaystyle{D_s:=-i\frac{d}{ds}}$.
Actually, this conjugation is performed on the normal bundle of
$\gamma$ and the result is governed by the spectrum of $P_\gamma$.
In the elliptic case for example, one has all the action variables
in the form of oscillators
$$ \hat{I}_k = \alpha_k(D^2_{y_k}+y^2_k)$$
where $\alpha_k$ are the exponent factors of the linear
Poincar\'{e} map spectrum and $(s,y_k)$ are the Fermi normal
coordinates around $\gamma$.\\
In the real hyperbolic case one has alternatively $$ \hat{I}_k =
\mu_k(D_{y_k}y_k+y_k D_{y_k})$$ Under some additional
assumptions that we are going to specify later, it turns out that
for a non-degenerate closed geodesic $\gamma$ there always exists
a microlocal $FIO$, $W: T^* N_\gamma/{0}\rightarrow
T^*(S^1\times\mathbb{R}^n)$ which conjugates $ \sqrt{\triangle}$
in the form
\begin{equation}
W\sqrt{\triangle}W^{-1} =
D_s+\frac{H_\alpha}{L}+\sum{\frac{\tilde{p}_k(\hat{I_1},\ldots\hat{I_n},L)}{D_s^k}}
\end{equation}
where $\tilde{p}_k$ are polynomials of degree $k+1$ in the
action variables and  $2H_\alpha = \sum{\alpha_k \hat{I}_k}$.\\
Next, one hopes to establish a one-to-one correspondence between the
normal form coefficients and the wave trace invariants. In order to
do so, it is helpful to use residual calculus techniques to show
that $$ a_{\gamma,k} = Res_{z=0}{{\rm Tr}D_t^k\,
e^{it\sqrt{\triangle}}\sqrt{\triangle}\,^{-z}}$$ After putting
$\sqrt{\triangle}$ into normal form one hopefully obtains the
equality $$ a_{\gamma,k} = \Phi_k (\tilde{p})\frac{i^\sigma}{\sqrt{|\det(\mathbb{I}-P_\gamma)|}}$$
where the $\Phi_k$'s form a special class of polynomials in the $\tilde{p}_j$'s and
hence in the $\hat{I}_l$'s, and the last factor to the right is a character of a
metaplectic representation. The important thing we need to know
about the above formula is that it is in our case invertible,
which means that the Birkhoff normal form coefficients can be
derived from the wave trace invariants.\\

\paragraph{First Integrals}
The last step in the procedure consists in determining the metric
form the normal form. This is actually deriving the metric from
the hamiltonian function, which is quite clear conceptually, but
demands some additional efforts. Therefore we restrict ourselves to the case of simple real analytic surfaces of revolution with simple length spectra.\\
The metric on such a surface may be written in the form
$$ds^2 = dr^2 + a(r)^2d\Theta^2$$
and impose the additional condition that there is precisely one critical point at which $a''<0$ that corresponds to equatorial geodesic. Moreover, we demand the Poincar\'{e} map to be of 'twist' type\footnote{which means, roughly speaking, that its differential rotates all tangent vectors in the same direction}.\\
Surfaces in this class have completely integrable geodesic flows and the corresponding hamiltonians posses global real analytic action-angle variables. The first action variable in our case obviously can be chosen to be the angular momentum $I_1 = p_\theta$ and then the second takes the form
$$I_2 = \frac{1}{\pi}\int_{r_-}^{r_+}{\sqrt{E^2 - \frac{I_1^2}{a(r)^2}}dr}+|I_1| $$
with $r_{\pm}$ being the two extremals of $r$ lying on the annulus determined by $I_1$. From the first part of $I_2$, one may retrieve that (see \cite{Ze2} for calculations) both  $\sum_\pm{|a'(r_\pm(x))|}$ and  $\sum_\pm{|a'(r_\pm(x))|^{-1}}$ are spectral invariants\footnote{note that the integral of interest is Abel transform and hence invertible, we have chosen $x$ to denote $a(r)$} and hence $a'(r)$ is determined by the spectrum. But since we have $a(0)=0$, the shape of the surface is spectrally determined.

\subsection{The Balain-Bloch Approach}
The idea of the method is
essentially using exact identities for the Dirichlet or Neumann
resolvent in terms of the free resolvent. It is far more
explicit than the other two methods and rather convenient to apply
for bounded domains. It has first been used by Balain and Bloch for the Robin problem in a three-dimensional ball.\\
Suppose for the time being that we are dealing with a convex analytic region in the plane and consider an $r$-fold bouncing-ball orbit $\gamma^r$ of length $L_\gamma$ along the $y$-axis. Then denote by $f_\pm^{(j)}$ the Taylor coefficients of the boundary, considered as a function of one variable and calculated at both ends of the orbit.\\
In this case we have the expression
$$ \det({\rm Id}-P_\gamma) =  L_\gamma^{2r}\det{h} $$
where $h = {\rm Hess}(f)$ and let $h^{ij}_{\pm}$  denote the matrix entries of its inverse\footnote{note that one has in addition $h^{ij}_{-} = h^{i-1,j-1}_{+}$}. Then, under the assumption that the domain is symmetric with respect to reflections in the $x$-direction (so that we may drop the '$\pm$' subscript) and $\gamma,\,\gamma^{-1}$ are the only geodesics whose length is $L_\gamma$, one may express the wave invariants at $\gamma$ as polynomials of the $f^{(j)}$'s. \\
In order to see why this is so, let us first consider the boundary-value problem for the wave equation and construct a regularized resolvent $R_\rho$ using a cut-off
$\rho$ around the geodesic length $L_\gamma$. Under the above assumptions, this resolvent admits a complete asymptotic expansion of the form
\begin{equation}
{\rm Tr}\,\mathbb{I}_\Omega R_\rho(k+i\tau)\sim e^{(ik-\tau)L_\gamma}
\sum_{j=1}^\infty{(a_{\gamma j}+a_{\gamma^{-1}j})k^{-j}},\quad k\rightarrow\infty
\end{equation}
which is essentially the same as the wave trace asymptotic at $t = L_\gamma$. In order to find it, one reduces the problem to the boundary, using exact identities from
potential theory. The key formula in the so-called \emph{Fredholm-Neumann} reduction yields
\begin{equation}
R_\Omega(k+i\tau) = R_0(k+i\tau) - {D\ell}(k+i\tau)\circ({\rm Id}+
N(k+i\tau))^{-1}\circ\mathcal{B}\circ\,S\ell^{\,t}(k+i\tau)
\end{equation}
where $R_0$ is the free-space resolvent in $\mathbb{R}^2$ and $\mathcal{B}: H^s(\Omega)\rightarrow H^{s-1/2}(\partial\Omega)$ is the restriction to the boundary. The single and double-layer potential operators ${S\ell},\,{D\ell}:\, H^s(\partial\Omega)\rightarrow H_{loc}^{s+1/2}(\Omega)$ are constructed as in (\ref{layer}), with the help of the free-space Green function $G_0$ and its normal derivative respectively. Note that here ${S\ell}$ and ${D\ell}$ are considered as Fourier integral operators in this context ($S\ell^{\,t}$ denoting the transpose), and finally $N(k+i\tau)$ stands for the the \emph{Fredholm resolvent}, that is the boundary integral operator, induced by $D\ell\in\Psi^{-1}(\partial\Omega)$(see \cite{Ze}). It is classical that ${\rm Id} + N(k+i\tau)$ is an automorphism of the Sobolev space $H^s(\partial\Omega)$.\\
The above formula is in close analogy to the \emph{Grushin} result, that relates the trace of the wave group to traces, concerning the monodromy operator (see \cite{Ze} for example).\\
Following this analogy a bit farther, under the assumption that $L_\gamma$ is the only length in ${\rm supp}\hat{\rho}$ one obtains
\begin{equation}
\int_\mathbb{R}{\rho(k-\lambda)\frac{d}{d\lambda}\ln\det({\rm Id}+ N(k+i\tau))d\lambda}
\sim\sum_{j=1}^\infty{a_{\gamma j}k^{-j}}
\end{equation}
where the $a_{\gamma j}$'s are the wave invariants at $\gamma$ and ${\bf \det}$ denotes the usual Fredholm determinant.\\
Next one examines semi-classically $N(k+i\tau)$ and the geometric series expansion for $({\rm Id}+ N(k+i\tau))^{-1}$. The former clearly has a WKB approximation away from the diagonal, where it has singularity of order $-1$, which\footnote{it is actually order $-2$ in the two-dimensional case} represents it by a FIO with phase equal to the boundary distance function\footnote{from the off-diagonal asymptotic of $G_0$ it follows that as long as $|q(\phi)-q(\phi')|\geq |k|^{1-\epsilon}$ for some $\epsilon < 1$, $N(k+i\tau)$ is a semi-classical FIO with a phase function given by the boundary distance function} $d(\phi,\phi') = |q(\phi)-q(\phi')|$. Then substituting the explicit form of the free Green function in the plane into the definition of $N$, one obtains for its integral kernel
\begin{equation}
N(k+i\tau, q(\phi), q(\phi')) = -2(k+i\tau)H^{(1)}_1(k+i\tau d(\phi,\phi'))\cos\alpha(\phi,\phi')
\end{equation}
where $\alpha(\phi,\phi') = \angle (\overline{q(\phi)-q(\phi')}, \nu_{q(\phi)})$ with $\nu_{q(\phi)}$ being the unit normal at the boundary point $q(\phi)$. Note that this is exactly the angle of incidence at $q(\phi)$ for a billiard trajectory passing through $q(\phi)$ and $q(\phi')$. Then keeping in mind that for convex domains the boundary distance function $d(\phi,\phi')$ generates the billiard dynamics, it is natural to regard $N(k+i\tau)$ as a global quantization of the billiard ball map.\\
Due to the singularity at the diagonal it is not possible to apply directly the stationary-phase approximation scheme and we need some kind of regularization. So the next step is to perform (at least formally) a finite geometric series operator expansion with a remainder. After introducing a suitable cut-off, one separates $N(k+i\tau)$ into a tangential and transversal part. The former is irrelevant to the semi-classical quantization of the billiard map and behaves like an Airy operator, classically associated to gliding (or 'creeping') rays at the boundary. Its contribution turns out to be possible to factor out, which is one of the main features of the Balain-Bloch approach. Another feature is the possibility to introduce a cut-off at $\gamma$ that ensures the good behavior of the estimate.\\
After having done all this, one ends up with oscillatory integral
series which demand a stationary phase method to evaluate.
Therefore a kind of Feynman graph technique\footnote{a diagrammatic technique used as a means of computing integrals
appearing as quantum corrections in perturbation theory} is incorporated to determine all the coefficients. \\
Calculations show that the coefficients $f_\pm^{(2j)}(0)$ and $f_\pm^{(2j-1)}(0)$ of the Taylor expansion for the boundary\footnote{considered as a graph of a function of one variable} appear in the stationary-phase expansion for the first time in the term of order $k^{-j+1}$. In the presence of a symmetry axis,  the leading part of this term has the form\footnote{recall the notations from the beginning of the paragraph} $$a_{\gamma,j-1}\sim2rL_\gamma(h^{11})^jf^{(2j)}(0)+\ldots$$ where the dots stand for lower order derivatives. And in addition, if the domain has the symmetries of the ellipse, since the odd-dimensional coefficients disappear, one may iteratively determine the $f^{(2j)}$'s and hence the whole geometry. This gives a strong result for analytic convex plane domains with two symmetries - they can be determined by the wave invariants at a bouncing-ball orbit along the second symmetric axis.

\subsection{Variation on a Theme by Guillemin and Melrose}
Actually, although we revise it in the context of the Balain-Bloch strategy, the results obtained in \cite{GM} and \cite{GM2}, rely on the construction of a microlocal paramatrix.\\
The main result in \cite{GM} is that the spectrum of the Robin problem in the interior of the ellipse, together with the pure Neumann spectrum completely determine the function $\mathcal{S}$, taking part in the boundary condition. More precisely, the difference of the wave traces for the modified and pure Neumann case respectively:
$$\mathcal{\chi} = {\rm Tr}\cos{\tau\sqrt{\triangle_\mathcal{S}}}- {\rm Tr}\cos{\tau\sqrt{\triangle_0}}$$
is shown to be $\Psi D O$ of order $-1$ (the most singular parts cancel each other) which admits an asymptotic expansion in fractional powers of $t-T$ at $t=T$, around a closed geodesic with period $T$  and leading symbol
\begin{equation}
\sigma_L(\mathcal{\chi}) \sim \sum_{j}{\frac{\mathcal{S}(s_j)}{\sin{\theta_j}}}
\end{equation}
where summation is taken over all points of reflection at the boundary.\\
Guillemin and Melrose menaged to rewrite this spectral invariant as an integral over the set of closed orbits\footnote{which is a smoothly imbedded manifold since the billiard map is \emph{clean} and its action is conjugate to rotation} with period $T$ ( $\sigma_T$  below denotes the induced invariant measure)
\begin{equation}
\sigma_L(\mathcal{\chi}) = \int_{Y_T}{\frac{\mathcal{S}}{\sin{\theta}}\,d\sigma_T}
\end{equation}
Using some simple arguments from the geometry of the ellipse, the authors rewrite the above in the form
\begin{equation} \label{expl}
\int_0^{2\pi}{\frac{\mathcal{S}(s)h(s)ds}{\sqrt{2M(s)+Z - b}}}
\end{equation}
where the function $Z$ actually defines the confocal caustics by the equation
$$ \frac{x^2}{e+Z}+\frac{y^2}{Z} = 1$$
($e=a-b$ as before) and $M(s)$ denotes the function
$$M(s)  = \frac{a+b}{2}-\frac{ab}{a+b-e\cos{2s}}$$
Moreover we have $$\sin{\theta} = \sqrt{\frac{Z-b}{M}}$$
and the rest follows from considerations concerning the invariance of the integral measure. More precisely, the volume form $\omega = \sin{\theta}d\theta\wedge h(s)ds$ can be written also as $$\omega = dZ\wedge f(s)ds $$
where the second multiplier to the right is the invariant \emph{Leray} form on the curves $Z=const$, that is exactly the one we need. We have from the above $$f(s) = \sin{\theta}\frac{\partial\theta}{\partial Z}h(s)$$
and easily find
$$ \frac{\partial Z}{\partial\theta} = -2\sqrt{(Z-b)(2M-Z+b)}$$
Using the symmetry of the ellipse $D\cong\mathbb{Z}_2\otimes\mathbb{Z}_2$, one may write the above integral as a sum of four components and the function $\mathcal{S}$ is determined at the endpoints of each interval, that could be easily checked, expanding over two 'special' orbits - the major and minor axes. Moreover, since the remaining factor in the integral is strictly monotonous in each of the intervals, one sees that the boundary function is determined itself.\\
Evaluating the significance of the result of Guillemin and Melrose, we should admit that it is not very likely that in practice anyone will ever desperately need to 'guess' the elasticity of the boundary of a vibrating elliptical region by its spectrum, but this is one of the very few exactly solved inverse results that makes it of particular interest for further investigation. The articles \cite{PT} and \cite{Zee} provide some insight about this. \\

\noindent In this section we offer an easy way to obtain explicit formulae for the spectral variations with respect to varying the boundary function. It is worth saying that the method is completely identical to the one offered in the second chapter of \cite{MF}, where it was applied to the pure Neumann case only. As a matter of fact the same investigation has been cited in \cite{Zee}, but using a different method, and obtaining, as it seems, a different result, which we consider a minor misunderstanding.\\
To start with, we regard the spectral problem
\begin{equation}
\left (\triangle+\lambda \right )\phi (x) = 0 \quad x\in\Omega
\end{equation}\label{de}
in a plane domain  $\Omega$ , enclosed by the ellipse $\Sigma    = \partial\Omega$.\\
Let $\phi_{\,0}$ be a solution satisfying the homogeneous Robin boundary condition
\begin{equation}
\left ( \frac{\partial}{\partial n}+\mathcal{S}_0(s)\right )\phi_{\,0}(s) = 0\quad s\in \Sigma
\end{equation}
and let $G_0$ denote the corresponding Green function. By definition we have
\begin{eqnarray}
\left (\triangle+\lambda \right )G^0_\lambda(x,y) & = & \delta(x-y) \\
\left ( \frac{\partial}{\partial n}+\mathcal{S}_0(x)\right )G^0_\lambda(x,y) & = &  0,\quad x\in \Sigma  \nonumber
\end{eqnarray}
Now let us perturb the boundary condition with a small correction
$$\mathcal{S}_0\rightarrow \mathcal{S}= \mathcal{S}_0+\epsilon \mathcal{S}\,'+ \mathcal{O}(\epsilon^2)$$
and denote with $\phi$ and $G$ the solution and the Green function for $\mathcal{S}$ respectively.\\
Then we take the spectral equation for $ G^0_\lambda$ and multiply it by $\phi$, and the one for $\phi$ we multiply by $ G^0_\lambda$. After subtracting the latter from the former and integrating over the $\Omega$ we get
\begin{equation} \phi(x) = \int_\Omega{\left [\phi(y)\triangle G^0_\lambda(x,y)- G^0_\lambda(x,y)\triangle \phi(y)\right ]}d\omega
\end{equation}
Applying the Green theorem to the above leads to
\begin{equation}  \label{phi}
\phi(x)= \epsilon\int_{\Sigma}{\mathcal{S}\,'\,G^0_\lambda\,\phi\,}d\sigma
\end{equation}
from which $\phi$ is determined. \\
In order to find $\lambda'$ we choose to work in the basis spanned by the eigenfunctions $\phi_{\,0}$ of the non-perturbed system. We expand in this basis
\begin{equation}\phi = \sum{c_\alpha \phi_{\,0}^\alpha} \label{eigen}\end{equation}
and take the usual spectral expansion of the Green function  \begin{equation}
G^0_\lambda(x,y)=\sum_\alpha{\frac{\bar{\phi}_{\, 0}^\alpha(x)\phi_{\, 0}^\alpha(y)}{\lambda - \lambda^0_k}}
\end{equation}
which, substituted in (\ref{phi}), together with (\ref{eigen}), leads to
\begin{equation} \label{main}
\sum_\mu{\phi_0^\mu\,c_\mu} = \epsilon\sum_{\mu,\nu}{\phi_0^\mu \frac{\mathcal{S}\,'_{\mu\nu}}{\lambda - \lambda^0_\mu}\, c_\nu}\qquad \forall \mu
\end{equation}
where $\mathcal{S}_{\mu\nu}'$ are the $\mathbb{L}_2(\Sigma)$ matrix entries of the boundary function derivative in the sense that
$$\mathcal{S}'_{\mu\nu} = \int_{\partial\Omega}{\mathcal{S}\,'(\sigma)\bar{\phi}_\mu(\sigma)\phi_\nu(\sigma)\,d\sigma} $$
Then (\ref{main}) turns into a system of homogenous matric equations (for all $\mu$) for the unknown coefficients $c_\nu$. More precisely
\begin{equation}
(\lambda-\lambda^0_\mu)c_\mu = \epsilon\sum{\mathcal{S}_{\mu\nu}'c_\nu}
\end{equation}
Now we restrict ourselves to the solution $\phi_\mu$ of the perturbed problem, corresponding to eigenvalue $\lambda_\mu$ . If we would like to solve the integral equation (\ref{phi}) with the method of consequent approximations, we need to substitute as a first 'guess' $\phi_0^i$ under the integral sign. Then, up to first order we have $c_\nu=0,\: \nu\neq \mu$ and certainly $c_\mu=1$. Then, this first approximation gives
$$\lambda_\mu - \lambda^0_\mu = \epsilon \mathcal{S}\,'_{\mu\mu}+ \mathcal{O}(\epsilon^2)$$
meaning that the spectrum is being shifted up to first order in $\epsilon$ namely by the diagonal element of the boundary function correction, exactly as in the standard stationary perturbation theory, only with the difference that this time the role of 'potential' is played by a single layer distribution instead of a volume term.\\
In more explicit terms, we have for the first correction to the Robin spectrum the expression
\begin{equation}
\dot{\lambda}_\nu=\int_{\Sigma}{\mathcal{S}\,'(\sigma)|\,\phi_0^\nu|^2\,d\sigma}
\end{equation}
The analogy to the old quantum mechanical perturbation-theory problem spreads even farther and allows for writing the higher-order terms as well. For instance one has
$$\lambda'' = \epsilon^2\sum_{i\neq j}{\frac{\mathcal{S}\,'_{\mu\nu}\mathcal{S}_{\nu\mu}'}{\lambda - \lambda^0_j - \mathcal{S}\,'_{\nu\nu}}}$$
and so forth. \\
From the above considerations it also becomes clear how to vary the trace of the wave propagator
$$ {\rm Tr}U(\tau) = {\rm Tr}{\,\exp{(i\tau\sqrt{\triangle})}} = \sum_\nu {e^{i\tau\sqrt{\lambda_\nu}}}$$
According to the Guillemin and Melrose's result, we have\footnote{note that the classical momenta $p=\cos{\theta}$ and thus the denominator $\sin\theta = \sqrt{1-p^2}$ are not altered by changing the boundary condition, nor is the integral measure}
\begin{equation}\label{WT} \delta{\sigma_L}_{|\tau = T} = \epsilon\int_{Y_T}{\frac{\mathcal{S}\,'(s)}{\sin{\theta(s)}}d\sigma_T}
\end{equation}
According to \cite{GM} the dependence of the wave trace on the boundary function appears only in the second term, so it is not sensitive to varying $\mathcal{S}$ until the third term. Or, in other words, the $\Psi DO$ corresponding to the variation, is of order $-2$.\\
\noindent We note once more that the results obtained by Zelditch in \cite{Zee} (see page 14) are slightly different - for instance he obtains a variation of the wave trace of the form\footnote{that obviously contradicts with our result}
$$\delta{\rm Tr}U(\tau) = i\tau\sum_\nu {\delta \lambda_\nu\, e^{i\tau\sqrt{\lambda_i}}}$$
and this is used (again in cooperation with the result of Guillemin and Melrose) as a mens of evaluation of \emph{boundary traces}\footnote{the definition of the boundary trace of a function in the context of a certain boundary-value problem is $f^b = (\mathbb{I} - \mathcal{B})f_{|\partial\Omega}$, where $\mathcal{B}$ is the corresponding to the condition boundary operator} such as
$$\sum_{\lambda_k\leq\lambda}|\phi^k_{_\Sigma}|^2 \sim C_n\lambda^{\frac{n}{2}} + \mathcal{O}(\lambda^\frac{n-1}{2})$$
for the $n$-dimensional case. Note that this power dependence suffers slight modification according to the present considerations. \\
Now let us see what happens if we formally differentiate directly the trace (\ref{WT}). According to the simple rule for composite functions we have
\begin{equation}
\frac{d}{d\epsilon}_{|\epsilon=0}{\rm Tr}U(\tau) = i\tau\sum_\nu \frac{\dot{\lambda}_\nu}{2\sqrt{\lambda_\nu}}\,{e^{i\tau\sqrt{\lambda_\nu}}}
\end{equation}
Substitution into the earlier obtained form of $\dot{\lambda}_\nu$ yields
\begin{equation}\label{express}
\frac{d}{d\epsilon}_{|\epsilon=0}{\rm Tr}U(\tau) = \frac{i\tau}{2}\int_{\partial{\Omega}}{\sum_\nu \frac{e^{i\tau\sqrt{\lambda_\nu}}}{\sqrt{\lambda_\nu}}
\mathcal{S}\,'(s)|\,\phi_0^\nu(s)|^2\,d\sigma}
\end{equation}
Note also that from the above expression one may easily find the first terms in
the power series expansion for the wave trace perturbation. Namely
\begin{equation} \label{two}
\frac{d}{d\epsilon}_{|\epsilon=0}{\rm Tr}U(\tau) = {i\tau}\zeta(1/2)-\frac{\tau^2}{2}\dot{\zeta}(-1)
-\frac{i\tau^3}{6}\sum{\dot{\lambda}_\nu\sqrt{\lambda_\nu}}
+\frac{\tau^4}{16}\dot{\zeta}(-2)+\ldots\nonumber
\end{equation}
Then using again the representation of the zeta function as a Mellin transform of the heat kernel, we obtain after cancelation of individual terms, an expression for the linear term
\begin{equation}
\frac{d}{d\epsilon}_{|\epsilon=0}{\rm Tr}U(\tau)\sim \frac{iL(\partial\Omega)} {4\pi^{3/2}}\int_{0}^\infty{dt\left ( 2\mathcal{S}' t^{-1/2} +\frac{1}{4}( \kappa \mathcal{S}' + 2 \mathcal{S}\mathcal{S}')+\mathcal{O}(\sqrt{t}) \right )},\quad \tau\rightarrow 0 \nonumber
\end{equation}
and may also fight for expressing higher-order terms in a similar manner.\\
To the contrary, one may express $\zeta$-invariants in terms of derivatives of wave trace coefficients.\\
Note also that the above formulae hold in the strict sense only away from the singularities, present due to periodic orbits, that is for $\tau\neq T_\gamma$. However, one would like to use it together with (\ref{WT}) in order to obtain a formal expression for the first derivative of $\sigma_L$.

\end{document}